\documentclass[12pt]{article}
\usepackage{amssymb}
\usepackage{amsfonts}
\usepackage{indentfirst}
\usepackage{amsopn}
\usepackage{amsmath}
\usepackage{amsthm}
\usepackage{amscd}
\usepackage{graphicx}
\usepackage[dvips]{epsfig}
\usepackage{color,xcolor}

\makeatletter\@addtoreset {equation}{section}\makeatother

\begin{document}
	
	\title{Rogue waves on the background of periodic standing waves in the derivative NLS equation}
	
	\author{Jinbing Chen$^1$ and Dmitry E Pelinovsky$^2$ \\
		{\small $^1$ School of Mathematics, Southeast University, Nanjing, Jiangsu 210096, P.R. China}\\
	{\small $^2$ Department of Mathematics, McMaster University, Hamilton, Ontario, Canada, L8S 4K1} }

	\maketitle

\begin{abstract}
	The derivative nonlinear Schr\"{o}dinger (DNLS) equation is the canonical model for dynamics of nonlinear waves in plasma physics and optics. We study exact solutions describing rogue waves on the background of periodic standing waves in the DNLS equation. We show that the space-time localization of a rogue wave is only possible if the periodic standing wave is modulationally unstable. If the periodic standing wave is modulationally stable, the rogue wave solutions degenerate into algebraic solitons propagating along the background and interacting with the periodic standing waves. Maximal amplitudes of rogue waves are found analytically and confirmed numerically.
\end{abstract}


\section{Introduction}

Fundamental models for dynamics of waves in fluids, plasmas, and optical systems
are written in terms of integrable systems such as the nonlinear Schr\"{o}dinger (NLS) equation, the derivative nonlinear Schr\"{o}dinger (DNLS) equation,
and their multi-component generalizations. These models simplify the complicated dynamics by accounting of only two mechanisms in the wave evolution: dispersion and nonlinearity.

The focusing NLS equation is the most studied model of this class. Periodic waves in complex physical systems are modeled by the constant-amplitude
waves of the NLS equation, which are known to be modulationally unstable \cite{ZakOst}.
Related to the modulational instability, rogue waves (localized waves of large amplitude appearing from nowhere and disappearing without a trace)
emerge on the constant-amplitude wave background. These rogue waves
are described by the exact solutions of the NLS equation \cite{Akh,Matveev,JYang,BoYang}.

In the context of the multi-component NLS models, it was discovered
in \cite{Wab1,Wab2} that rogue waves only emerge on the modulationally
unstable constant-amplitude wave background. If waves are modulationally
stable in a subset of the parameter region, no rogue wave solutions
can be constructed and numerical simulations do not show the occurrence of localized waves of large amplitude \cite{Wab3}.

The concept of {\em modulational instability} is introduced to describe
instability of the constant-amplitude wave with respect to
perturbations of increasingly large periods \cite{ZakOst}. If
the constant-amplitude wave
is unstable with respect to perturbations of smaller periods
but is stable with respect to perturbations of increasingly larger periods,
the constant-amplitude wave is said to be {\em modulationally stable},
even though it is still unstable in the time evolution of the
governing model \cite{BHJ}.
Note that the concept of modulational instability was dubbed in \cite{Wab1,Wab2} as ``baseband modulational instability".

Multi-periodic wave patterns in complex physical systems are modelled
by the periodic standing wave solutions of the NLS equation.
These periodic standing waves are also modulationally unstable \cite{DS}
and rogue waves on their background exist as exact solutions
of the NLS equation \cite{CPnls,Feng} and are observed
in the numerical simulations \cite{AZ} and optical and
hydrodynamical experiments \cite{XuKibler}.

A special relation between the modulational instability
of periodic standing waves
and the existence of rogue wave solutions was discovered in \cite{CPW}.
If the unstable spectral band intersects
the origin in the complex spectral plane tangentially to the imaginary axis,
the corresponding rogue wave solution degenerates into a propagating algebraic soliton on the periodic standing wave background.
Similarly, it was shown for the sine--Gordon equation \cite{PelWright}
that rogue waves are localized in space-time for modulationally
unstable librational waves but degenerate into propagating
algebraic solitons for modulationally stable rotational
waves (which are still unstable with respect to perturbations of
shorter periods).

{\em The purpose of this work} is to study rogue waves in the DNLS equation which models long weakly nonlinear Alfv\'{e}n waves propagating along the constant magnetic field
in cold plasmas \cite{Mio,M}. Rogue waves on the constant-amplitude wave background were constructed recently in \cite{YangYang}. Here we investigate properties of the rogue waves arising on the background of the periodic standing waves.

The precise characterization of modulational instability of periodic standing waves in the DNLS equation was completed in our previous work \cite{CPU},
where we implemented the algebraic method of nonlinearization of Lax equations
after the previous works in \cite{Qiao,Ma2,Zhou,Cao2,Chen}. Compared to the previous studies of periodic standing waves in the DNLS equation
in \cite{Kam1,Kam2,DU2,Hakaev}, the results of \cite{CPU} gave a full
picture of three possible types of the periodic standing waves.
The first type is modulationally unstable with two figure-eight unstable bands,
the second type is modulationally unstable with one figure-eight unstable band,
and the third type is modulationally stable (in fact, stable with respect
to perturbations of any period).

{\em The main outcomes of this work} are summarized as follows:

\begin{itemize}
	\item The periodic standing waves
	with two figure-eight unstable bands admits two rogue waves
	localized in space-time;
	\item The periodic standing waves
	with one figure-eight unstable band admits one rogue wave
	and two algebraic solitons propagating along its background;
	\item The periodic standing waves
	with no unstable bands admits four algebraic solitons
	and no rogue waves.
\end{itemize}

Our main results are in agreement with the previous observations
in \cite{Wab1,Wab2} and in \cite{CPW,PelWright} that the space-time localization of
rogue waves on the constant-amplitude and periodic standing wave background
is possible if and only if the background is modulationally unstable
(once again, unstable with respect to perturbations of increasingly long
period). There is currently no mathematical proof of this result
for a general integrable wave system.

The regular way to construct the rogue wave solutions on the background
of the periodic standing waves is to use the Darboux transformation.
Darboux transformations for the DNLS equation (\ref{dnls})
are well-known after the previous works in \cite{Imai,Steudel,Xu,GuoLingLiu}. Applications of Darboux transformation to generate breathers on the background of constant-amplitude solutions can be found in \cite{Geng2}. Darboux transformations are also useful to add solitons in the mathematical analysis of existence of solutions to the initial-value problem  \cite{PSS17}.

We use the Darboux transformation with the non-periodic solutions
of the Lax equations associated with the periodic standing waves 
of the DNLS equation. Although the algebraic method from
\cite{CPU} only gives periodic solutions of the Lax equations,
the non-periodic solutions can be explicitly characterized in terms
of integrals of the periodic standing wave solutions, as we show here.
Since the computations are long and technical, our presentation 
includes only main results and numerical visualizations, whereas computational details are given as appendices in the supplementary material. 

In addition to numerical visualization of rogue wave solutions
(either localized in space-time or propagating as algebraic solitons),
we also compute maximal amplitudes of the rogue waves.
These maximal amplitudes are important for future experiments with
the rogue waves on the periodic standing wave background. Some progress
on analysis of maximal amplitudes for more general quasi-periodic
solutions of the DNLS equation was recently obtained in \cite{WrightDNLS}.
Other recent studies of quasi-periodic solutions of the DNLS equation
can be found in \cite{Chen} and also in \cite{Geng1,ZhaoFan}.

The paper is organized as follows. Section \ref{sec-2} reviews
the construction of periodic standing waves based on our previous work \cite{CPU}. We show in Section \ref{sec-3} that Darboux transformation with
the periodic eigenfunctions of the Lax equations
transform the class of periodic standing waves into itself.
Section \ref{sec-4} reports on rogue waves obtained after
Darboux transformation with the non-periodic solutions of the Lax  equations. Section \ref{sec-5} concludes the paper with the discussion of
further directions.

\section{Periodic standing waves}
\label{sec-2}

We consider the DNLS equation in the following normalized form:
\begin{equation}
i \psi_t + \psi_{xx} + i (|\psi|^2 \psi)_x = 0,
\label{dnls}
\end{equation}
where $i = \sqrt{-1}$ and $\psi(x,t) : \mathbb{R}\times \mathbb{R}\mapsto \mathbb{C}$. As is shown in \cite{KN-1978}, the DNLS equation
is a compatibility condition $\phi_{xt} = \phi_{tx}$
of the following Lax equations
\begin{equation}
\label{lax-equations}
\phi_x = U(\psi,\lambda) \phi, \quad \phi_t = V(\psi,\lambda) \phi,
\end{equation}
where
\begin{equation}
\label{matrix-U}
U(\psi,\lambda) = \left(\begin{array}{cc}
-i \lambda^2 & \lambda \psi\\
-\lambda \bar{\psi} & i \lambda^2\\
\end{array}
\right)
\end{equation}
and
\begin{equation}
\label{matrix-V}
V(\psi,\lambda) = \left(\begin{array}{cc}
-2 i \lambda^4 + i \lambda^2 |\psi|^2 & 2 \lambda^3 \psi
+ \lambda (i \psi_x - |\psi|^2 \psi) \\
-2 \lambda^3 \bar{\psi} + \lambda (i \bar{\psi}_x + |\psi|^2 \bar{\psi}) & 2 i \lambda^4 - i \lambda^2 |\psi|^2\\
\end{array}
\right).
\end{equation}

The standing wave reduction of the DNLS equation (\ref{dnls}) takes the form
\begin{equation}
\label{trav-wave}
\psi(x,t) = u(x+2ct)e^{4ibt},
\end{equation}
where $b$ and $c$ are real parameters and $u$ satisfies the second-order equation:
\begin{equation}\label{2.9}
\frac{d^2 u}{d x^2} + i \frac{d}{dx} (|u|^2 u) + 2i c \frac{d u}{d x} - 4 b u = 0.
\end{equation}
The second-order equation (\ref{2.9}) is integrable with the following two first-order invariants:
\begin{equation}\label{2.22}
2i \left( \bar{u} \frac{d u}{dx} - u \frac{d \bar{u}}{dx} \right) - 3|u|^4
- 4c|u|^2 = 4a
\end{equation}
and
\begin{equation}\label{2.23}
2\left|\frac{d u}{dx}\right|^2 -|u|^6 -2c |u|^4 - 4(a+2b)|u|^2 = 8d,
\end{equation}
where $a$ and $d$ are real parameters. As shown in \cite{Kam1} (see also \cite{CPU}), the standing wave solutions of the differential equations
(\ref{2.9}), (\ref{2.22}) and (\ref{2.23})
are related to four pairs of roots of the following polynomial:
\begin{equation}
\label{polynomial}
P(\lambda) = \lambda^8 -2c \lambda^6 + \lambda^4 (a+2b+c^2) + \lambda^2\left[d-c(a+2b)\right] + b^2.
\end{equation}
In the remainder of this section, we review details in the construction
of periodic standing waves based on our previous work \cite{CPU}.

\subsection{Eigenvalues and eigenvectors of the Lax equations}

Roots of $P(\lambda)$ given by (\ref{polynomial})
determine eigenvalues of the Kaup--Newell (KN) spectral
problem
\begin{equation}\label{lax-1}
\varphi_x = U(u,\lambda) \varphi,\quad
\end{equation}
for either periodic or anti-periodic eigenvectors $\varphi = (p,q)^T$
related to the periodic potential $u$. Eigenvector $\varphi$ arises in the separation of variable
\begin{equation}
\label{trav-wave-eigenfunction}
\phi(x,t) = e^{2i b t \sigma_3} \varphi(x+2ct,t),
\end{equation}
where $\phi$ is a solution of the Lax equations (\ref{lax-equations})
associated with the standing wave $\psi$ of the DNLS equation (\ref{dnls})
given by (\ref{trav-wave}) and $\sigma_3 = {\rm diag}(1,-1)$. Therefore,
the eigenvector $\varphi$ of the KN spectral problem (\ref{lax-1})
also satisfies the time-evolution equation
\begin{equation}\label{lax-2}
\varphi_t + 2 c \varphi_x + 2 i b \sigma_3 \varphi = V(u,\lambda) \varphi.
\end{equation}

Roots of $P(\lambda)$ arise either as
complex quadruplets or double real pairs or purely imaginary pairs \cite{CPU}:
\begin{itemize}
	\item If $P(\lambda)$ admits a quadruplet of complex roots $\{ \lambda_1,\bar{\lambda}_1,-\lambda_1,-\bar{\lambda}_1\}$, then
	the standing wave $u$ and the eigenvector $\varphi = (p_1,q_1)^T$ for the eigenvalue $\lambda_1$ are related by
\begin{equation}\label{constraint-red-2}
u = \lambda_1 p_1^2 + \bar{\lambda}_1 \bar{q}_1^2.
\end{equation}

\item If $P(\lambda)$ admits a pair of double real roots $\{ \lambda_1, -\lambda_1 \}$ with two linearly independent eigenvectors $\varphi = (p_1,q_1)^T$ and
$\varphi = (\bar{q}_1,-\bar{p}_1)^T$ for the eigenvalue $\lambda_1$, then the same expression (\ref{constraint-red-2}) holds.

\item If $P(\lambda)$ admits two pairs of purely imaginary eigenvalues $\{ i \beta_1, -i \beta_1, i \beta_2, -i\beta_2\}$, then
the standing wave $u$ and the eigenvectors $\varphi = (p_1,-i\bar{p}_1)^T$ and $\varphi = (p_2,-i\bar{p}_2)$ for the eigenvalues $\lambda_1 = i \beta_1$ and $\lambda_2 = i \beta_2$ are related by
\begin{equation}\label{constraint-red-1}
u = i \beta_1 p_1^2 + i \beta_2 p_2^2.
\end{equation}
\end{itemize}

\subsection{Complex Hamiltonian systems}

Eigenvectors $\varphi = (p_1,q_1)^T$ and $\varphi = (p_2,q_2)^T$
of the KN spectral problem (\ref{lax-1}) with eigenvalues
$\lambda_1$ and $\lambda_2$ satisfy a finite-dimensional complex Hamiltonian
system, which is integrable \cite{Chen}. This complex Hamiltonian system is equivalent to the Lax equation
\begin{equation}\label{2.11}
\frac{d}{dx} \Psi = [\mathcal{U},\Psi],
\end{equation}
where $\mathcal{U}$ is obtained from $U(u,\lambda)$ in (\ref{matrix-U}) with $u$ given by either  (\ref{constraint-red-2}) or (\ref{constraint-red-1}).
The $2$-by-$2$ Lax matrix $\Psi$ is given by
\begin{equation}\label{2.12}
\Psi := \left(\begin{array}{cc}
\Psi_{11} & \Psi_{12} \\
\Psi_{21} & -\Psi_{11}
\end{array} \right),
\end{equation}
where
\begin{eqnarray*}
\Psi_{11} &=& - i - \frac{\lambda_1^2 p_1 q_1}{\lambda^2 - \lambda_1^2} -
\frac{\lambda_2^2 p_2 q_2} {\lambda^2- \lambda_2^2}, \label{2.13}\\
\Psi_{12} &=& \lambda\left(
\frac{\lambda_1 p_1^2}{\lambda^2 - \lambda_1^2} + \frac{\lambda_2 p_2^2} {\lambda^2-\lambda_2^2}
\right), \label{2.14}\\
\Psi_{21} &=& -\lambda\left(
\frac{\lambda_1 q_1^2}{\lambda^2 - \lambda_1^2} + \frac{\lambda_2 q_2^2} {\lambda^2-\lambda_2^2}
\right). \label{2.15}
\end{eqnarray*}
Entries of the Lax matrix $\Psi$ can be rewritten in terms of $u$ by
\begin{eqnarray*}
\label{2.16}
\Psi_{11} &=& \frac{-i}{(\lambda^2 - \lambda_1^2)(\lambda^2-\lambda_2^2)} \left[\lambda^4 - \lambda^2 \left(c+ \frac12\left|u\right|^2\right) +b
\right], \\
\label{2.17}
\Psi_{12} &=& \frac{\lambda}{(\lambda^2 - \lambda_1^2)(\lambda^2-\lambda_2^2)}
\left[\lambda^2 u + \frac{i}{2} \frac{d u}{d x}-\frac{1}{2} u\left| u\right|^2 -c u \right], \\
\Psi_{21} &=& \frac{-\lambda}{(\lambda^2 - \lambda_1^2)(\lambda^2-\lambda_2^2)}
\left[\lambda^2 \bar{u} - \frac{i}{2} \frac{d \bar{u}}{d x}-\frac{1}{2} \bar{u} \left| u\right|^2 - c \bar{u} \right], \label{2.18}
\end{eqnarray*}
so that
\begin{equation}\label{2.20}
\det \Psi = - \Psi_{11}^2 - \Psi_{12} \Psi_{21} = \frac{P(\lambda)}{(\lambda^2 - \lambda_1^2)^2(\lambda^2-\lambda_2^2)^2},
\end{equation}
where $P(\lambda)$ is given by (\ref{polynomial}).
Eigenvalues $\lambda_1$ and $\lambda_2$ arising in the poles of $\Psi$ must be chosen from the roots of $P(\lambda)$.

Let the four pairs of roots of the polynomial $P(\lambda)$ in (\ref{polynomial}) be denoted by $\{ \pm \lambda_1, \pm\lambda_2, \pm\lambda_3, \pm \lambda_4\}$. The polynomial can be factorized by its roots:
\begin{equation}\label{3.22}
P(\lambda) = (\lambda^2  - \lambda_1^2) (\lambda^2-\lambda_2^2) (\lambda^2  - \lambda_3^2) (\lambda^2-\lambda_4^2).
\end{equation}
Comparison of (\ref{polynomial}) and (\ref{3.22}) yields the relations
\begin{equation}\label{3.23}
\left\{\begin{array}{l}
\lambda_1^2+\lambda_2^2+\lambda_3^2+ \lambda_4^2=2c,\\
(\lambda_1^2+\lambda_2^2)(\lambda_3^2+ \lambda_4^2)+\lambda_1^2 \lambda_2^2 +\lambda_3^2 \lambda_4^2=a+2b+c^2,\\
\lambda_1^2 \lambda_2^2 (\lambda_3^2+ \lambda_4^2) +
\lambda_3^2\lambda_4^2 (\lambda_1^2+\lambda_2^2)=ac+2bc-d,\\
\lambda_1^2 \lambda_2^2 \lambda_3^2 \lambda_4^2 =b^2,
\end{array}
\right.
\end{equation}
which allow to express parameter $(a,b,c,d)$ in terms of $\{ \lambda_1, \lambda_2, \lambda_3, \lambda_4\}$ explicitly:
\begin{equation}\label{3.25}
\left\{
\begin{array}{lll}
a &=& -\frac{1}{4} [ (\lambda_1 + \lambda_2)^2 - (\lambda_3 + \lambda_4)^2]
[ (\lambda_1 - \lambda_2)^2 - (\lambda_3 - \lambda_4)^2],\\
b&=& \lambda_1 \lambda_2 \lambda_3 \lambda_4, \\
c&=& \frac12 (\lambda_1^2+\lambda_2^2+\lambda_3^2+\lambda_4^2),\\
d&=& -\frac 18 (\lambda_1^2+\lambda_2^2 -\lambda_3^2-\lambda_4^2)
(\lambda_1^2-\lambda_2^2+\lambda_3^2-\lambda_4^2)
(\lambda_1^2-\lambda_2^2-\lambda_3^2+\lambda_4^2).
\end{array}\right.
\end{equation}

\subsection{Characterization of the periodic standing waves}

Explicit solutions for the periodic standing waves satisfying
(\ref{2.9}), (\ref{2.22}), and (\ref{2.23}) are obtained after using the polar form $u(x)=R(x) e^{i\theta(x)}$ with
\begin{equation}
\label{3.5}
\frac{d \theta}{d x} =-\frac{a}{R^2}  -\frac 34 R^2 - c
\end{equation}
and
\begin{equation}\label{3.6}
\left(\frac{d R}{d x}\right)^2+ \frac{a^2}{R^2} + \frac{1}{16} R^6 + \frac{c}{2} R^4
+ R^2 \left(c^2 - 4b -\frac{a}{2}\right) + 2ac - 4d = 0.
\end{equation}
The transformation
$\rho = \frac{1}{2} R^2$ brings (\ref{3.6}) to the form
\begin{equation}\label{3.8}
\left(\frac{d \rho}{d x}\right)^2 + Q(\rho) = 0,
\end{equation}
where
\begin{equation}
\label{quartic-Q}
Q(\rho) = \rho^4 +4 c \rho^3 +2(2c^2 -a -8b)\rho^2 +4(ac-2d)\rho +a^2.
\end{equation}
Let the four roots of the polynomial $Q(\rho)$
be denoted by $\{ u_1, u_2, u_3, u_4 \}$ so that
\begin{equation}
Q(\rho) = (\rho - u_1) (\rho - u_2) (\rho - u_3) (\rho - u_4).
\label{factorization-Q}
\end{equation}
Comparison of (\ref{quartic-Q}) with (\ref{factorization-Q}) yields
\begin{equation}\label{3.10}\left\{\begin{array}{l}
u_1+u_2+u_3+u_4=- 4c,\\
u_1u_2+u_1u_3+u_1u_4+u_2u_3+u_2u_4+u_3u_4=2(2c^2 -a -8b),\\
u_1u_2u_3+u_1u_2u_4+u_1u_3u_4+u_2u_3u_4=4(2d- ac),\\
u_1u_2u_3u_4=a^2.\\
\end{array}
\right.
\end{equation}
The roots $\{u_1,u_2,u_3,u_4\}$ of $Q$ are related to the roots
$\{ \pm \lambda_1, \pm \lambda_2, \pm \lambda_3, \pm \lambda_4 \}$
of $P$ due to (\ref{3.25}) and (\ref{3.10}).
It was shown in \cite{Kam1} and more recently in \cite{CPU}
that the relations are expressed by
\begin{equation}\label{3.26}
\left\{\begin{array}{lll}
u_1&=&-\frac12 (\lambda_1-\lambda_2+\lambda_3-\lambda_4)^2,\\
u_2&=&-\frac12 (\lambda_1-\lambda_2-\lambda_3+\lambda_4)^2,\\
u_3&=&-\frac12 (\lambda_1+\lambda_2-\lambda_3-\lambda_4)^2,\\
u_4&=&-\frac12 (\lambda_1+\lambda_2+\lambda_3+\lambda_4)^2.
\end{array}\right.
\end{equation}
Two families of periodic standing waves
are obtained from the quadrature (\ref{3.8}) with (\ref{factorization-Q}).

\subsubsection{Four roots of $Q$ are real}

Assume the following ordering for the four real roots of $Q$:
\begin{equation}
\label{order-roots}
u_4\leq u_3 \leq u_2 \leq u_1.
\end{equation}
Periodic solutions to the quadrature (\ref{3.8})
are expressed explicitly (see, e.g., \cite{CPgardner}) by
\begin{equation}\label{3.14}
\rho(x) = u_4 +
\frac{(u_1-u_4)(u_2-u_4)}{(u_2-u_4)+(u_1-u_2){\rm sn}^2 (\nu x;k)},
\end{equation}
where positive parameters $\nu$ and $k$ are uniquely expressed by
\begin{equation}\label{3.13}
\nu = \frac{1}{2} \sqrt{(u_1-u_3)(u_2-u_4)}, \quad
k = \frac{\sqrt{(u_1-u_2)(u_3-u_4)}}{\sqrt{(u_1-u_3)(u_2-u_4)}}.
\end{equation}
The periodic solution $\rho$ is located in the interval $[u_2,u_1]$ and has the period $L = 2 K(k) \nu^{-1}$.
The solution (\ref{3.14}) with (\ref{3.13}) is meaningful for $\rho = \frac{1}{2} R^2 \geq 0$ in two cases:
\begin{equation}
\label{configuration-1}
\lambda_1 = \bar{\lambda}_3 = \alpha_1 + i \beta_1, \qquad
\lambda_2 = \bar{\lambda}_4 = \alpha_2 + i \beta_2
\end{equation}
and
\begin{equation}
\label{configuration-2}
\lambda_1 = i \beta_1, \qquad
\lambda_2 = i \beta_2, \qquad
\lambda_3 = i \beta_3, \qquad
\lambda_4 = i \beta_4.
\end{equation}
The roots ordered as (\ref{order-roots}) satisfy the more precise ordering
\begin{equation}
\label{order-roots-precise}
u_4 \leq u_3 \leq 0 \leq u_2 \leq u_1
\end{equation}
in the case of (\ref{configuration-1}) and
\begin{equation}
\label{order-roots-precise-1}
0 \leq u_4 \leq u_3  \leq u_2 \leq u_1
\end{equation}
in the case of (\ref{configuration-2}). Note that in the case
of (\ref{order-roots-precise-1}), another periodic solution is obtained
from (\ref{3.14}) by exchanging $u_1$ with $u_3$ and $u_2$ with $u_4$:
\begin{equation}\label{3.14-another}
\rho(x) = u_2 -
\frac{(u_2-u_3)(u_2-u_4)}{(u_2-u_4)-(u_3-u_4){\rm sn}^2 (\nu x;k)},
\end{equation}
where the values of $\nu$ and $k$ are the same as in (\ref{3.13}).

In the case of (\ref{configuration-1}),
if $\alpha_1,\alpha_2,\beta_1,\beta_2$ are all positive, it follows from (\ref{3.26}) that
\begin{equation}
\label{3.27}
\left\{ \begin{array}{l}
\alpha_1 = \frac{1}{2\sqrt{2}}(\sqrt{-u_4} + \sqrt{-u_3}), \\
\alpha_2 = \frac{1}{2\sqrt{2}}(\sqrt{-u_4} - \sqrt{-u_3}), \end{array}
\right.
\qquad
\left\{ \begin{array}{l}
\beta_1 = \frac{1}{2\sqrt{2}}(\sqrt{u_1} + \sqrt{u_2}), \\
\beta_2 = \frac{1}{2\sqrt{2}}(\sqrt{u_1} - \sqrt{u_2}), \end{array}
\right.
\end{equation}
so that $0 \leq \alpha_2 \leq \alpha_1$ and $0 \leq \beta_2 \leq \beta_1$.

In the case of (\ref{configuration-2}),
it follows from (\ref{3.26}) that
\begin{equation}
\label{3.27a}
\left\{ \begin{array}{l}
\beta_1 = \frac{1}{2\sqrt{2}}(\sqrt{u_1} + \sqrt{u_2} + \sqrt{u_3} + \sqrt{u_4}), \\
\beta_2 = \frac{1}{2\sqrt{2}}(- \sqrt{u_1} - \sqrt{u_2}
+ \sqrt{u_3} + \sqrt{u_4}), \\
\beta_3 = \frac{1}{2\sqrt{2}}(\sqrt{u_1} - \sqrt{u_2} - \sqrt{u_3}
+ \sqrt{u_4}), \\
\beta_4 = \frac{1}{2\sqrt{2}}(- \sqrt{u_1} + \sqrt{u_2} - \sqrt{u_3}
+ \sqrt{u_4}),\end{array}
\right.
\end{equation}
so that $\beta_2 \leq \beta_4 \leq \beta_3 \leq \beta_1$.

\subsubsection{Two roots of $Q$ are real and one pair of roots is complex-conjugate}

Assume that the two real roots are ordered as $u_2\leq u_1$ and
the complex-conjugate roots are given by $u_{3,4} = \gamma\pm i\eta$
so that
\begin{equation}
\label{order-roots-complex}
0 \leq u_2 \leq u_1, \quad u_3 = \gamma + i \eta, \quad u_4= \gamma - i \eta.
\end{equation}
Periodic solutions to the quadrature (\ref{3.8})
are expressed explicitly (see, e.g., \cite{CPgardner}) by
\begin{equation}\label{3.20}
\rho(x)= u_1 + \frac{(u_2-u_1)(1-{\rm cn} (\mu x;k))}{1+\delta +(\delta-1){\rm cn} (\mu x;k)},
\end{equation}
where positive parameters $\delta$, $\mu$, and $k$ are uniquely expressed by
\begin{equation}\label{3.18}
\delta = \frac{\sqrt{(u_2-\gamma)^2 + \eta^2}}{\sqrt{(u_1-\gamma)^2 + \eta^2}},\qquad
\mu = \sqrt[4]{\left[(u_1 - \gamma)^2+\eta^2 \right]\left[(u_2- \gamma)^2+\eta^2 \right]},
\end{equation}
and
\begin{equation}\label{3.18-1}
2 k^2 = 1- \frac{(u_1 - \gamma)(u_2 - \gamma)+ \eta^2}{\sqrt{\left[(u_1 - \gamma)^2+\eta^2 \right]\left[(u_2- \gamma)^2+\eta^2 \right]}}.
\end{equation}
The periodic solution $\rho$ is located in the interval $[u_2,u_1]$ and has the period $L = 4 K(k) \mu^{-1}$.
The periodic solution (\ref{3.20}) with (\ref{3.18}) and (\ref{3.18-1}) is meaningful for $\rho = \frac{1}{2} R^2 \geq 0$ if
\begin{equation}
\label{configuration-3}
\lambda_1 = \bar{\lambda}_2 = \alpha_1 + i \beta_1, \qquad
\lambda_3 = i \beta_3, \qquad
\lambda_4 = i \beta_4
\end{equation}
with the following relations
\begin{equation}
\label{a3.16}
\left\{ \begin{array}{l}
\alpha_1 = \frac{1}{2} \sqrt{\sqrt{\gamma^2 + \eta^2}-\gamma}, \\
\beta_1 = \frac{1}{2\sqrt{2}}(\sqrt{u_1} + \sqrt{u_2}), \end{array}
\right.
\qquad
\left\{ \begin{array}{l}
\beta_3 = \frac{\eta}{2\sqrt{\sqrt{\gamma^2 + \eta^2}-\gamma}} + \frac{1}{2\sqrt{2}}(\sqrt{u_1} - \sqrt{u_2}), \\
\beta_4 = \frac{\eta}{2\sqrt{\sqrt{\gamma^2 + \eta^2}-\gamma}} - \frac{1}{2\sqrt{2}}(\sqrt{u_1} - \sqrt{u_2}).
\end{array}
\right.
\end{equation}
where $\alpha_1 \geq 0$, $\beta_1 \geq 0$, and $\beta_4 \leq \beta_3$.

\section{Darboux transformations}
\label{sec-3}

Darboux transformations for the DNLS equation (\ref{dnls}) were
constructed previously in \cite{Imai,Steudel,Xu,GuoLingLiu}. Here we use the exact formulas for the one-fold and two-fold Darboux transformations.
Validity of the transformation formulas is verified in Appendices \ref{appendix-a} and \ref{appendix-b}.

Let $\psi$ be a solution of the DNLS equation (\ref{dnls}) and let $\phi$ be a solution of the Lax equations (\ref{lax-equations}) for the potential $\psi$ with a fixed value of $\lambda$. If $\psi$ is the standing wave solution in the form (\ref{trav-wave}), then the solution $\phi$ can be expressed in the form (\ref{trav-wave-eigenfunction}). We denote
$\varphi = (p_1,q_1)^T$ for $\lambda_1$ and $\varphi = (p_2,q_2)^T$ for $\lambda_2$.

Darboux transformations generate new solutions
to the DNLS equation (\ref{dnls}) in the form
\begin{equation}
\label{new-solution}
\hat{\psi}(x,t) = \hat{u}(x+2ct,t) e^{4ibt}.
\end{equation}
The following three basic Darboux transformations will be used.

\begin{itemize}
\item If $\lambda_1 = i \beta_1$ and $q_1 = -i \bar{p}_1$, then the
new solution is given by
\begin{equation}\label{5.2}
\hat{u} = -\frac{\bar{p}_1^2}{p_1^2} \left[ u + 2 i \beta_1 \frac{p_1}{\bar{p}_1} \right] e^{-8ibt}.
\end{equation}

\item If $\lambda_{1,2} = i \beta_{1,2}$ and $q_{1,2} = -i \bar{p}_{1,2}$, then the new solution is given by
\begin{equation}\label{5.8a}
\hat{u} = \left( \frac{\beta_1 \bar{p}_1 p_2 -  \beta_2 \bar{p}_2 p_1}{\beta_1 \bar{p}_2 p_1 - \beta_2 \bar{p}_1 p_2} \right)^2 \left( u + \frac {2 i (\beta_1^2 - \beta_2^2) p_1 p_2} {\beta_1 \bar{p}_1 p_2 -  \beta_2 \bar{p}_2 p_1}\right).
\end{equation}

\item If $\lambda_1 \in \mathbb{C}$ with ${\rm Re}(\lambda_1), {\rm Im}(\lambda_1) \neq 0$, then the new solution is given by
\begin{equation}\label{5.8}
\hat{u} = \left(\frac{\bar{\lambda}_1|p_1|^2 + \lambda_1|q_1|^2}{\lambda_1 |p_1|^2 + \bar{\lambda}_1 |q_1|^2}\right)^2 \left[ u -
\frac{2i (\lambda_1^2 - \bar{\lambda}_1^2) p_1\bar{q}_1}
{\bar{\lambda}_1 |p_1|^2 + \lambda_1 |q_1|^2} \right].
\end{equation}
\end{itemize}

Here we consider eigenvectors of the KN spectral problem (\ref{lax-1}) found from the complex Hamiltonian system (\ref{2.11}). We show that the Darboux transformations with these eigenvectors produce new solutions of the DNLS equation which are translated versions of the same periodic standing waves.

Comparing expressions for $\Psi_{12}$ and $\Psi_{21}$ yields the following relations for the squared components of the eigenvectors:
\begin{equation}\label{4.1}
\left\{\begin{array}{l}
\lambda_1 p_1^2= \frac{1}{\lambda_1^2 - \lambda_2^2}\left[\frac{1}{2} \left( i \frac{du}{dx} -  |u|^2 u \right) + (\lambda_1^2-c) u\right],\vspace{2mm}\\
\lambda_1 q_1^2= \frac{1}{\lambda_1^2 - \lambda_2^2} \left[-\frac{1}{2} \left( i \frac{d \bar{u}}{dx} +  |u|^2 \bar{u} \right) + (\lambda_1^2-c) \bar{u} \right],
\end{array}
\right.
\end{equation}
whereas comparing expressions for $\Psi_{11}$ yields
\begin{equation}\label{4.1a}
\lambda_1^2 p_1 q_1= \frac{i}{\lambda_1^2 - \lambda_2^2}\left[
b - c \lambda_1^2 + \lambda_1^4 - \frac12 \lambda_1^2 |u|^2 \right].
\end{equation}
By using the polar form decompositions
\begin{equation}\label{4.2}
u(x) = R(x) e^{i\theta(x)}, \quad
p_1(x) = P_1(x) e^{\frac{i}{2}\theta(x)}, \quad
q_1(x) = Q_1(x) e^{-\frac{i}{2}\theta(x)}
\end{equation}
and the phase equation (\ref{3.5}), we can rewrite relations (\ref{4.1}) and (\ref{4.1a}) in the form:
\begin{equation}\label{4.4}
\left\{\begin{array}{l}
\lambda_1 P_1^2= \frac{1}{\lambda_1^2- \lambda_2^2}\left[\frac{i}{2} \frac{dR}{dx} - \frac{1}{8} R^3 + \frac{a}{2R}
+ \left(\lambda_1^2- \frac{c}{2}\right) R \right],\vspace{2mm}\\
\lambda_1 Q_1^2= \frac{1}{\lambda_1^2- \lambda_2^2} \left[-\frac{i}{2} \frac{dR}{dx} - \frac{1}{8} R^3 + \frac{a}{2R}
+ \left(\lambda_1^2- \frac{c}{2}\right) R \right],
\end{array}
\right.
\end{equation}
and
\begin{equation}\label{4.4a}
\lambda_1^2 P_1 Q_1= \frac{i}{\lambda_1^2- \lambda_2^2}\left[
b - c \lambda_1^2 + \lambda_1^4 - \frac12 \lambda_1^2 R^2 \right].
\end{equation}
The periodic standing waves are given by either (\ref{3.14}) or (\ref{3.20})
for $\rho := \frac{1}{2} R^2$. Depending on parameters
$(a,b,c,d)$, roots of $P(\lambda)$ include either pairs of 
purely imaginary eigenvalues or complex quadruplets.

\subsection{One-fold transformation (\ref{5.2}) for the periodic wave (\ref{3.14})}

Let us consider the case (\ref{configuration-2}) for the periodic wave (\ref{3.14}) with $0 \leq u_4 \leq u_3 \leq u_2 \leq u_1$.
Four pairs of purely imaginary eigenvalues exist. Without loss of
generality, we pick one eigenvalue $\lambda_1 = i \beta_1$.
It is shown in Appendix \ref{app-C} that the new solution can be expressed 
in the form $\hat{\rho} := \frac{1}{2} |\hat{u}|^2$ with 
\begin{eqnarray}
\hat{\rho} = \frac{\beta_1^2 a (b - c \beta_1^2 - \beta_1^4) + 2 d \beta_1^4 + (b^2 + \beta_1^4 (2b + a - c^2) + \beta_1^8) \rho + \beta_1^2 (b - c\beta_1^2 - \beta_1^4) \rho^2}{(b + c \beta_1^2 + \beta_1^4 + \beta_1^2 \rho)^2}. \label{b7.6}
\end{eqnarray}
For the periodic wave (\ref{3.14}), this expression reduces to the form:
\begin{equation}
\hat{\rho}(x)  = v_1 -
\frac{(v_1 - v_3) (v_1 - v_4)}{(v_1 - v_3) + (v_3 - v_4) {\rm sn}^2 (\nu x;k)},
\label{hat-rho}
\end{equation}
where the values of $\nu$ and $k$ are the same as in (\ref{3.13})
and the new roots $0 \leq v_4 \leq v_3 \leq v_2 \leq v_1$ are given by
\begin{equation}
\label{roots-v1-v4}
\left\{ \begin{array}{l}
v_1 = \frac{1}{4}(\sqrt{u_1} + \sqrt{u_2} + \sqrt{u_3} - \sqrt{u_4})^2, \\
v_2 = \frac{1}{4}(\sqrt{u_1} + \sqrt{u_2} - \sqrt{u_3} + \sqrt{u_4})^2, \\
v_3 = \frac{1}{4}(\sqrt{u_1} - \sqrt{u_2} + \sqrt{u_3} + \sqrt{u_4})^2, \\
v_4 = \frac{1}{4}(- \sqrt{u_1} + \sqrt{u_2} + \sqrt{u_3} + \sqrt{u_4})^2.
\end{array}
\right.
\end{equation}
The new solution (\ref{hat-rho}) is obtained from the periodic solution (\ref{3.14}) after $u_{1,2,3,4}$ are replaced by $v_{1,2,3,4}$
and the transformation $v_1 \leftrightarrow v_4$ and $v_2 \leftrightarrow v_3$ 
is used. The new periodic solution has four pairs of purely imaginary eigenvalues
$\{ \pm i \tilde{\beta}_1, \pm i \tilde{\beta}_2, \pm i \tilde{\beta}_3, \pm i \tilde{\beta}_4 \}$ related to $v_{1,2,3,4}$ by (\ref{3.27a}) after
the transformations above. It is easy to verify that the location of the purely imaginary eigenvalues is invariant under the transformation (\ref{roots-v1-v4}) 
with $\tilde{\beta}_1 = \beta_1$ and $\tilde{\beta}_{2,3,4} = -\beta_{2,3,4}$ 
if $2 \sqrt{v_4} = -\sqrt{u_1} + \sqrt{u_2} + \sqrt{u_3} + \sqrt{u_4} \geq 0$.

The new periodic solution (\ref{hat-rho}) satisfies the same differential equation (\ref{3.8}) with (\ref{quartic-Q}) having parameters
$\tilde{a}$, $\tilde{b}$, $\tilde{c}$, and $\tilde{d}$ related to eigenvalues $\{ \pm i \tilde{\beta}_1, \pm i \tilde{\beta}_2, \pm i \tilde{\beta}_3, \pm i \tilde{\beta}_4 \}$ by (\ref{3.25})
and to turning points $v_{1,2,3,4}$ by (\ref{3.10}).
Since $\tilde{\beta}_1 = \beta_1$ and $\tilde{\beta}_{2,3,4} = -\beta_{2,3,4}$,
it follows from (\ref{3.25}) and (\ref{3.10})
that
\begin{equation}
\label{transformation-abcd}
\tilde{d} = d, \quad \tilde{c} = c, \quad
\tilde{b} = -b, \quad \mbox{\rm and} \quad \tilde{a} = a + 4b.
\end{equation}
Thus, the one-fold Darboux transformation (\ref{5.2}) transforms one periodic solution (\ref{3.14}) of the differential equation (\ref{3.8}) with given parameters $(a,b,c,d)$ to
another periodic solution (\ref{hat-rho}) 
of the same equation with different parameters
$(a+4b,-b,c,d)$. The new and old solutions are related to the same four pairs of purely imaginary eigenvalues $\{ \pm i \beta_1, \pm i \beta_2, \pm i \beta_3, \pm i \beta_4 \}$. Note that the transformation $\tilde{b} = -b$ also follows
from comparison of (\ref{new-solution}) and (\ref{5.2}).

\subsection{Two-fold transformation (\ref{5.8a}) for the periodic wave (\ref{3.14})}

Let us now pick two eigenvalues $\lambda_{1,2} = i \beta_{1,2}$ in
the case (\ref{configuration-2}) with $0 \leq u_4 \leq u_3 \leq u_2 \leq u_1$.
The new solution (\ref{5.8a}) can be reduced after long symbolic computations to the form $\hat{\rho} := \frac{1}{2} |\hat{u}|^2$ with 
\begin{equation}
\hat{\rho}(x) = u_3 + \frac{(u_1 - u_3)(u_2 - u_3)}{(u_1 - u_3) - (u_1 - u_2) {\rm sn}^2 (\nu x;k)}, \label{b3.30}
\end{equation}
The expression
(\ref{b3.30}) is obtained from the expression (\ref{3.14}) by means of the transformation $u_1\leftrightarrow u_2$ and $u_3 \leftrightarrow u_4$. This transformation generates the same periodic wave (\ref{3.14}) but translated by half-period:
\begin{eqnarray*}
	\hat{\rho} \left(x + K(k) \nu^{-1}\right) &=& \frac {u_2 (u_1-u_3) - u_3(u_1 - u_2) {\rm sn}^2(\nu x + K(k);k)} {(u_1-u_3) - (u_1 - u_2) {\rm sn}^2 (\nu x + K(k);k)} \\
	&=& \frac {u_1 (u_2 - u_4) + u_4 (u_1 -u_2)
		{\rm sn}^2 (\nu x;k)} {(u_2 - u_4) + (u_1 -u_2) {\rm sn}^2 (\nu x;k)} = \rho(x),
\end{eqnarray*}
where we have used formulas
$$
{\rm sn}(x+K(k);k) = \frac{{\rm cn}(x;k)}{{\rm dn}(x;k)}, \quad
{\rm dn}^2(x;k) = 1 - k^2 {\rm sn}^2(x;k), \quad
{\rm cn}^2(x;k) = 1 - {\rm sn}^2(x;k)
$$
together with the definition of $k$ in (\ref{3.13}).

This recurrence of the periodic solution (\ref{3.14}) after the two-fold transformation (\ref{5.8a}) can be explained as follows.
It follows from (\ref{roots-v1-v4}) that
$$
\left\{ \begin{array}{l}
\sqrt{v_1} + \sqrt{v_2} = \sqrt{u_1} + \sqrt{u_2}, \\
\sqrt{v_1} - \sqrt{v_2} = \sqrt{u_3} - \sqrt{u_4}, \\
\sqrt{v_3} + \sqrt{v_4} = \sqrt{u_3} + \sqrt{u_4}, \\
\sqrt{v_3} - \sqrt{v_4} = \sqrt{u_1} - \sqrt{u_2}. \end{array} \right.
$$
A composition of two transformations (\ref{roots-v1-v4})
restore the original roots $u_{1,2,3,4}$:
\begin{equation}
\label{roots-inverse-u1-u4}
\left\{ \begin{array}{l}
u_1 = \frac{1}{4}(\sqrt{v_1} + \sqrt{v_2} + \sqrt{v_3} - \sqrt{v_4})^2, \\
u_2 = \frac{1}{4}(\sqrt{v_1} + \sqrt{v_2} - \sqrt{v_3} + \sqrt{v_4})^2, \\
u_3 = \frac{1}{4}(\sqrt{v_1} - \sqrt{v_2} + \sqrt{v_3} + \sqrt{v_4})^2, \\
u_4 = \frac{1}{4}(- \sqrt{v_1} + \sqrt{v_2} + \sqrt{v_3} + \sqrt{v_4})^2,
\end{array}
\right.
\end{equation}
Similarly, parameters $(a,b,c,d)$ are invariant after the composition
of two transformations (\ref{transformation-abcd}):
\begin{equation}
\label{transformation-composition}
(a,b,c,d) \mapsto (a + 4b,-b, c, d) \mapsto (a + 4b + 4(-b),-(-b),c,d) = (a,b,c,d).
\end{equation}
As a result, the new solution (\ref{b3.30}) satisfies the quadrature (\ref{3.8}) with the same values of parameters $(a,b,c,d)$ as in (\ref{quartic-Q}).

\subsection{Two-fold transformation (\ref{5.8}) for the periodic wave (\ref{3.14})}
	
Let us consider the case (\ref{configuration-1}) for the periodic wave (\ref{3.14}) with $u_4 \leq u_3 \leq 0 \leq u_2 \leq u_1$.
Two complex quadruplets exist. We pick one eigenvalue $\lambda_1$ from the two quadruplets. It is shown in Appendix \ref{app-E} that the new solution (\ref{5.8}) can be written in the form $\hat{\rho} := \frac{1}{2} |\hat{u}|^2$ with 
\begin{equation}
\hat{\rho}(x) = u_3 + \frac {(u_1-u_3) (u_2-u_3)} {(u_1-u_3) - (u_1 - u_2) {\rm sn}^2 (\nu x;k)},
\label{5.10}
\end{equation}
which is the same as (\ref{b3.30}). Thus, the two-fold transformation (\ref{5.8}) with the complex quadruplet produces the same result as 
the two-fold transformation (\ref{5.8a})
with two purely imaginary eigenvalues.

\subsection{One-fold transformation (\ref{5.2}) for the periodic wave (\ref{3.20})}

Let us consider the case (\ref{configuration-3}) for the periodic wave (\ref{3.20}) with $0 \leq u_2 \leq u_1$ and $u_{3,4} = \gamma \pm i \eta$.
Two pairs of purely imaginary eigenvalues exist and a quadruplet of
complex eigenvalues. Without loss of
generality, we pick one eigenvalue $\lambda_3 = i \beta_3$.
It is shown in Appendix \ref{app-CC} that the new solution 
(\ref{5.2}) can be written in the form $\hat{\rho} := \frac{1}{2} |\hat{u}|^2$ with 
\begin{equation}\label{3.3.7}
\hat{\rho}(x) = \tilde{v}_2 + \frac {(\tilde{v}_1 - \tilde{v}_2)(1 - {\rm cn} (\mu x;k))}
{1 + \tilde{\delta} + (\tilde{\delta}-1) {\rm cn} (\mu x;k)},
\end{equation}
where the new roots $0\leq \tilde{v}_2 \leq \tilde{v}_1$ and $\tilde{v}_{3,4} = \tilde{\gamma} \pm i \tilde{\eta}$ are given by
\begin{equation}\label{3.3.6}
\left\{\begin{array}{lll}
\tilde{v}_1 &=& \frac14 \left(\sqrt{u_1} + \sqrt{u_2} + \sqrt{2(\sqrt{\gamma^2 + \eta^2} + \gamma)}\right)^2,\\
\tilde{v}_2 &=& \frac14 \left(\sqrt{u_1} + \sqrt{u_2} - \sqrt{2(\sqrt{\gamma^2 + \eta^2} + \gamma)}\right)^2,\\
\tilde{v}_3 &=& \frac14 \left(\sqrt{u_1} - \sqrt{u_2} + i \sqrt{2(\sqrt{\gamma^2 + \eta^2} - \gamma)}\right)^2,\\
\tilde{v}_4 &=& \frac14 \left(\sqrt{u_1} - \sqrt{u_2} - i \sqrt{2(\sqrt{\gamma^2 + \eta^2} - \gamma)}\right)^2,\\
\end{array}
\right.
\end{equation}
the values of $\mu$ and $k$ are the same as in \eqref{3.18} and \eqref{3.18-1},
and
\begin{eqnarray}
\label{3.3.6a}
\tilde{\delta} = \frac{\sqrt{(\tilde{v}_1- \tilde{v}_3)(\tilde{v}_1 - \tilde{v}_4)}}
{\sqrt{(\tilde{v}_2 - \tilde{v}_3)(\tilde{v}_2 - \tilde{v}_4)}}.
\end{eqnarray}
Note that the new periodic solution \eqref{3.3.7} coincides with the periodic solution (\ref{3.20}) after $u_{1,2,3,4}$ are replaced by $\tilde{v}_{1,2,3,4}$ and $\tilde{v}_1$ is exchanged with $\tilde{v}_2$. We have also confirmed 
the validity of the transformation (\ref{transformation-abcd})
for the case of (\ref{3.3.7}). Thus, the one-fold Darboux transformation \eqref{5.2} transforms the periodic solution (\ref{3.20}) of the differential equation \eqref{3.8} with given parameters $(a,b,c,d)$ to another solution (\ref{3.3.7}) 
of the same equation with different parameters ($a +4b$,$-b$,$c$,$d$).

\subsection{Two-fold transformation (\ref{5.8a}) for the periodic wave (\ref{3.20})}

Let us pick now two purely imaginary eigenvalues $\lambda_{3,4} = i \beta_{3,4}$ in the case of \eqref{configuration-3}. The new solution (\ref{5.8a}) can be reduced after long symbolic computations to the form $\hat{\rho} := \frac{1}{2} |\hat{u}|^2$ with 
\begin{eqnarray}
\hat{\rho}(x) = u_2 + \frac{(u_1- u_2)(1 - {\rm cn} (\mu x;k))}{(1 - {\rm cn} (\mu x;k)) + \delta^{-1} (1 + {\rm cn} (\mu x;k))}, \label{3.5.1}
\end{eqnarray}
which is the periodic wave (\ref{3.20}) after the transformation $u_1\leftrightarrow u_2$. The latter transformation yields the same periodic wave (\ref{3.20}) but translated by half-period:
\begin{eqnarray*}
	\hat{\rho} \left(x + 2K(k) \mu^{-1}\right) &=& u_2 + \frac {(u_1 - u_2) (1 + {\rm cn} (\mu x;k))} {1 + {\rm cn} (\mu x;k) + \delta^{-1} (1 - {\rm cn} (\mu x;k))} \\
	&=& u_1 + \frac{(u_2-u_1)(1-{\rm cn} (\mu x;k))}{1+\delta +(\delta-1){\rm cn} (\mu x;k)}=\rho(x),
\end{eqnarray*}
where we have used formulas ${\rm cn}(x+2K(k);k) = - {\rm cn} (x;k)$.
Thus, the two-fold transformation (\ref{5.8a}) applied to the periodic solution (\ref{3.20}) produces a translation of the same periodic solution (\ref{3.20}).
This is explained again by the fact that the composition
of two transformations (\ref{transformation-abcd}) in  (\ref{transformation-composition}) returns the original parameters
$(a,b,c,d)$ of the quadrature (\ref{3.8}).

\subsection{Two-fold transformation (\ref{5.8}) for the periodic wave (\ref{3.20})}

Finally, we pick eigenvalue $\lambda_1$ from the complex quadruplet 
in the case (\ref{configuration-3}). It is shown in Appendix \ref{app-F} that 
the new solution (\ref{5.8}) can be written in the form $\hat{\rho} := \frac{1}{2} |\hat{u}|^2$ with 
\begin{equation}
\label{a7.5}
\hat{\rho} = u_2 + \frac {(u_1 - u_2) (1 - {\rm cn} (\mu x;k))} {1 - {\rm cn} (\mu x;k) + \delta^{-1} (1 + {\rm cn} (\mu x;k))},
\end{equation}
which is the same as (\ref{3.5.1}). Thus, the two-fold transformation (\ref{5.8}) with the complex eigenvalue produces again the same outcome as the two-fold transformation (\ref{5.8a}) with two purely imaginary eigenvalues.

\section{Rogue wave solutions}
\label{sec-4}

Here we construct rogue wave solutions to the DNLS equation (\ref{dnls}) by using
Darboux transformations with the second solution of the Lax equations
(\ref{lax-equations}) for the same eigenvalues given by roots of
the polynomial $P(\lambda)$. The first solution of the Lax equations
are periodic for these eigenvalues, whereas the second solution is generally
non-periodic. We use transformations (\ref{5.2}) and (\ref{5.8a})
if $P(\lambda)$ admits pairs of  purely imaginary roots
and transformation (\ref{5.8}) if $P(\lambda)$ admits
quadruplets of complex roots.

It is known from \cite{CPU} that two pairs of purely imaginary roots of
$P(\lambda)$ are related to the stable spectrum in the linearization
of the DNLS equation (\ref{dnls}) at the periodic standing wave solution (\ref{trav-wave}), whereas a quadruplet of complex roots of $P(\lambda)$ is
related to the modulationally unstable spectrum.
In full agreement with the modulational stability analysis, we show that
the new solutions related to two pairs of purely imaginary roots
describe algebraic solitons propagating
on the background of the periodic standing wave, whereas
the new solutions related to a quadruplet of complex eigenvalues describe rogue waves  localized in space and time on the background of the periodic standing wave.

From a technical point of view, we construct the second solution to the
Lax equations differently for the purely imaginary roots
and for the complex roots. Similar differences were previously
discovered for the periodic travelling waves in the sine--Gordon equation \cite{PelWright}.

\subsection{Periodic wave (\ref{3.14}) with $\lambda_1$ being purely imaginary}

Let $\lambda_1 = i \beta_1 \in i \mathbb{R}$ be an eigenvalue of the KN spectral
problem (\ref{lax-1}). We use the decomposition (\ref{trav-wave}) and (\ref{trav-wave-eigenfunction}) with the eigenvector $\varphi=(p_1,q_1)^\mathrm{T}$,
where $q_1 = -i \bar{p}_1$ and $p_1$ satisfies the linear system
\begin{eqnarray}
\label{lax-combined-1}
\frac{\partial p_1}{\partial x} = i \beta_1^2 p_1 + \beta_1 u \bar{p}_1, \end{eqnarray}
and
\begin{eqnarray}
 \frac{\partial p_1}{\partial t}
+ 2c \frac{\partial p_1}{\partial x} + 2 i b p_1 = -i (2 \beta_1^4 + \beta_1^2 |u|^2) p_1
- 2 \beta_1^3 u \bar{p}_1 + \beta_1 (i u_x - |u|^2 u) \bar{p}_1.
\label{lax-combined-2}
\end{eqnarray}
This system follows from the Lax equations (\ref{lax-1}) and (\ref{lax-2})
due to the reduction $q_1 = -i \bar{p}_1$
for $\lambda_1 = i \beta_1$. The second, linearly independent solution $\varphi=(\hat{p}_1,\hat{q}_1)^\mathrm{T}$
of the system (\ref{lax-combined-1}) and (\ref{lax-combined-2}) can be written in the form:
\begin{equation}\label{4.5}
\hat{p}_1= p_1\chi_1 - \frac{1}{2 q_1}, \qquad
\hat{q}_1= q_1\chi_1 + \frac{1}{2 p_1},
\end{equation}
where $\chi_1$ is assumed to be a real-valued function of $x$ and $t$. Wronskian between the two solutions is normalized by $p_1\hat{q}_1-\hat{p}_1 q_1=1$.
If $q_1 = -i \bar{p}_1$ and $\chi_1$ is real, then $\hat{q}_1 = -i \bar{\hat{p}}_1$.

Substituting (\ref{4.5}) into
(\ref{lax-combined-1}) and (\ref{lax-combined-2}) written for $\hat{p}_1$
and using the same equations for $p_1$ yields the following equations for $\chi_1$:
\begin{equation}\label{4.6}
\frac{\partial \chi_1}{\partial x} = \frac{i \beta_1}
{2 |p_1|^4} \left( u \bar{p}_1^2 - \bar{u} p_1^2 \right)
\end{equation}
and
\begin{eqnarray}
\frac{\partial \chi_1}{\partial t} + 2c \frac{\partial \chi_1}{\partial x} = \frac{i \beta_1}
{2 |p_1|^4} \left( \bar{u} p_1^2 - u \bar{p}_1^2 \right) (|u|^2 + 2 \beta_1^2)
- \frac{\beta_1}{2|p_1|^4} (\bar{u}_x p_1^2 + u_x \bar{p}_1^2).
\label{4.7}
\end{eqnarray}
In particular, we confirm that the function $\chi_1(x,t)$ is real.

By using the decomposition (\ref{4.2}) and the representations (\ref{4.4}) and (\ref{4.4a}), we deduce from (\ref{4.6}) that
\begin{equation}\label{4.6a}
\frac{\partial \chi_1}{\partial x} = \frac{\beta_1^4 (\beta_1^2-\beta_2^2) (a - 2(c+2\beta_1^2) \rho - \rho^2)}
{2(b + c \beta_1^2 + \beta_1^4 + \beta_1^2 \rho)^2},
\end{equation}
where we substituted $\lambda_2 = i \beta_2$ for another purely imaginary eigenvalue. Similarly, we deduce from (\ref{4.7}) that
\begin{eqnarray}
\frac{\partial \chi_1}{\partial t} = -\frac{2 \beta_1^4 (\beta_1^2-\beta_2^2) (a (c+\beta_1^2) - d - 2(c \beta_1^2 + \beta_1^4 + b) \rho -\beta_1^2 \rho^2)}
{(b + c \beta_1^2 + \beta_1^4 + \beta_1^2 \rho)^2},
\label{4.7a}
\end{eqnarray}
where we have used (\ref{3.5}), (\ref{3.8}), and (\ref{quartic-Q})
in order to express $u_x$. Equations (\ref{4.6a}) and (\ref{4.7a}) are compatible with $\chi_{1xt} = \chi_{1tx}$ if and only if the right-hand side of (\ref{4.7a}) is constant because $\rho$ depends on $x$ only. It is shown in Appendix \ref{app-I} that substituting  (\ref{3.25}), \eqref{3.14}, and (\ref{3.27a}) into (\ref{4.7a}) yields the following simple equation:
\begin{equation}\label{4.7b}
\frac{\partial \chi_1}{\partial t} = 2 \beta_1^2 (\beta_1^2 - \beta_2^2),
\end{equation}
from which we obtain
\begin{equation}
\label{chi}
\chi_1(x,t) = c_1 + k_1 x + f(x) + 2 \beta_1^2 (\beta_1^2-\beta_2^2) t,
\end{equation}
where $c_1 \in \mathbb{R}$ is an arbitrary constant of integration,
$$
k_1 = \frac{\nu \beta_1^4 (\beta_1^2-\beta_2^2)}{4K(k)} \int_0^{2K(k) \nu^{-1}} \frac{ (a - 2(c+2\beta_1^2) \rho - \rho^2)}
{(b + c \beta_1^2 + \beta_1^4 + \beta_1^2 \rho)^2} dx
$$
is the mean value of $\frac{\partial \chi_1}{\partial x}$
over the period $L = 2 \nu^{-1} K(k)$ of
the periodic wave $\rho$, and $f$ is the $L$-periodic function with the zero mean. The function $\chi_1$ remains bounded on the $(x,t)$ plane along the line
\begin{equation}
\label{direction-soliton}
k_1 (x + 2ct) + 2 \beta_1^2 (\beta_1^2-\beta_2^2) t = 0,
\end{equation}
where we have recalled the transformation
(\ref{trav-wave-eigenfunction}). The function
$\chi_1(x,t)$ grows linearly in $|x| + |t| \to \infty$ in 
the direction transversal to the line (\ref{direction-soliton}).

Recall that the eigenvector $\varphi = (p_1,q_1)^T$ defines
the transformed periodic wave in the form $\psi_{\rm tr}(x,t) =
u_{\rm tr}(x+2ct)e^{4ib t}$ with
\begin{equation}\label{1foldDT}
u_{\rm tr} = -\frac{\bar{p}_1^2}{p_1^2} \left[ u + 2 i \beta_1 \frac{p_1}{\bar{p}_1} \right] e^{-8ibt}.
\end{equation}
By using the second solution $\varphi = (\hat{p}_1,\hat{q}_1)^T$ given by
(\ref{4.5}), we define a new solution to the DNLS equation in the form
$\hat{\psi}(x,t) = \hat{u}(x+2ct)e^{4ib t}$ with
\begin{equation}\label{1foldDT-rogue}
\hat{u} = -\frac{\bar{\hat{p}}_1^2}{\hat{p}_1^2} \left[ u + 2 i \beta_1 \frac{\hat{p}_1}{\bar{\hat{p}}_1} \right] e^{-8ibt}.
\end{equation}

In order to illustrate the two solutions $u_{\rm tr}$ and $\hat{u}$, 
we consider the periodic standing wave (\ref{3.14})
with the particular choice of
$$
u_1 = 2, \quad u_2 = 1, \quad u_3 = 0.5, \quad u_4 = 0.
$$
This choice corresponds to parameters
$$
a = 0, \quad b = -\frac{7}{256}, \quad c = -\frac{7}{8}, \quad
d = \frac{1}{8}
$$ 
in the quadrature (\ref{3.8}) with (\ref{quartic-Q}). 
In particular, parameters satisfy $c^2 - 4b > 0$ \cite{CPU}.

Two periodic waves $\rho$ exist for the same values of parameters: the sign-definite wave in $[u_2,u_1]$ is given by (\ref{3.14})
and the sign-indefinite wave in $[u_4,u_3]$ is given by (\ref{3.14-another}). The sign-definite wave has the period $L = 2 \nu^{-1} K(k)$, whereas the sign-indefinite wave has the double period $L = 4 \nu^{-1} K(k)$. The sign-indefinite wave is smoothly represented in the original variable $R$ in the form:
\begin{equation}
\label{sign-indefinite}
R(x) = \frac{\sqrt{2 u_2 u_3} \; {\rm cn}(\nu x;k)}{\sqrt{u_2-u_3 {\rm sn}^2(\nu x;k)}}.
\end{equation}

\begin{figure}[htb!]
	\includegraphics[width=0.48\textwidth]{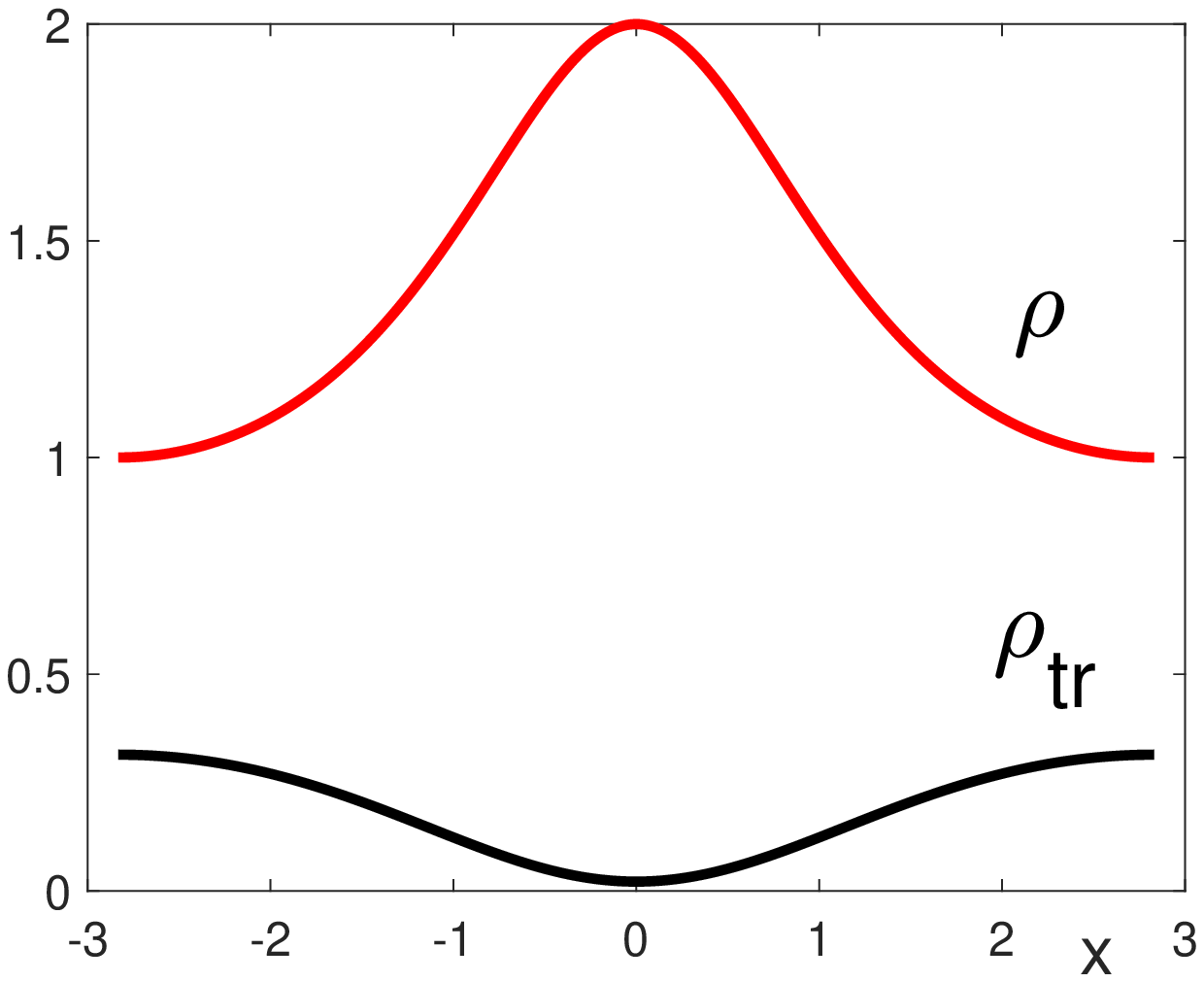}
	\includegraphics[width=0.48\textwidth]{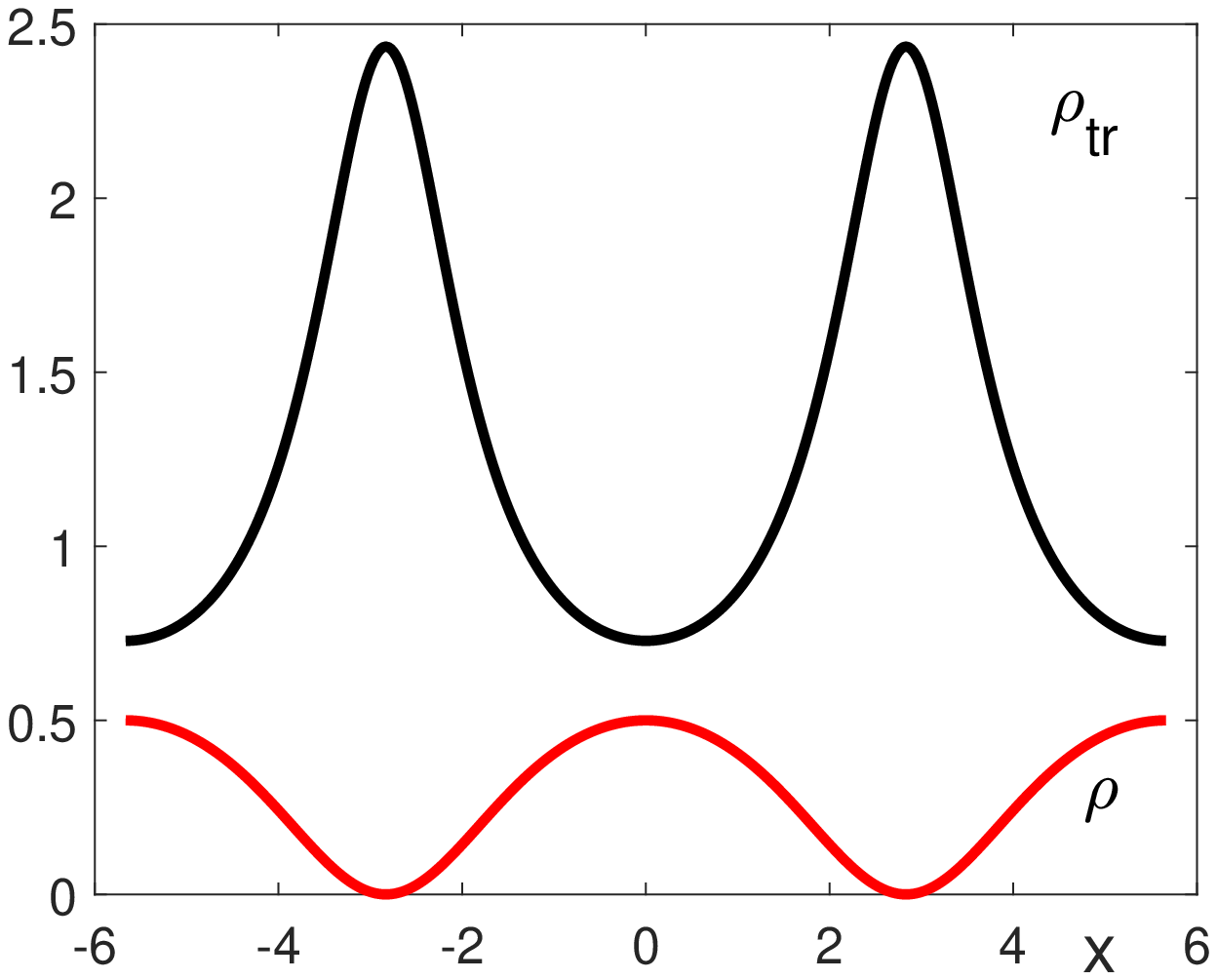}
	\caption{The periodic standing wave $\rho$ (red) and its transformed
		version $\rho_{\rm tr}$ (black) versus $x$. Left and right panels correspond to (\ref{3.14}) and (\ref{sign-indefinite}).} \label{fig-alg1}
\end{figure}

Figure \ref{fig-alg1} shows the plot of $\rho$
and $\rho_{\rm tr} := \frac{1}{2} |u_{\rm tr}|^2$
versus $x$. The left panel of Figure \ref{fig-alg1} shows transformation of the sign-definite wave (\ref{3.14}) and the right panel shows the same for the sign-indefinite wave (\ref{sign-indefinite}).

As is explained in Section \ref{sec-3}, the transformed wave $\rho_{\rm tr}$
in (\ref{1foldDT}) is different from a translation of the original wave $\rho$, in particular, $\rho$ has turning points $u_{1,2,3,4}$ but the transformed wave
$\rho_{\rm tr}$ has turning points $v_{1,2,3,4}$ given by (\ref{roots-v1-v4}).
For the sign-definite wave (left panel on Fig. \ref{fig-alg1}),
$\rho$ changes between $u_2 = 1$ and $u_1 = 2$, whereas
$\rho_{\rm tr}$ changes between
$v_4 \approx 0.02$ and $v_3 \approx 0.31$.
For the sign-indefinite wave (right panel on Fig. \ref{fig-alg1}),
$\rho$ changes between $u_4 = 0$ and $u_3 = 0.5$, whereas
$\rho_{\rm tr}$ changes between
$v_2 \approx 0.73$ and $v_1 \approx 2.44$.

\begin{figure}[htb!]
	\includegraphics[width=0.48\textwidth]{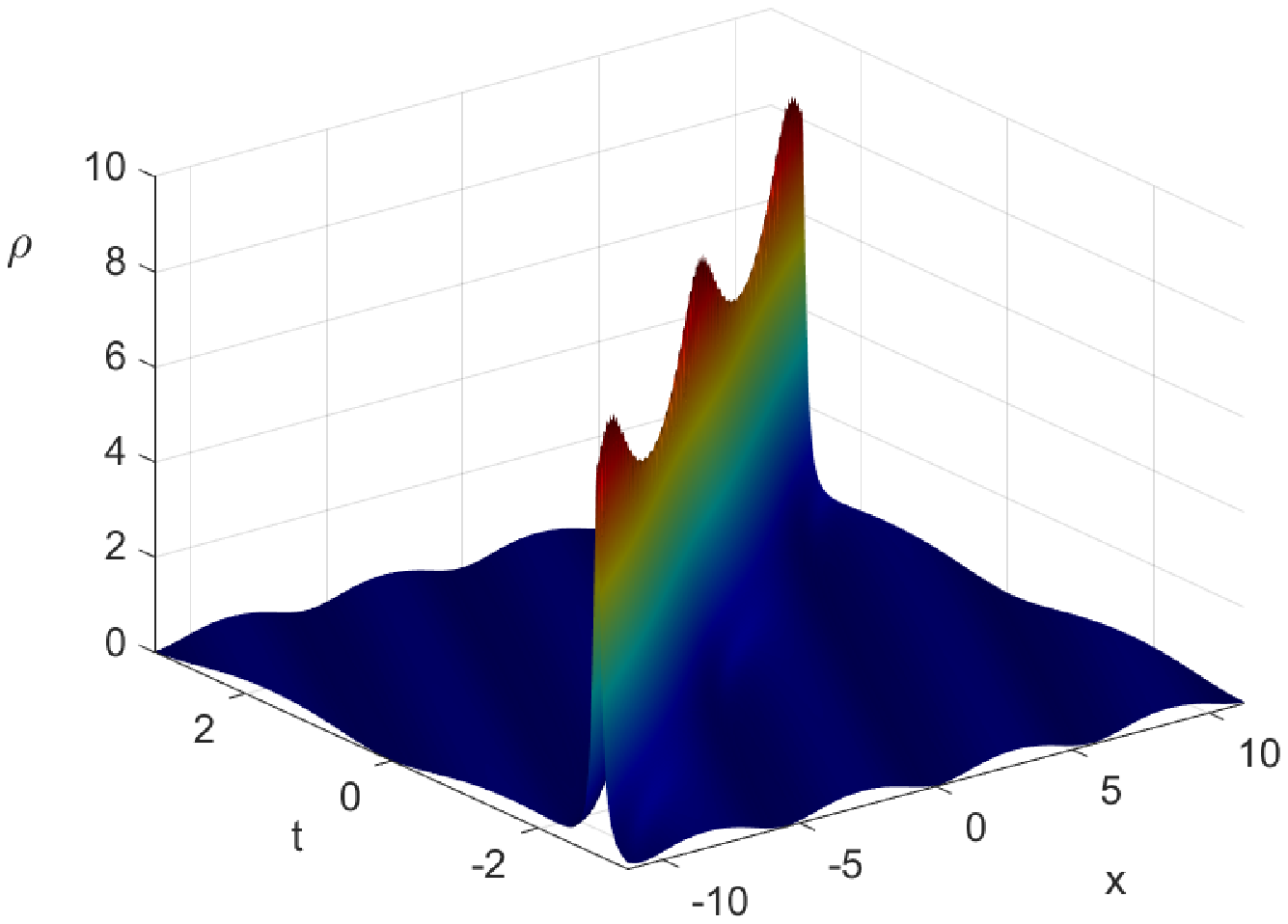}
	\includegraphics[width=0.48\textwidth]{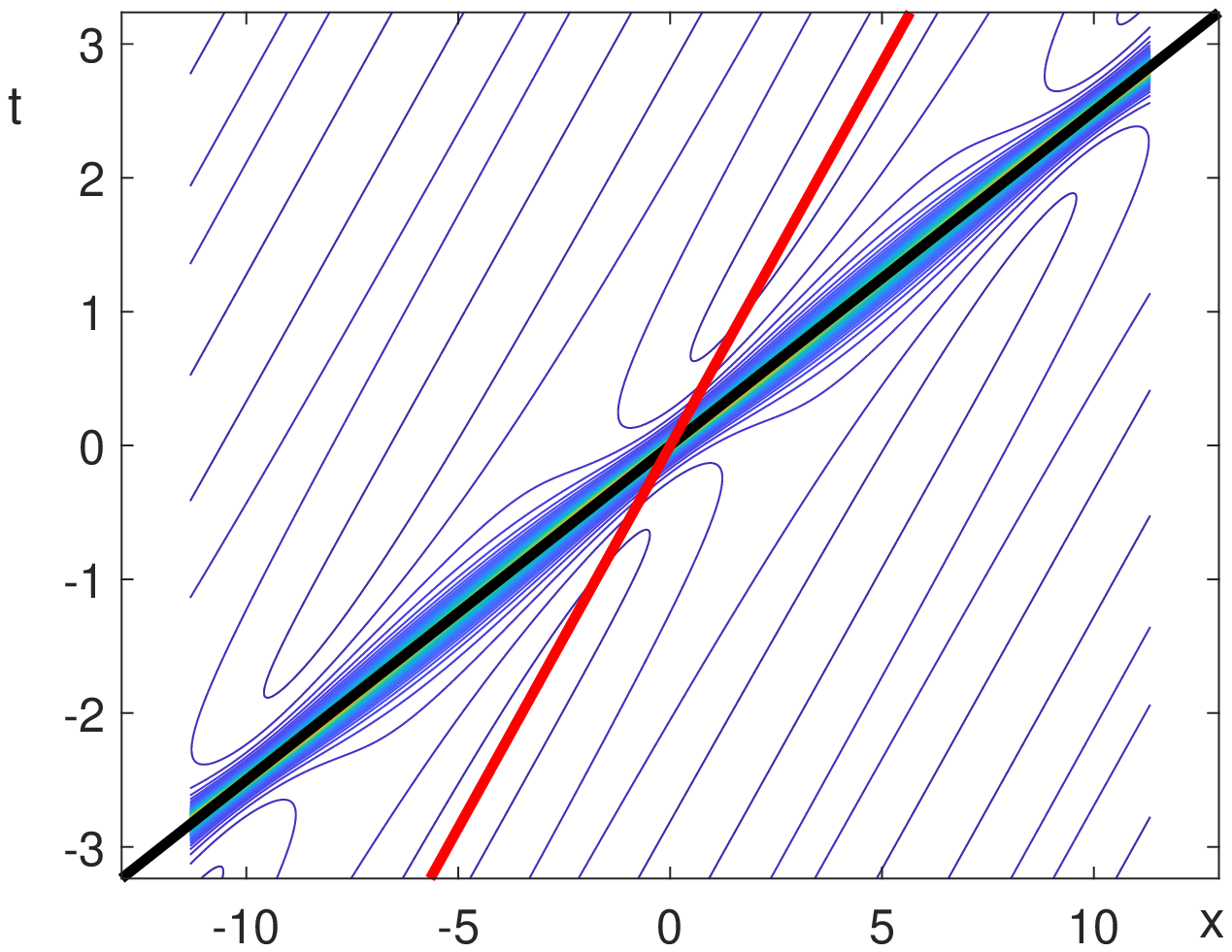} \\
	\includegraphics[width=0.48\textwidth]{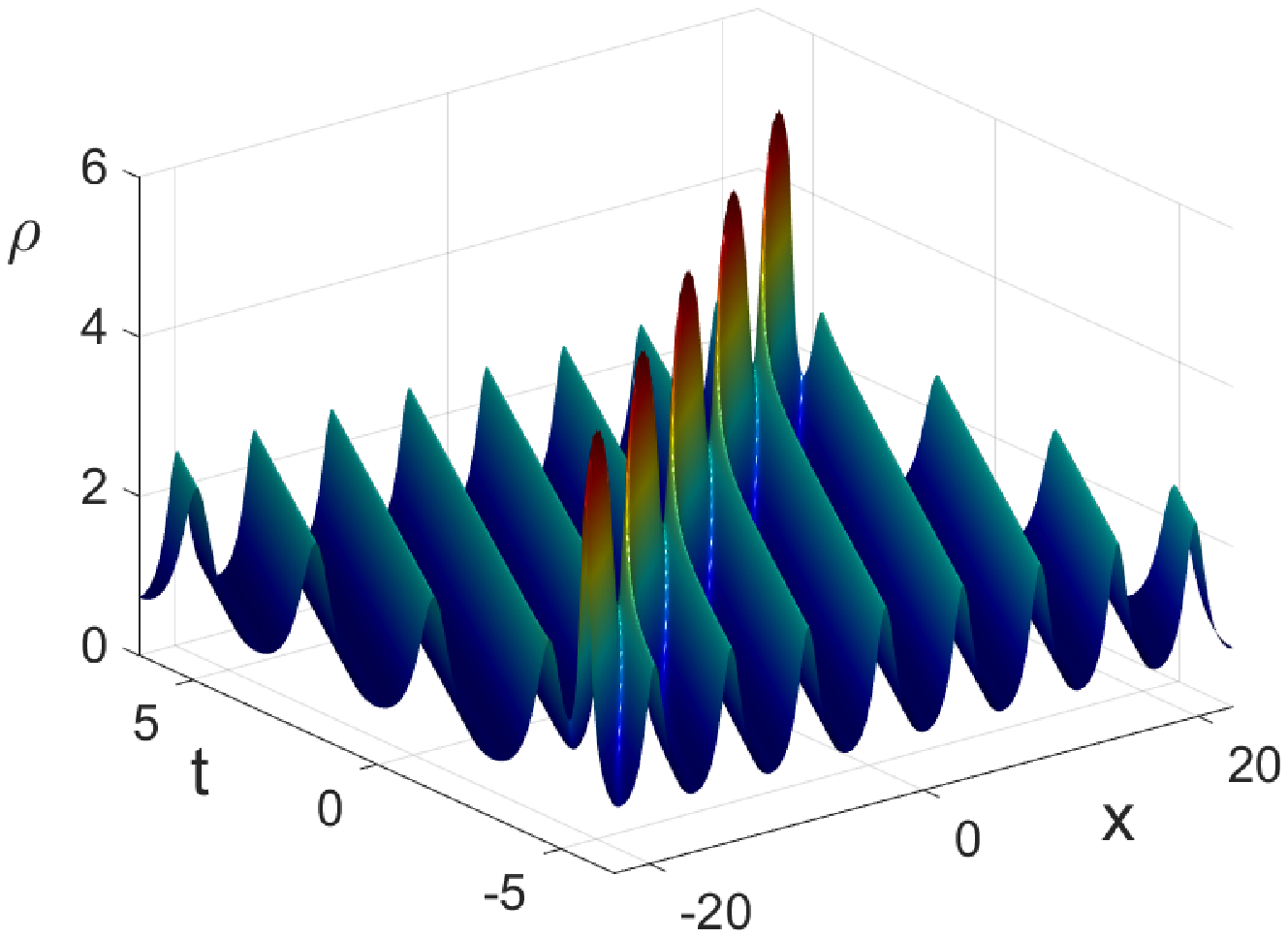}
	\includegraphics[width=0.48\textwidth]{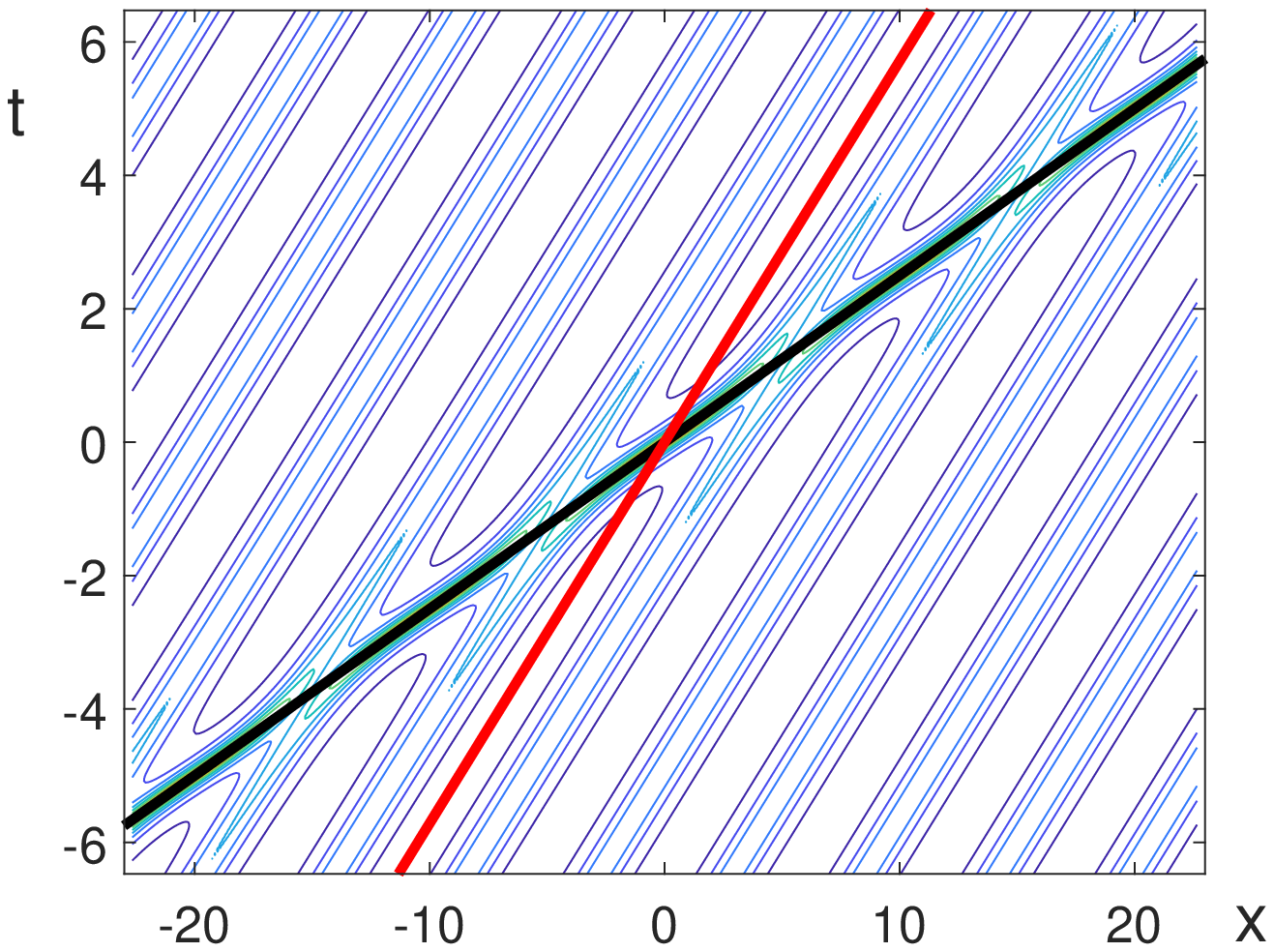}
	\caption{New solutions to the DNLS equation in variable $\hat{\rho}$
		which describe propagation of an algebraic soliton
		on the background of the periodic standing wave: solution surface
		(left) and contour plot (right).
		Top and bottom panels correspond to (\ref{3.14}) and (\ref{sign-indefinite}) respectively.} \label{fig-alg2}
\end{figure}

Figure \ref{fig-alg2} shows the surface plot of $\hat{\rho} := \frac{1}{2} |\hat{u}|^2$ on the left panel and the contour plot
on the right panel. We always use the choice $c_1 = 0$ in (\ref{chi}).
The red line on the contour plot shows the line $x + 2ct = 0$,
whereas the black line shows the line (\ref{direction-soliton}).

The new solution on Figure \ref{fig-alg2} describes propagation of the algebraic soliton along the direction (\ref{direction-soliton})
on the background of the periodic standing wave propagating along
the line $x + 2ct = 0$. The background periodic wave is given
in the limit $|\chi| \to \infty$, where
\begin{equation}
\label{background-wave}
\lim_{|\chi_1| \to \infty} \hat{u} = -\frac{\bar{p}_1^2}{p_1^2} \left[ u + 2 i \beta_1 \frac{p_1}{\bar{p}_1} \right] e^{-8ibt} = u_{\rm tr}
\end{equation}
which coincides with (\ref{1foldDT}). The maximum of the algebraic soliton is located at $\chi_1= 0$, where
\begin{equation}
\label{algebraic-wave}
\lim_{\chi_1 \to 0} \hat{u} = -\frac{\bar{p}_1^2}{p_1^2} \left[ u - 2 i \beta_1 \frac{p_1}{\bar{p}_1} \right] e^{-8ibt}.
\end{equation}
For the sign-definite periodic wave (\ref{3.14}),
it is shown in Appendix \ref{app-G} that
the algebraic soliton reaches its maximal walue given by 
\begin{equation}
\label{hat-rho-max}
\hat{\rho}_{\rm max} = \frac{1}{4} (3 \sqrt{u_1} + \sqrt{u_2} + \sqrt{u_3} + \sqrt{u_4})^2.
\end{equation}
We have computed $\hat{\rho}_{\rm max} \approx 8.85$ which coincides with the numerical values on the surface plot on the left top panel of Fig. \ref{fig-alg2}.

For the sign-indefinite periodic wave (\ref{sign-indefinite}),
similar computations give the following maximum of the algebraic soliton at
\begin{equation}
\label{hat-rho-max-2}
\hat{\rho}_{\rm max} = \frac{1}{4} (\sqrt{u_1} + \sqrt{u_2} + 3 \sqrt{u_3} + \sqrt{u_4})^2.
\end{equation}
We have computed $\hat{\rho}_{\rm max} \approx 5.14$ which coincides with the numerical values on the surface plot on the left bottom panel of Fig. \ref{fig-alg2}.

\begin{figure}[htb!]
	\includegraphics[width=0.45\textwidth]{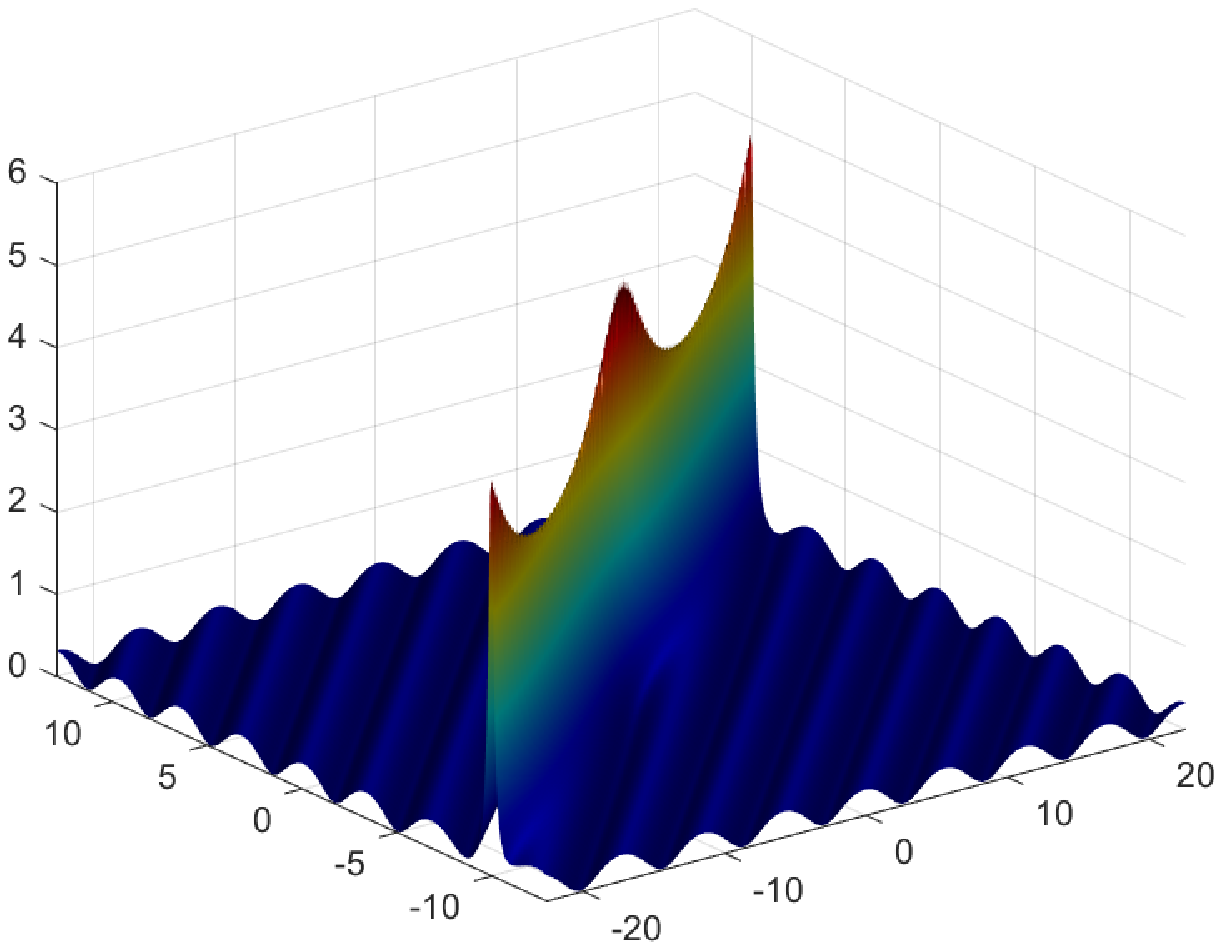}
	\includegraphics[width=0.45\textwidth]{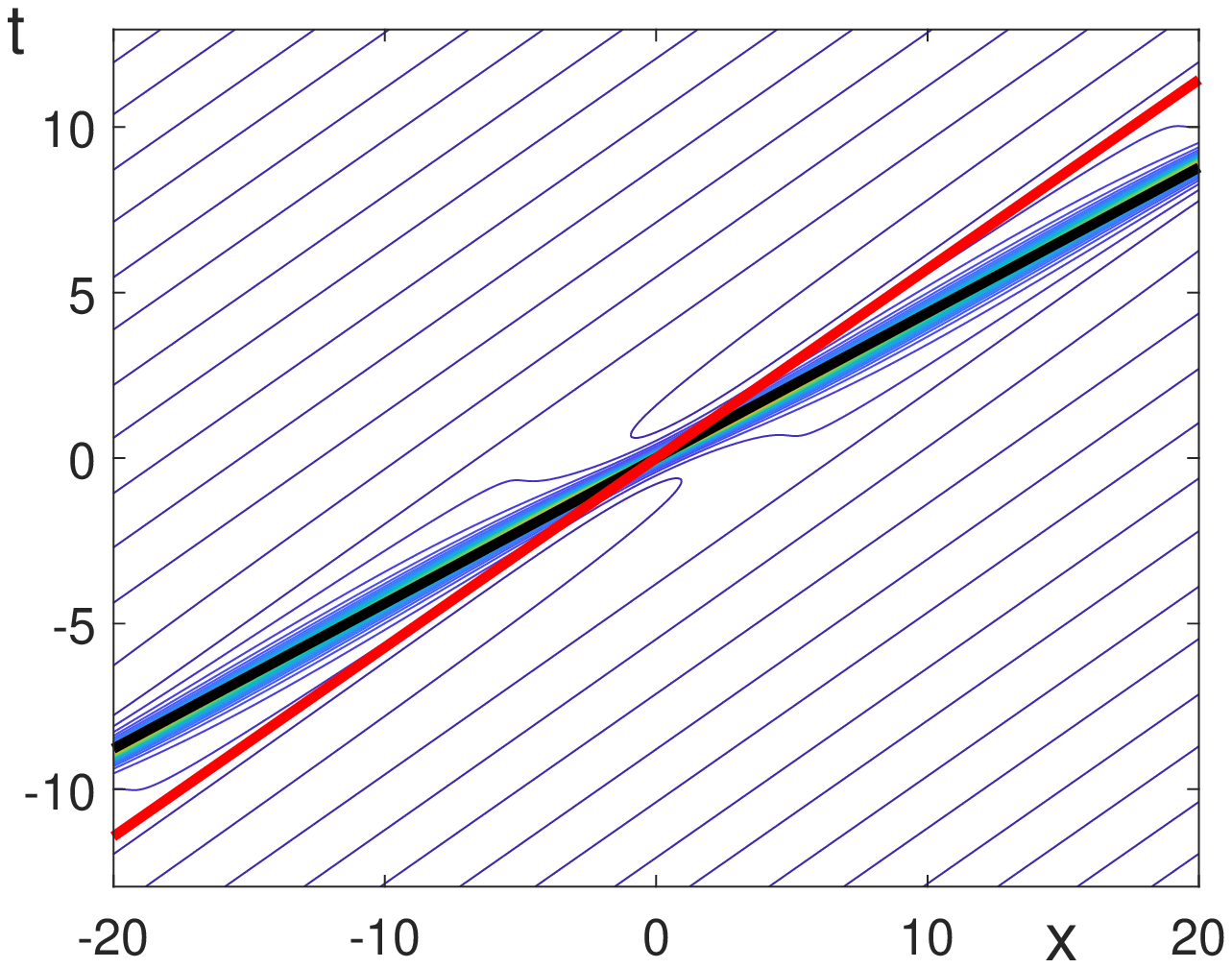} \\
	\includegraphics[width=0.45\textwidth]{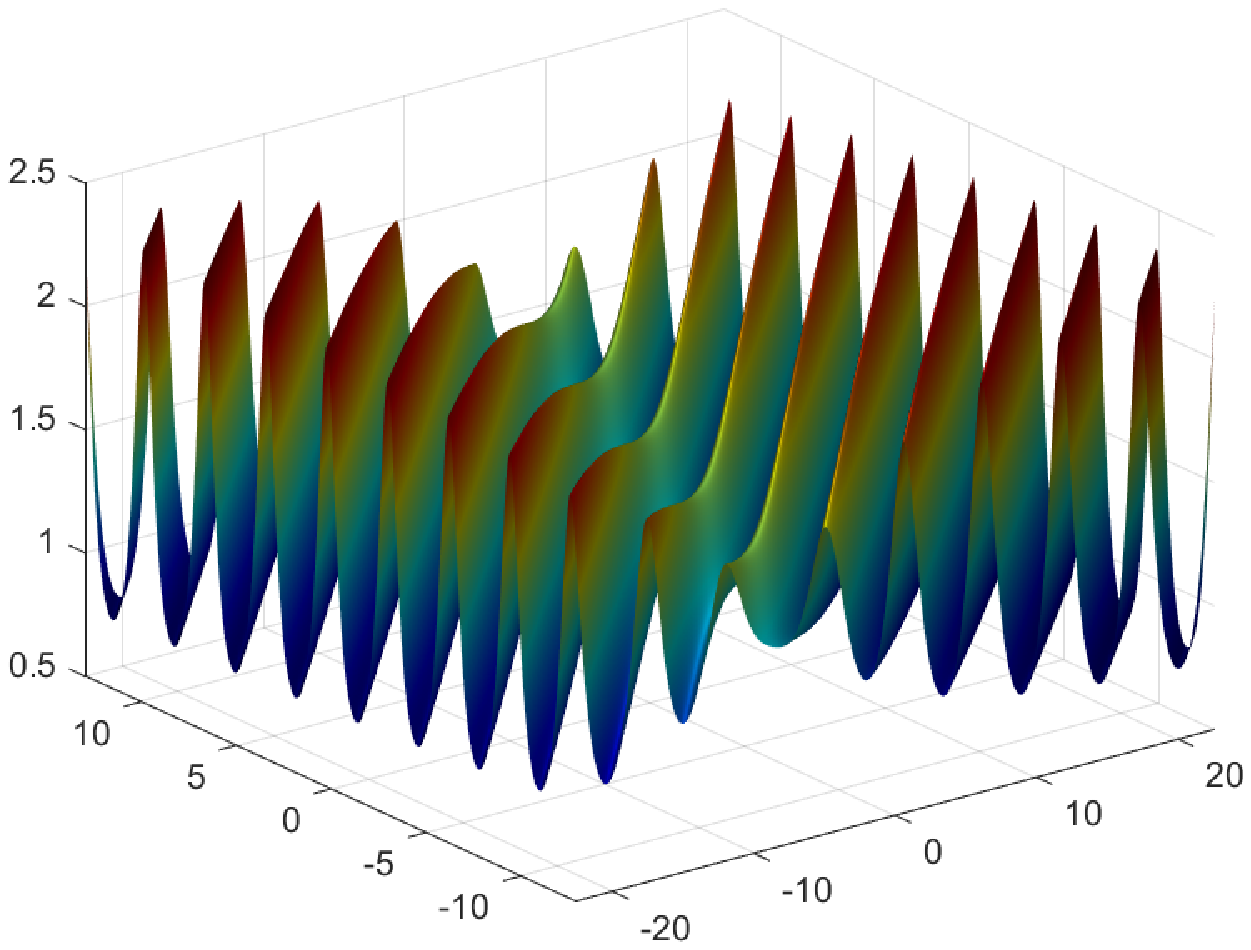}
	\includegraphics[width=0.45\textwidth]{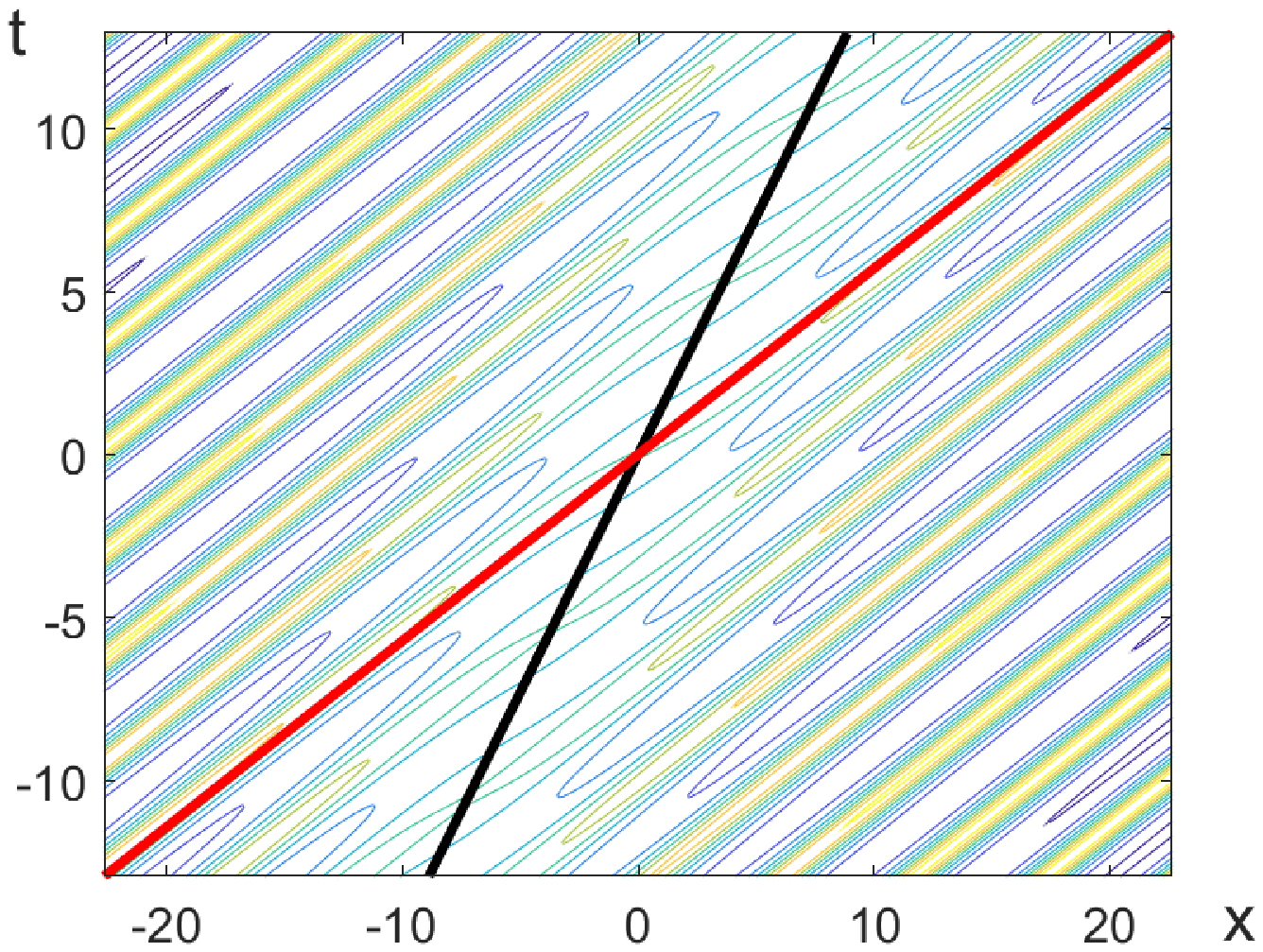} \\
	\includegraphics[width=0.45\textwidth]{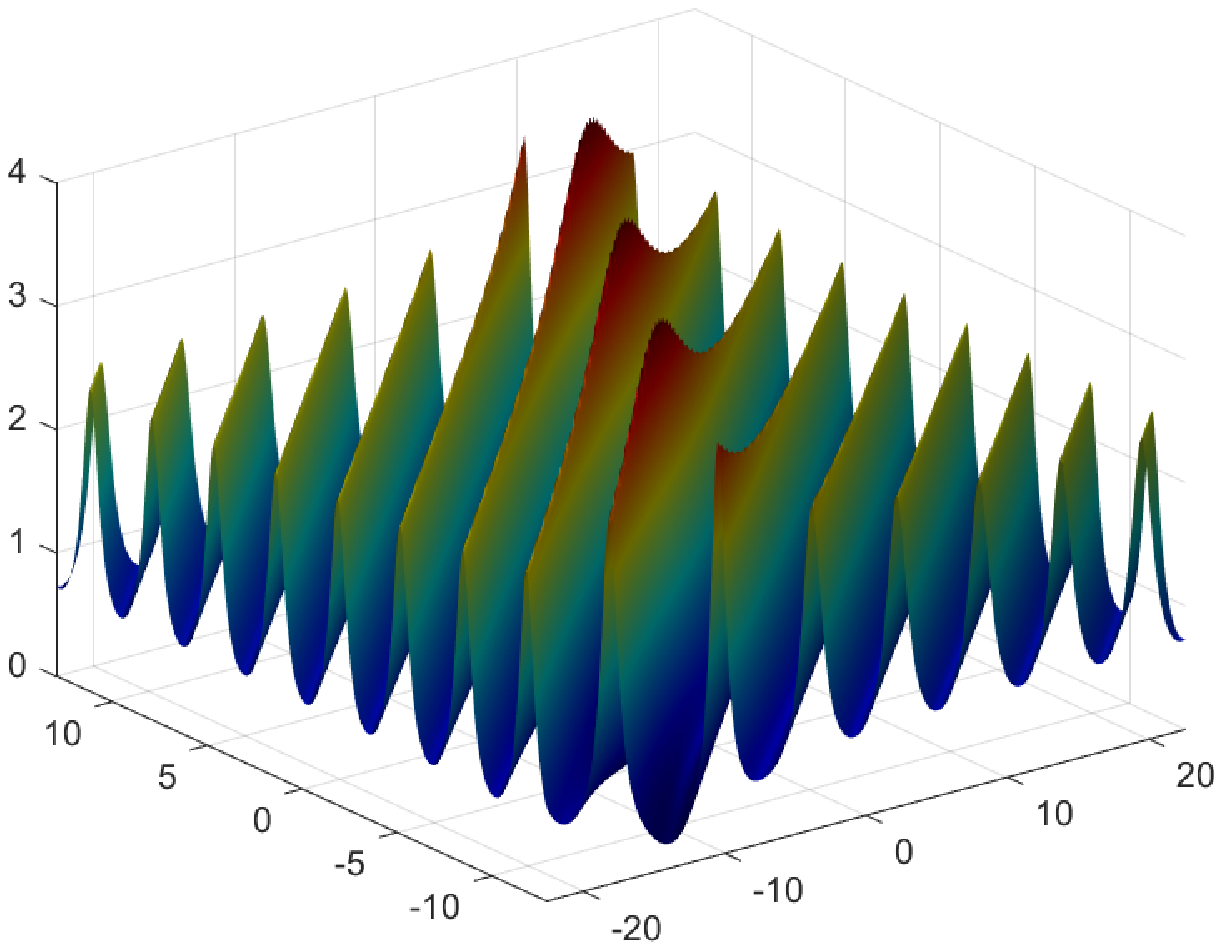}
	\includegraphics[width=0.45\textwidth]{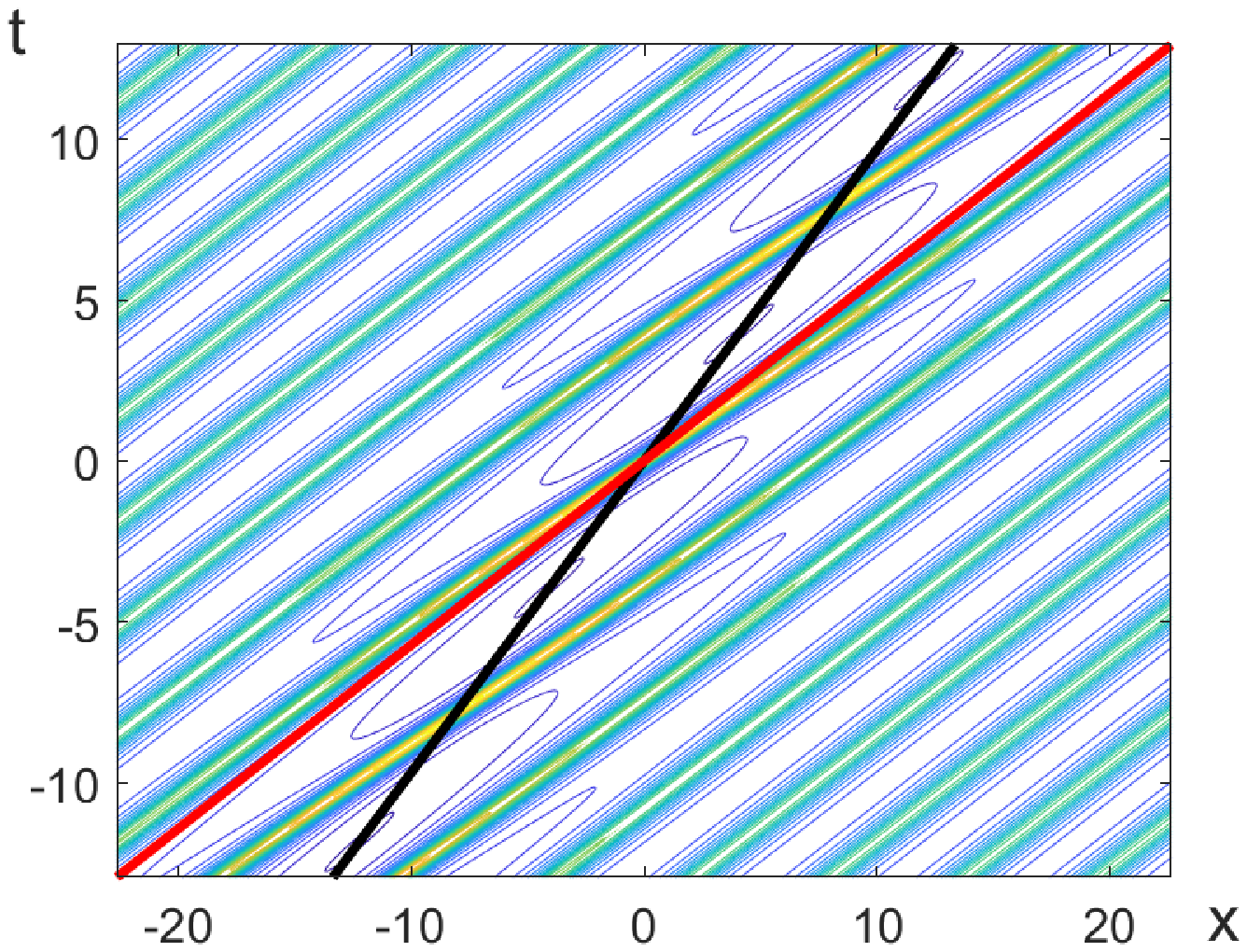}
	\caption{New solutions to the DNLS equation in variable $\hat{\rho}$
		obtained after the one-fold transformation
		of the periodic wave (\ref{3.14}) associated
		with the eigenvalues $i \beta_2$ (top), $i \beta_3$ (middle), and $i \beta_4$ (bottom).} \label{fig-alg3}
\end{figure}

Figure \ref{fig-alg2} corresponds to the case of $\lambda_1 = i \beta_1$.
Figure \ref{fig-alg3} presents similar results for the sign-definite periodic wave (\ref{3.14}) when the one-fold transformation is used with
the eigenvalues $\lambda_2 = i \beta_2$ (top),
$\lambda_3 = i \beta_3$ (middle), and $\lambda_4 = i \beta_4$ (bottom).
For the sign-definite periodic wave (\ref{3.14}), the transformed wave (\ref{1foldDT}) is the same for $\lambda_1$ and $\lambda_2$,
it is still located between $v_4 \approx 0.02$ and $v_3 \approx 0.31$
but it is translated by half-period between the two cases.
For the same sign-definite periodic wave (\ref{3.14}), the transformed wave (\ref{1foldDT}) is the same for $\lambda_3$ and $\lambda_4$, it is located between $v_2 \approx 0.73$ and $v_1 \approx 2.44$ but it is translated by half-period between the two cases. Note that $\beta_1 \approx 1.1$, $\beta_2 \approx -0.6$, $\beta_3 \approx -0.1$, and $\beta_4 \approx -0.4$. The algebraic soliton is largest in the case of the largest eigenvalue $\beta_1$ and smallest in the case of the smallest (in absolute value) eigenvalue $\beta_3$. In fact, it is a depression wave in the case of $\beta_3$.

The maximal values of $\hat{\rho}$ are computed similarly
to (\ref{hat-rho-max}) and (\ref{hat-rho-max-2}). They correspond to
$2 \hat{\beta}_{1,2,3,4}^2$ for each eigenvalue $\lambda_{1,2,3,4} = i \beta_{1,2,3,4}$,
where $\hat{\beta}_{1,2,3,4}$ is obtained from $\beta_{1,2,3,4}$
after $\sqrt{u}_1$ is replaced by $3 \sqrt{u}_1$ as in (\ref{hat-rho-max}) and $\sqrt{u_3}$ is replaced by $3 \sqrt{u_3}$ as in (\ref{hat-rho-max-2}).
As a result, we obtain 
$\hat{\rho}_{\rm max} \approx 5.14$ for $i \beta_2$, 
$\hat{\rho}_{\rm max} \approx 1.61$ for $i \beta_3$, 
and $\hat{\rho}_{\rm max} \approx 3.90$ for $i \beta_4$, 
in agreement with Fig. \ref{fig-alg3}. 

\begin{figure}[htb!]
	\includegraphics[width=0.45\textwidth]{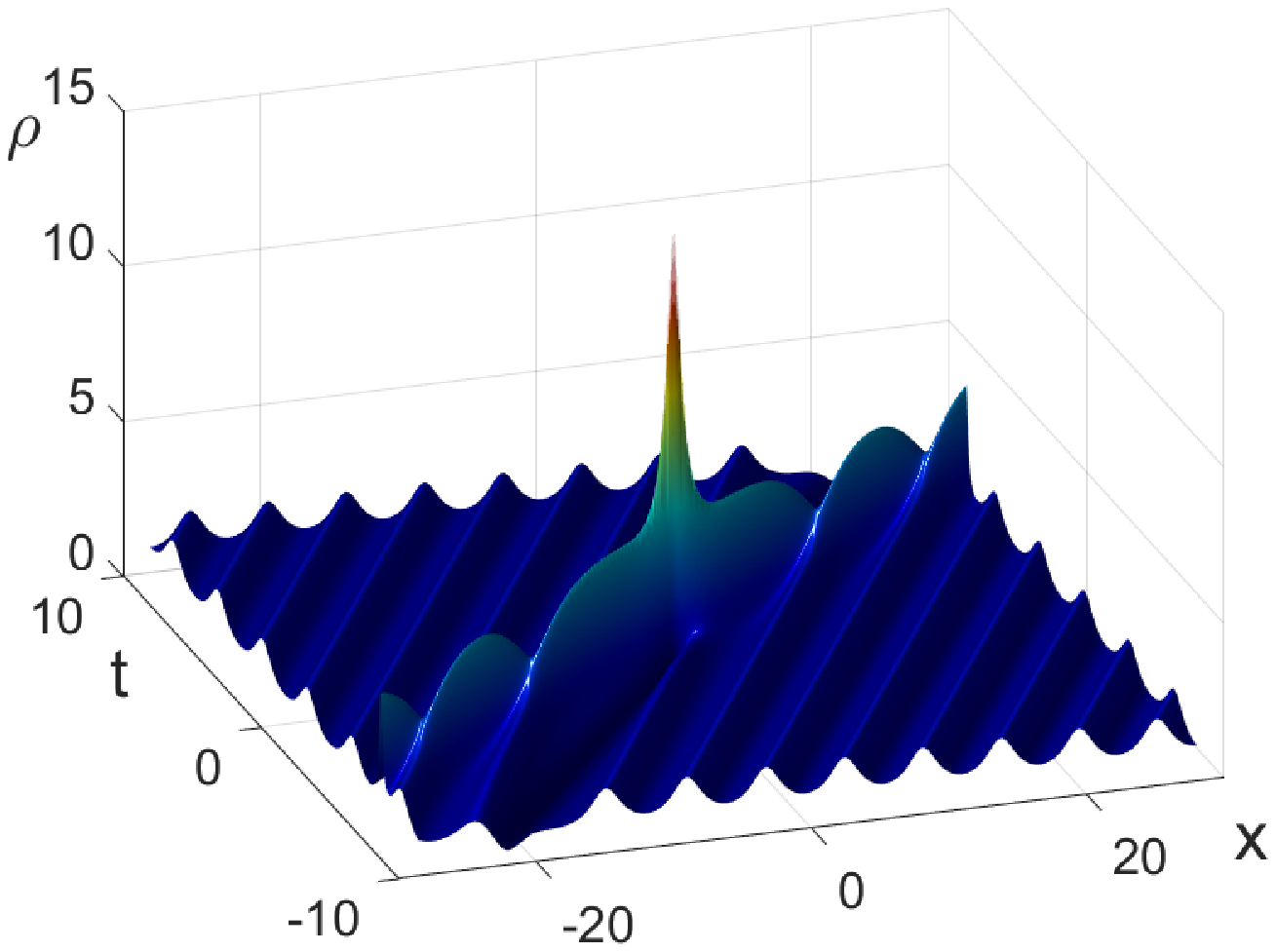}
	\includegraphics[width=0.45\textwidth]{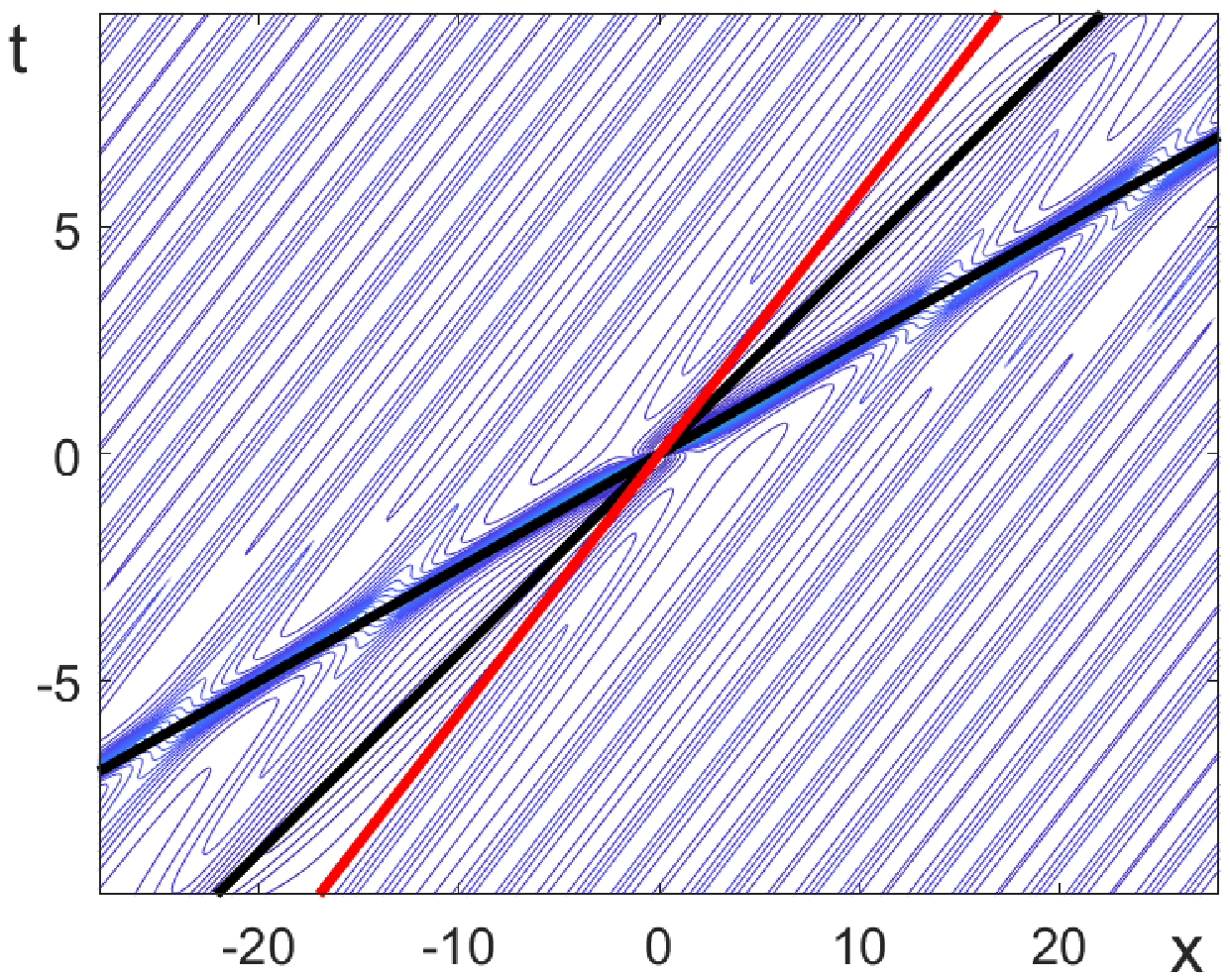} \\
	\includegraphics[width=0.45\textwidth]{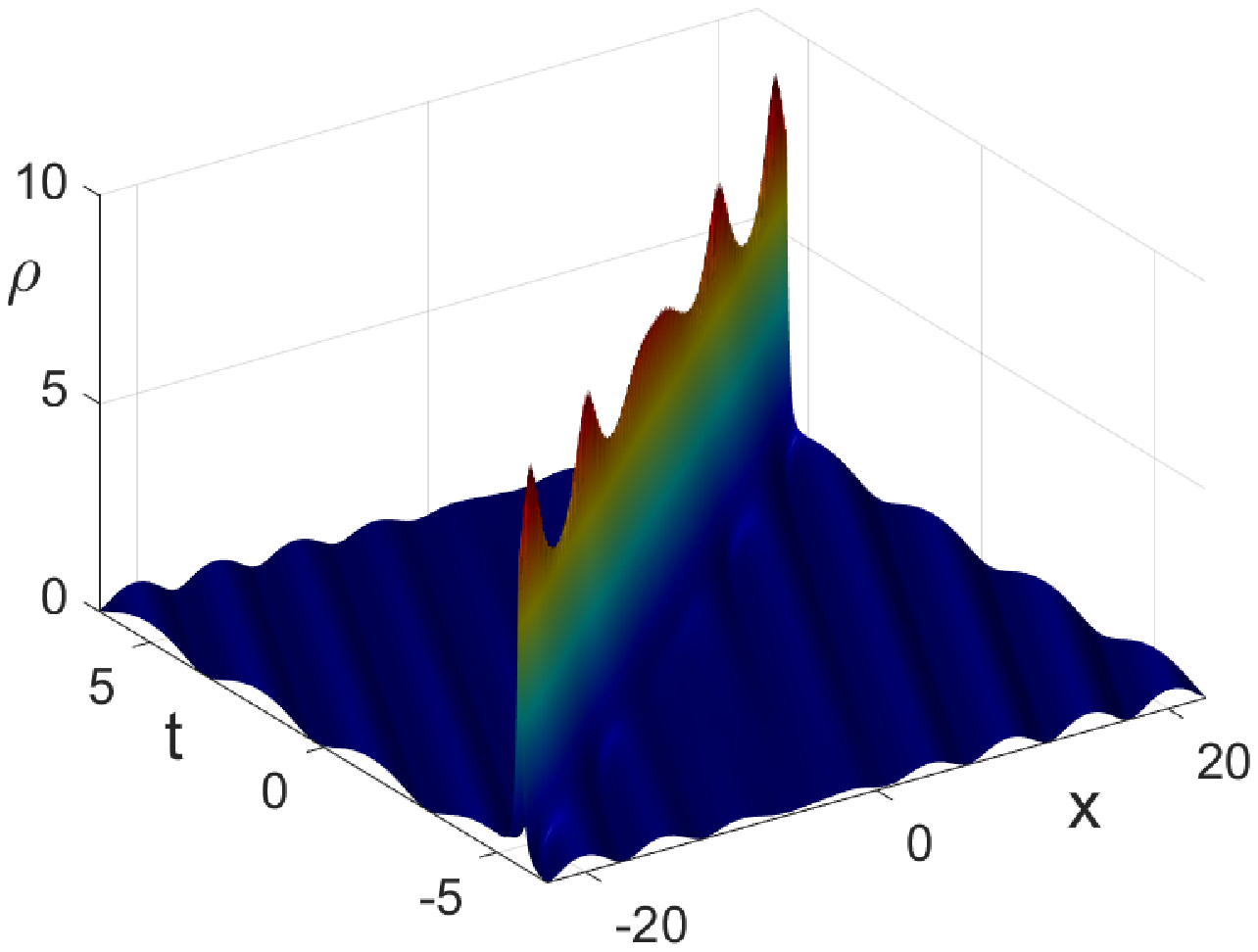}
	\includegraphics[width=0.45\textwidth]{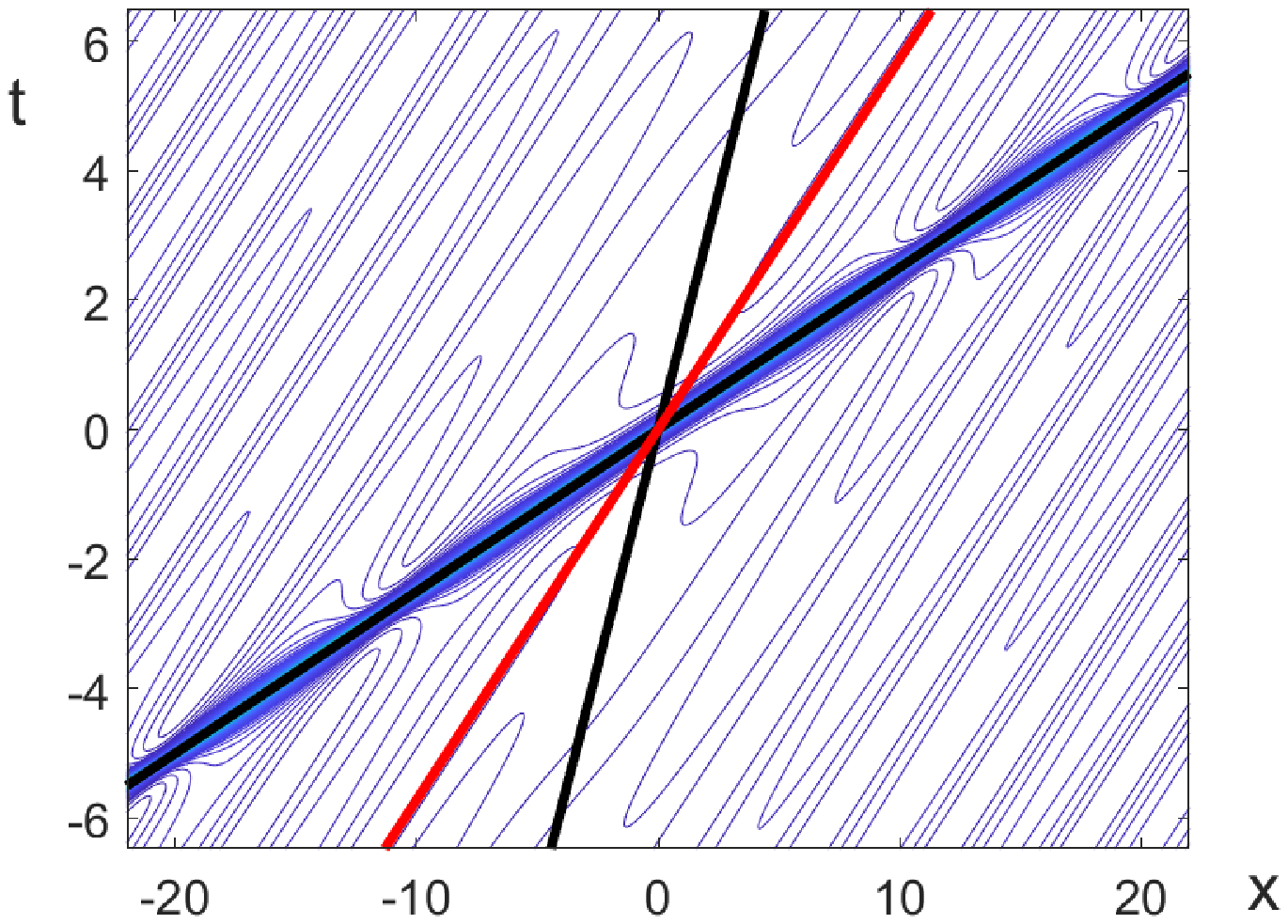} \\
		\includegraphics[width=0.45\textwidth]{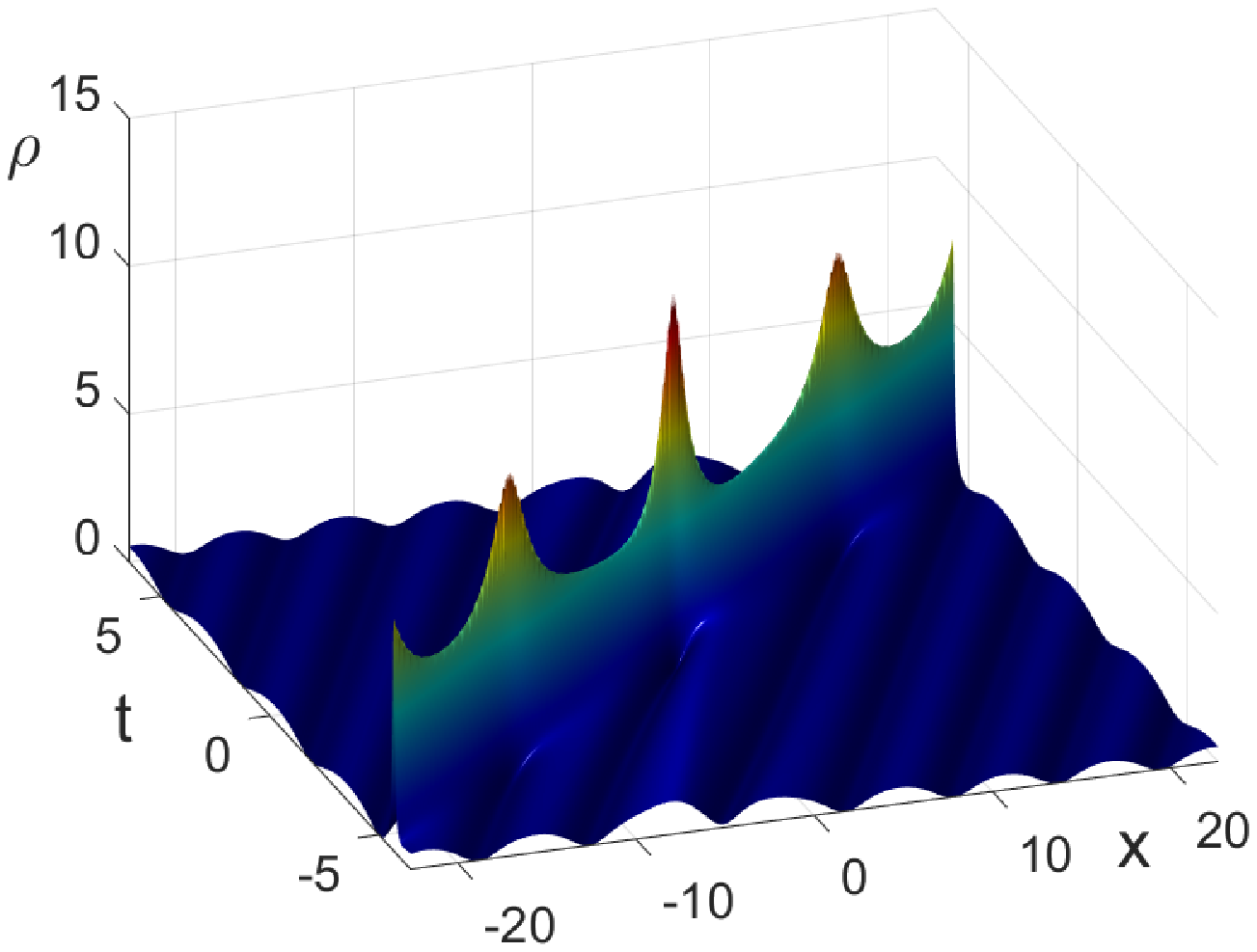}
	\includegraphics[width=0.45\textwidth]{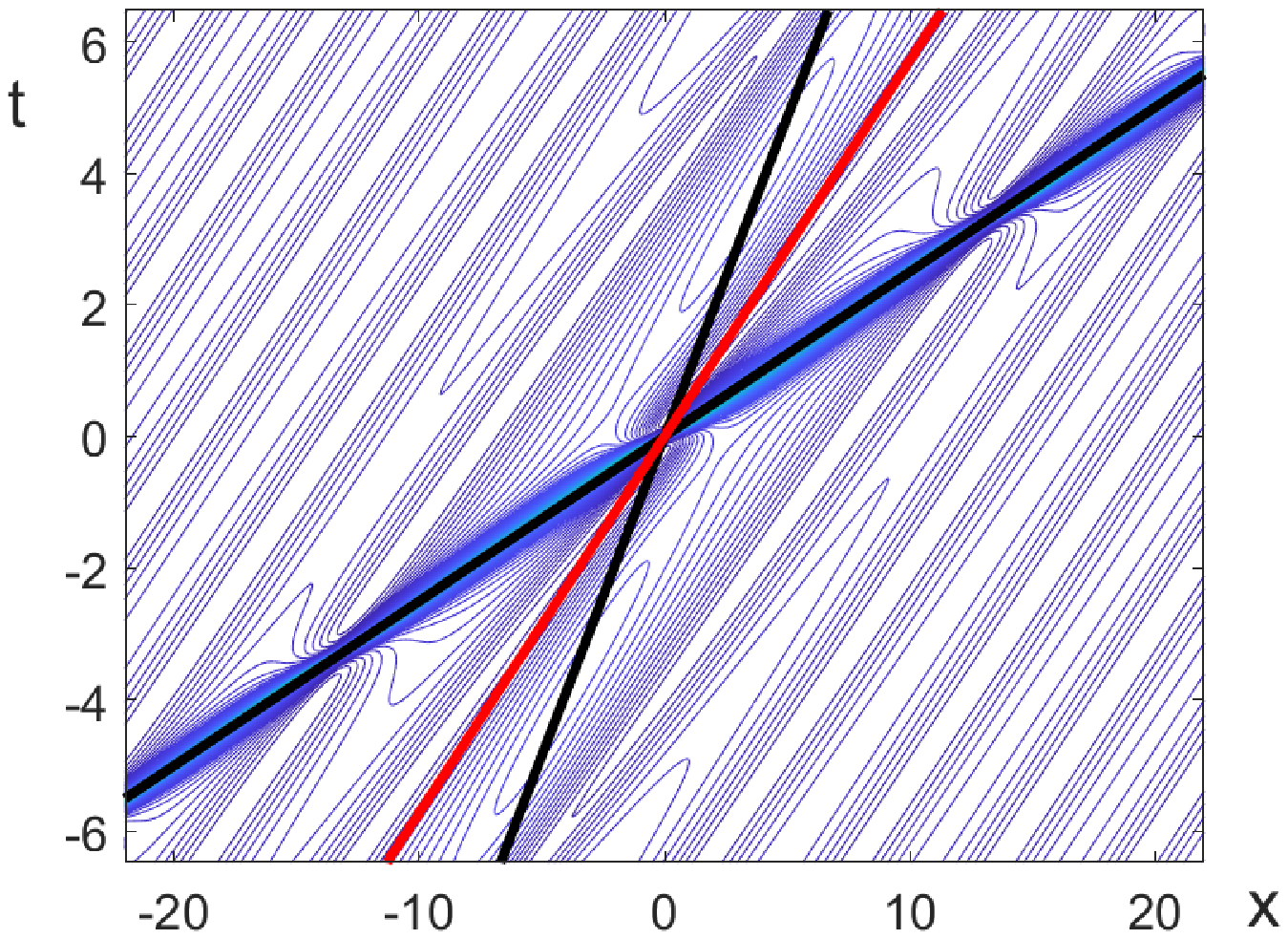}
	\caption{New solutions to the DNLS equation in variable $\hat{\rho}$
	obtained after the two-fold transformation
	of the periodic wave (\ref{3.14}): solution surface
(left) and contour plot (right).
		One eigenvalue is always $i \beta_1$, whereas the
		other eigenvalue is $i \beta_2$ (top), $i \beta_3$ (middle), and $i \beta_4$ (bottom).} \label{fig-alg4}
\end{figure}

Figure \ref{fig-alg4} presents the new solution to the DNLS equation
(\ref{dnls}) obtained with the two-fold transformation (\ref{5.8a})
for $\hat{p}_1$ and $\hat{p}_2$. Here we take one eigenvalue as
$\lambda_1 = i \beta_1$ and
the other eigenvalue being  $\lambda_2 = i \beta_2$ (top),
$\lambda_3 = i \beta_3$ (middle), and $\lambda_4 = i \beta_4$ (bottom).
The second solution $\varphi = (\hat{p}_1,\hat{q}_1)^T$ is defined
by (\ref{4.5}) with $\chi_1$ given by (\ref{chi}). We always take $c_1 = 0$
in (\ref{chi}) so that the algebraic solitons propagate along the corresponding lines 
(\ref{direction-soliton}). The line $x + 2 ct = 0$ is shown by red curve
and the two lines (\ref{direction-soliton}) for the two eigenvalues are shown by the black curves
on the right panels of Fig. \ref{fig-alg4}.

The new solution describes interaction of the two algebraic solitons
on the background of the transformed wave obtained with the two-fold transformation (\ref{5.8a}) for $p_1$ and $p_2$.
As is established earlier, for $\lambda_1$ and $\lambda_2$, the background wave is the same as the original sign-definite wave (\ref{3.14}) but translated
by half-period with $\rho_{\rm tr} \in [u_2,u_1]$. For $\lambda_1$ and either $\lambda_3$ or $\lambda_4$, the
background wave corresponds to the sign-indefinite  wave (\ref{sign-indefinite})
with $\rho_{\rm tr} \in [u_4,u_3]$. In the latter cases, the second
algebraic soliton is almost invisible on the middle and bottom panels
of Fig. \ref{fig-alg4}. We have checked by taking nonzero $c_1$ in
(\ref{chi}) that the two algebraic solitons become visible when they overlap at a point on the $(x,t)$ plane away from the origin. However, when we take $c_1 = 0$ in (\ref{chi}) for both eigenvalues, the two algebraic solitons overlap at the origin.

\subsection{Periodic wave (\ref{3.14}) with $\lambda_1$ in a complex quadruplet}

Let $\lambda_1 \in \mathbb{C}\backslash i \mathbb{R}$ be an eigenvalue of the KN spectral problem (\ref{lax-1}). We use the decomposition (\ref{trav-wave}) and (\ref{trav-wave-eigenfunction}) with
the eigenvector $\varphi = (p_1,q_1)^\mathrm{T}$
satisfying the linear system
\begin{eqnarray}
\label{lax-complex-1}
\left\{ \begin{array}{l}
\displaystyle \frac{\partial p_1}{\partial x} = -i \lambda_1^2 p_1 + \lambda_1 u q_1, \\
\displaystyle \frac{\partial q_1}{\partial x} = i \lambda_1^2 q_1 - \lambda_1 \bar{u} p_1,
 \end{array} \right.
 \end{eqnarray}
and
\begin{eqnarray}
\left\{ \begin{array}{l}
\displaystyle
\frac{\partial p_1}{\partial t}
+ 2c \frac{\partial p_1}{\partial x} + 2 i b p_1 = i (-2 \lambda_1^4 + \lambda_1^2 |u|^2) p_1
+ 2 \lambda_1^3 u q_1 + \lambda_1 (i u_x - |u|^2 u) q_1, \\
\displaystyle
\frac{\partial q_1}{\partial t}
+ 2c \frac{\partial q_1}{\partial x} - 2 i b q_1 = i (2 \lambda_1^4 - \lambda_1^2 |u|^2) q_1
- 2 \lambda_1^3 \bar{u} p_1 + \lambda_1 (i \bar{u}_x + |u|^2 \bar{u}) p_1.
 \end{array} \right.
\label{lax-complex-2}
\end{eqnarray}
The second, linearly independent solution $\varphi=(\hat{p}_1,\hat{q}_1)^\mathrm{T}$
of the system (\ref{lax-complex-1}) and (\ref{lax-complex-2})
can be written in the form:
\begin{equation}\label{z-4.5-1}
\hat{p}_1= p_1\chi_1 - \frac{\bar{q}_1}{|p_1|^2+|q_1|^2}, \qquad \hat{q}_1= q_1\chi_1 + \frac{\bar{p}_1}{|p_1|^2+|q_1|^2},
\end{equation}
where $\chi_1$ is a complex-valued function of $x$ and $t$. Wronskian between the two solutions is normalized by $p_1\hat{q}_1-\hat{p}_1 q_1=1$.

Substituting (\ref{z-4.5-1}) into
(\ref{lax-complex-1}) and (\ref{lax-complex-2}) written for $\varphi=(\hat{p}_1,\hat{q}_1)^\mathrm{T}$
and using the same equations for  $\varphi =( p_1,q_1)^\mathrm{T}$ yields the following equations for $\chi_1$:
\begin{equation}\label{z-4.6-1}
\frac{\partial \chi_1}{\partial x} = \frac{2i(\lambda_1^2 - \bar{\lambda}_1^2)\bar{p}_1\bar{q}_1 +
(\lambda_1 - \bar{\lambda}_1) (u\bar{p}_1^2 + \bar{u} \bar{q}_1^2)}
{(|p_1|^2 +|q_1|^2)^2}
\end{equation}
and
\begin{eqnarray}
\frac{\partial \chi_1}{\partial t} + 2c \frac{\partial \chi_1}{\partial x} &=&
\frac{1}{(|p_1|^2+|q_1|^2)^2} \nonumber \\
&& 
\left[2i\bar{p}_1\bar{q}_1(\lambda_1^2 - \bar{\lambda}_1^2)[2(\lambda_1^2+\bar{\lambda}_1^2) - |u|^2]
+ 2 (\lambda_1^3 - \bar{\lambda}_1^3) (u\bar{p}_1^2 + \bar{u} \bar{q}_1^2) \right. \nonumber\\ &&
\left.+ (\lambda_1 - \bar{\lambda}_1)[i (u_x \bar{p}_1^2 - \bar{u}_x \bar{q}_1^2) -
|u|^2(u\bar{p}_1^2 + \bar{u} \bar{q}_1^2)]\right]. \label{z-4.7-1}
\end{eqnarray}
Substituting  (\ref{4.4}) and (\ref{4.4a}) into (\ref{z-4.6-1}) yields
\begin{equation}\label{z-4.6-2}
\frac{\partial \chi_1}{\partial x} = \frac{\bar{\lambda}_1[\rho^2 + 2 (c - 2 \bar{\lambda}_1^2)\rho -a]
- 2(\lambda_1 +\bar{\lambda}_1)(b - c \bar{\lambda}_1^2 + \bar{\lambda}_1^4 - \bar{\lambda}_1^2 \rho)}
{\bar{\lambda}^2_1(\lambda_1 + \bar{\lambda}_1)(|p_1|^2 +|q_1|^2)^2}.
\end{equation}
Regarding (\ref{z-4.7-1}), it must again be constant. It is shown in Appendix \ref{app-I} that 
\begin{equation}\label{4.7b-2}
\frac{\partial \chi_1}{\partial t} = 2 \lambda_1^2 (\lambda_1^2 - \bar{\lambda}_1^2).
\end{equation}
It follows that (\ref{4.7b-2}) reduces to (\ref{4.7b}) if $\lambda_1 = i \beta_1$. 
We obtain from (\ref{z-4.6-2}) and (\ref{4.7b-2}) that 
\begin{equation}
\label{chi-2}
\chi_1(x,t) = c_1 + k_1 x + f(x) + 2 \lambda_1^2 (\lambda_1^2 - \bar{\lambda}_1^2) t,
\end{equation}
where $c_1 \in \mathbb{C}$ is an arbitrary constant of integration,
$k_1$ is the mean value of $\frac{\partial \chi_1}{\partial x}$
over the period $L = 2 \nu^{-1} K(k)$ of
the periodic wave $\rho$, and $f$ is the $L$-periodic function with the zero mean. The line equation
\begin{equation}
\label{direction-soliton-2}
k_1 (x+2ct) + 2 \lambda_1^2 (\lambda_1^2 - \bar{\lambda}_1^2) t = 0
\end{equation}
is now complex-valued, hence it defines two straight lines on the $(x,t)$
plane. If the straight lines have different slopes, they only intersect
at $(x,t) = (0,0)$ and this implies that the function
$\chi_1(x,t)$ grows linearly as  $|x| + |t| \to \infty$ everywhere in
the $(x,t)$ plane. Consequently, the new solution obtained with
the Darboux transformation (\ref{5.8}) at the second solution
$\varphi = (\hat{p}_1,\hat{q}_1)^T$ given by (\ref{z-4.5-1})
displays the rogue wave localized on the transformed periodic wave. 
The transformed periodic wave is 
obtained with the Darboux transformation (\ref{5.8}) at the first
solution $\varphi = (p_1,q_1)^T$.

In order to illustrate the two solutions, we consider the periodic standing wave (\ref{3.14})
with the particular choice of
$$
u_1 = 2, \quad u_2 = 1, \quad u_3 = 0, \quad u_4 = -0.5.
$$
This choice corresponds to parameters
$$
a = 0, \quad b = \frac{17}{256}, \quad c = -\frac{5}{8}, \quad d = -\frac{1}{8}
$$
in the quadrature (\ref{3.8}) with (\ref{quartic-Q}). Again, 
we preserve the constraint $c^2 - 4b > 0$.

\begin{figure}[htb!]
	\begin{center}
	\includegraphics[width=0.48\textwidth]{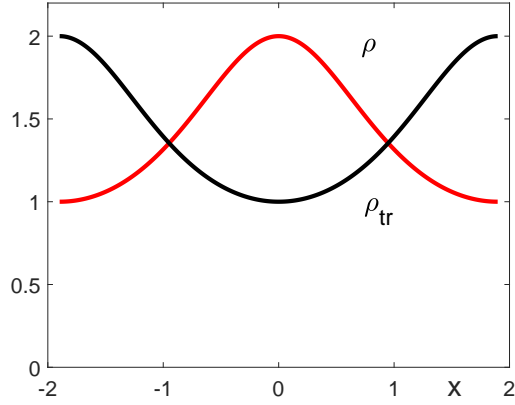}
	\end{center}
	\caption{The periodic standing wave $\rho$ (red) and its transformed
		version $\rho_{\rm tr}$ (black).} \label{fig-rogue1}
\end{figure}

Figure \ref{fig-rogue1} shows the periodic standing wave $\rho$ (red)
and its transformed version $\rho_{\rm tr}$ (black) after the two-fold transformation (\ref{5.8}) with $\varphi = (p_1,q_1)^T$. 	
In agreement with (\ref{5.10}), the transformed wave is a half-period translation of the original wave. Moreover, the same translation is true for both quadruplets in (\ref{configuration-1}).

\begin{figure}[htb!]
	\includegraphics[width=0.48\textwidth]{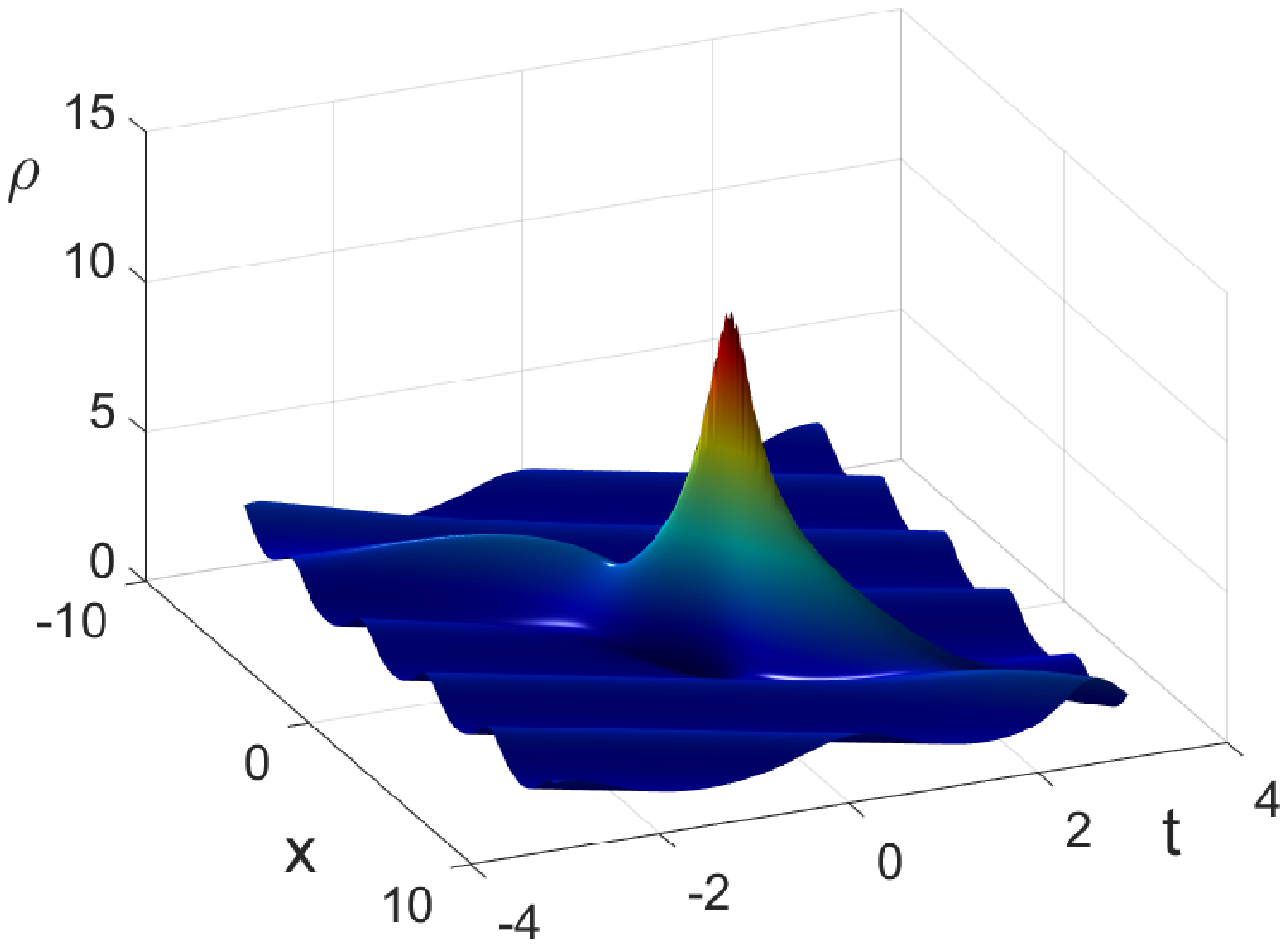}
	\includegraphics[width=0.48\textwidth]{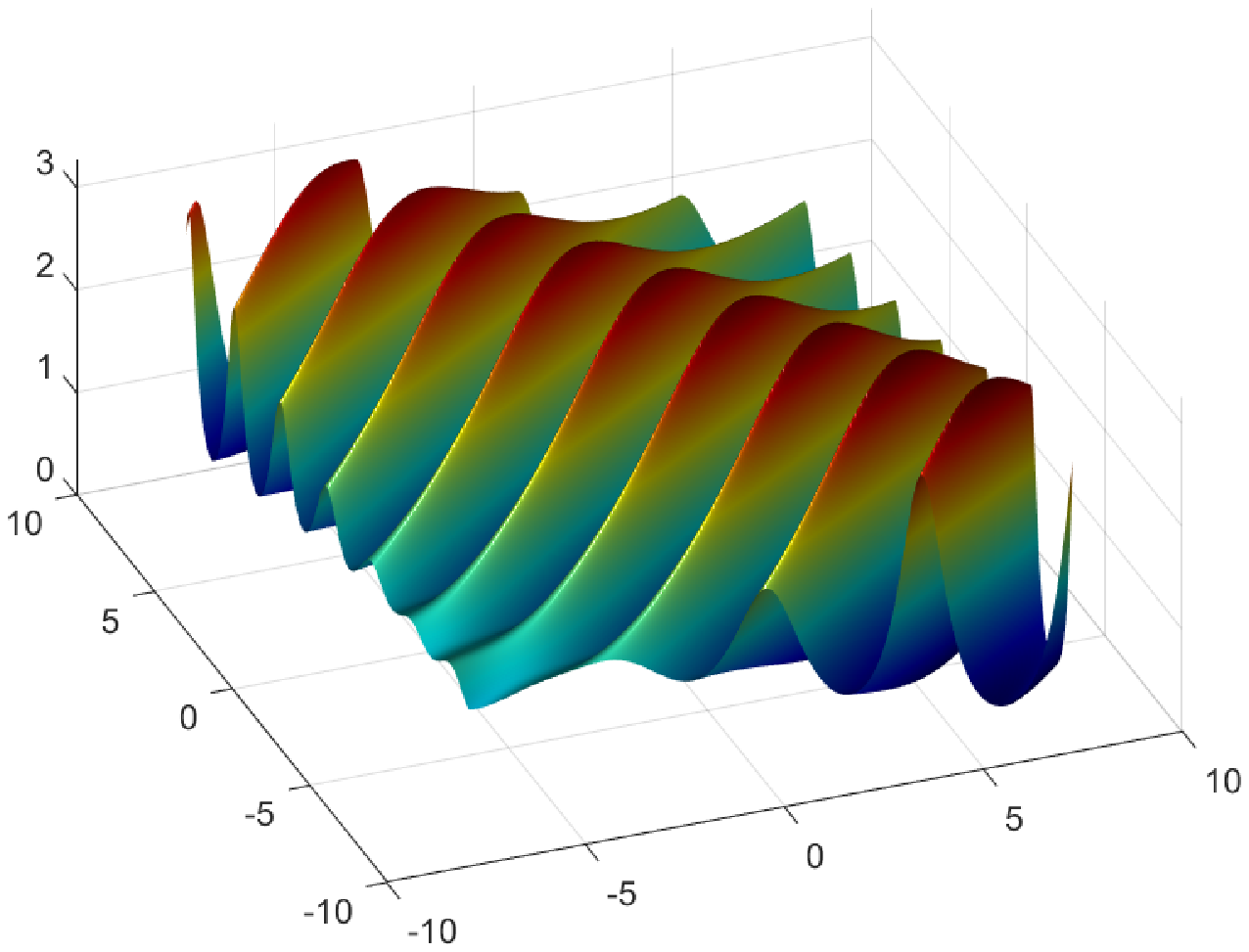}
	\caption{New solutions to the DNLS equation in variable $\hat{\rho}$
		which corresponds to the rogue waves
		on the background of the periodic standing wave.
		Left and right panels correspond to (\ref{3.14}) after the Darboux transformation
		(\ref{5.8}) with $\varphi = (\hat{p}_1,\hat{q}_1)^T$ for eigenvalues $\lambda_1$ and $\lambda_2$ respectively.} \label{fig-rogue2}
\end{figure}

Figure \ref{fig-rogue2} shows the rogue wave $\hat{\rho}$ on the background
of the periodic standing wave $\rho$ in (\ref{3.14}) with the same
parameters as in Figure \ref{fig-rogue1} after the two-fold transformation
(\ref{5.8}) with $\varphi = (\hat{p}_1,\hat{q}_1)^T$. The left panel corresponds to the quadruplet with $\lambda_1$
and the right panel corresponds to the quadruplet with $\lambda_2$.
Although the surface plot on the right panel does not show localization of the rogue wave on the scale displayed, we have checked that the real and imaginary part of equation (\ref{direction-soliton-2}) give two different lines intersecting at $(0,0)$ but the slopes of the two lines are close to each other.
As a result, ${\rm Re}(\chi)$ and ${\rm Im}(\chi)$ are bounded along two different directions, hence
the rogue wave $\hat{\rho}$ is still localized in space and time.

It is shown in Appendix \ref{app-J} that the maximum of the rogue wave occurs at the point $(0,0)$ if $c_1 = 0$ and it is given by
\begin{equation}
\label{max-rogue1}
\hat{\rho}_{\rm max} = (2 \sqrt{u_1} + \sqrt{u_2})^2
\end{equation}
for eigenvalue $\lambda_1$ and
\begin{equation}
\label{max-rogue2}
\hat{\rho}_{\rm max} = (2 \sqrt{u_1} - \sqrt{u_2})^2
\end{equation}
for eigenvalue $\lambda_2$. We have $\hat{\rho}_{\rm max} \approx 14.66$ for eigenvalue $\lambda_1$ and $\hat{\rho}_{\rm max} \approx 3.34$ for eigenvalue $\lambda_2$, which agree with the numerical values on Fig. \ref{fig-rogue2}.

\subsection{Darboux transformations for the periodic wave (\ref{3.20})}

We end this section with an example of the periodic standing wave (\ref{3.20})
for the particular choice:
$$
u_1 = 2, \quad u_2 = 0, \quad \gamma = 0.2, \quad \eta = 0.1.
$$
This choice corresponds to parameters
$$
a = 0, \quad b = 0.036, \quad c = -0.6, \quad d = 0.0125
$$
in the quadrature (\ref{3.8}) and (\ref{quartic-Q}) satisfying $c^2 - 4b > 0$.
The expression (\ref{3.20}) with $u_2 = 0$
can be written as
\begin{equation}\label{3.20-simple}
\rho(x) = u_1 \delta \frac{1+{\rm cn} (\mu x;k)}{1+\delta +(\delta-1){\rm cn} (\mu x;k)}.
\end{equation}

\begin{figure}[htb!]
	\includegraphics[width=0.48\textwidth]{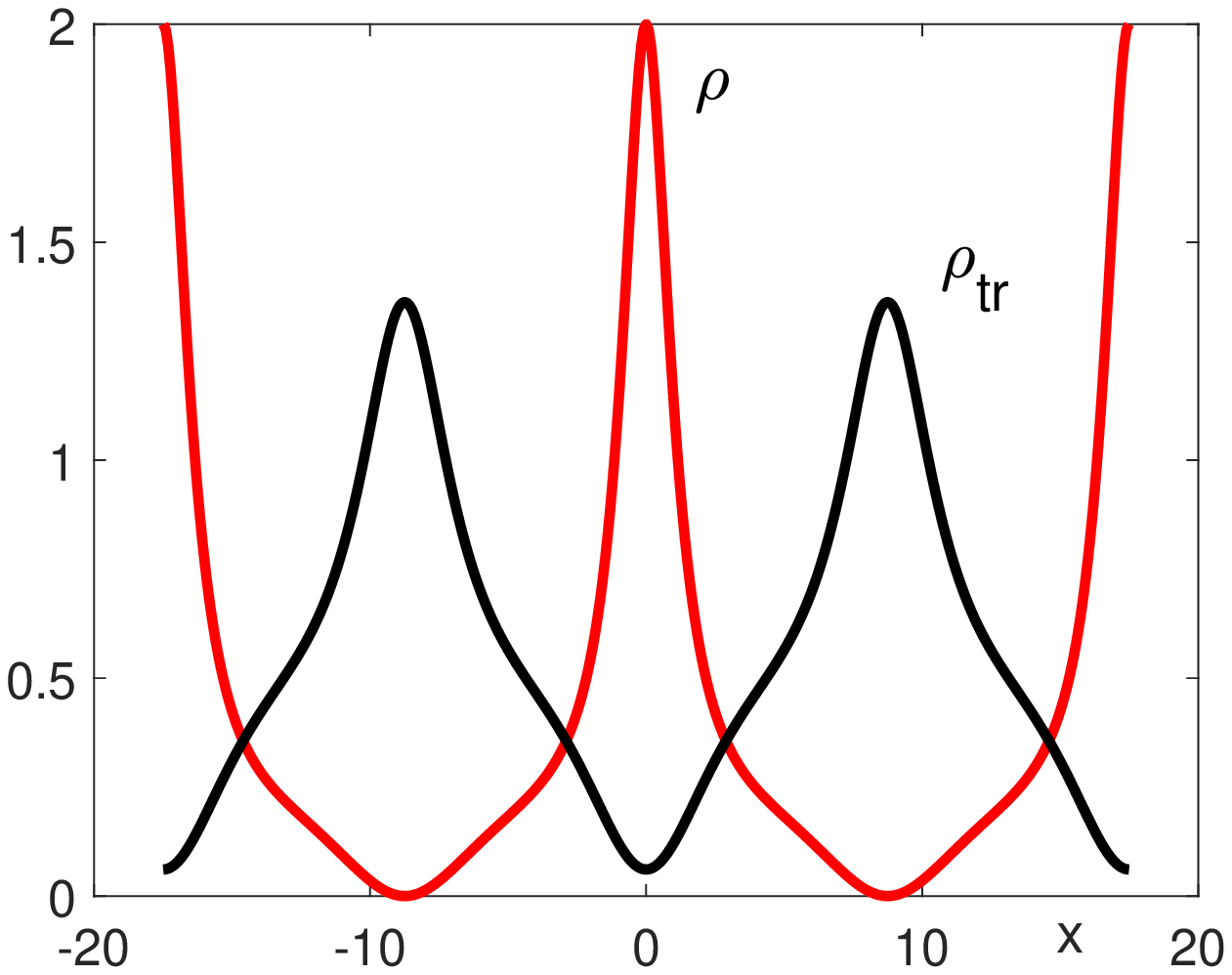}
	\includegraphics[width=0.48\textwidth]{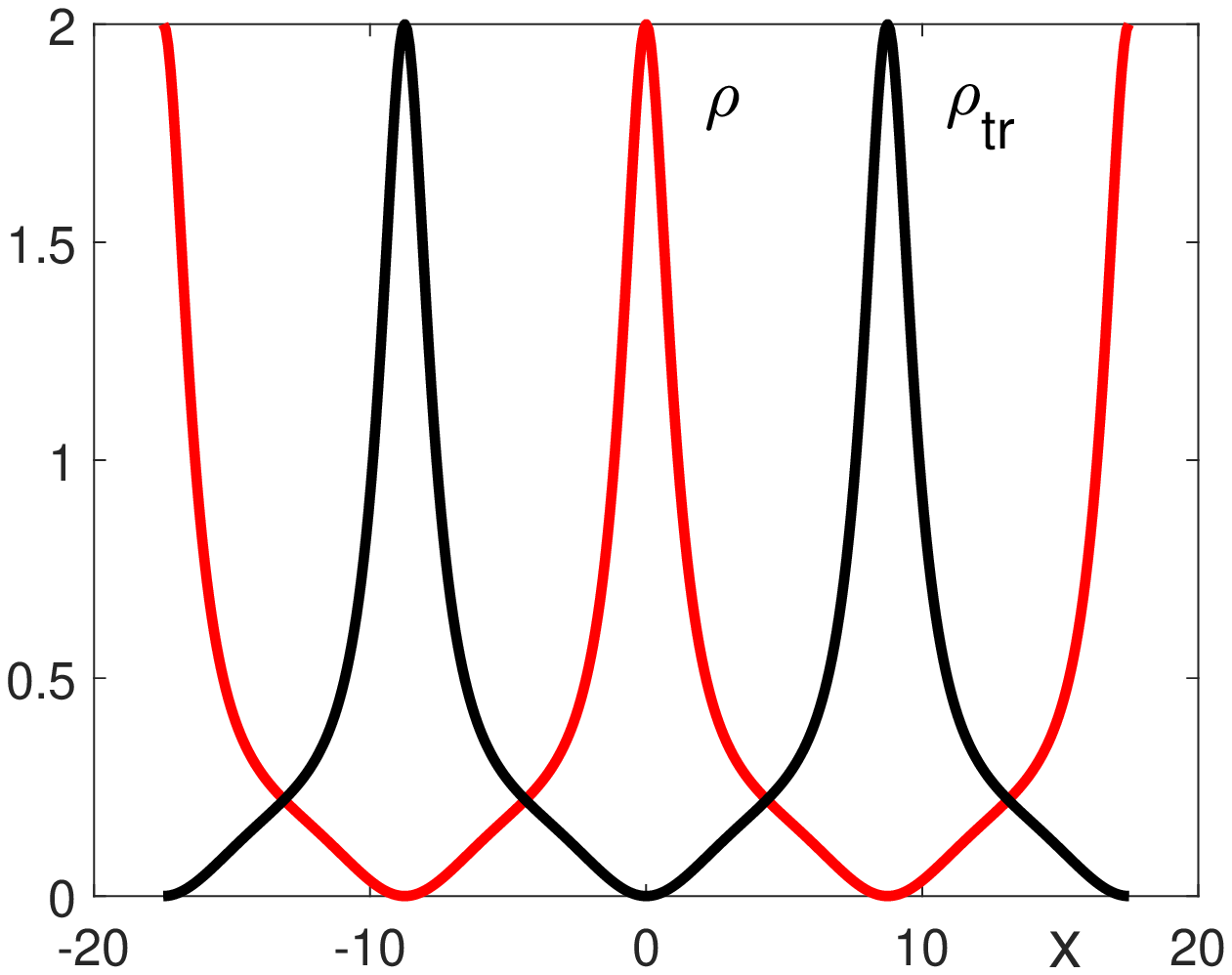}
	\caption{The periodic standing wave $\rho$ (red) and its transformed
		version $\rho_{\rm tr}$ (black) versus $x$ for the
		periodic standing wave (\ref{3.20-simple}). Left and right panels correspond to the eigenvalues $\lambda_3 = i \beta_3$ and $\lambda_1 = \alpha_1 + i \beta_1$ respectively.} \label{fig-mix1}
\end{figure}

The solution (\ref{3.20-simple}) is sign-indefinite since $\rho \in [0,u_1]$.
Extracting the square root analytically yields the exact
expression 
\begin{equation}\label{3.20-R}
R(x) = \frac{\sqrt{2u_1 \delta} {\rm cn} (\frac 12 \mu x;k)} {\sqrt{\delta {\rm cn}^2 (\frac 12 \mu x;k) +
		{\rm sn}^2 (\frac 12 \mu x;k) {\rm dn}^2 (\frac 12 \mu x;k)}}.
\end{equation}
The period of the periodic wave is  $L = 8 K(k) \mu^{-1}$.

The periodic wave (\ref{3.20-simple}) corresponds to the configuration
(\ref{configuration-3}) with one complex quadruplet $\{ \lambda_1, \bar{\lambda}_1, -\lambda_1, -\bar{\lambda}_1\}$ and
two pairs of purely imaginary eigenvalues $\{ \pm i \beta_3, \pm i \beta_4\}$.

\begin{figure}[htb!]
	\includegraphics[width=0.48\textwidth]{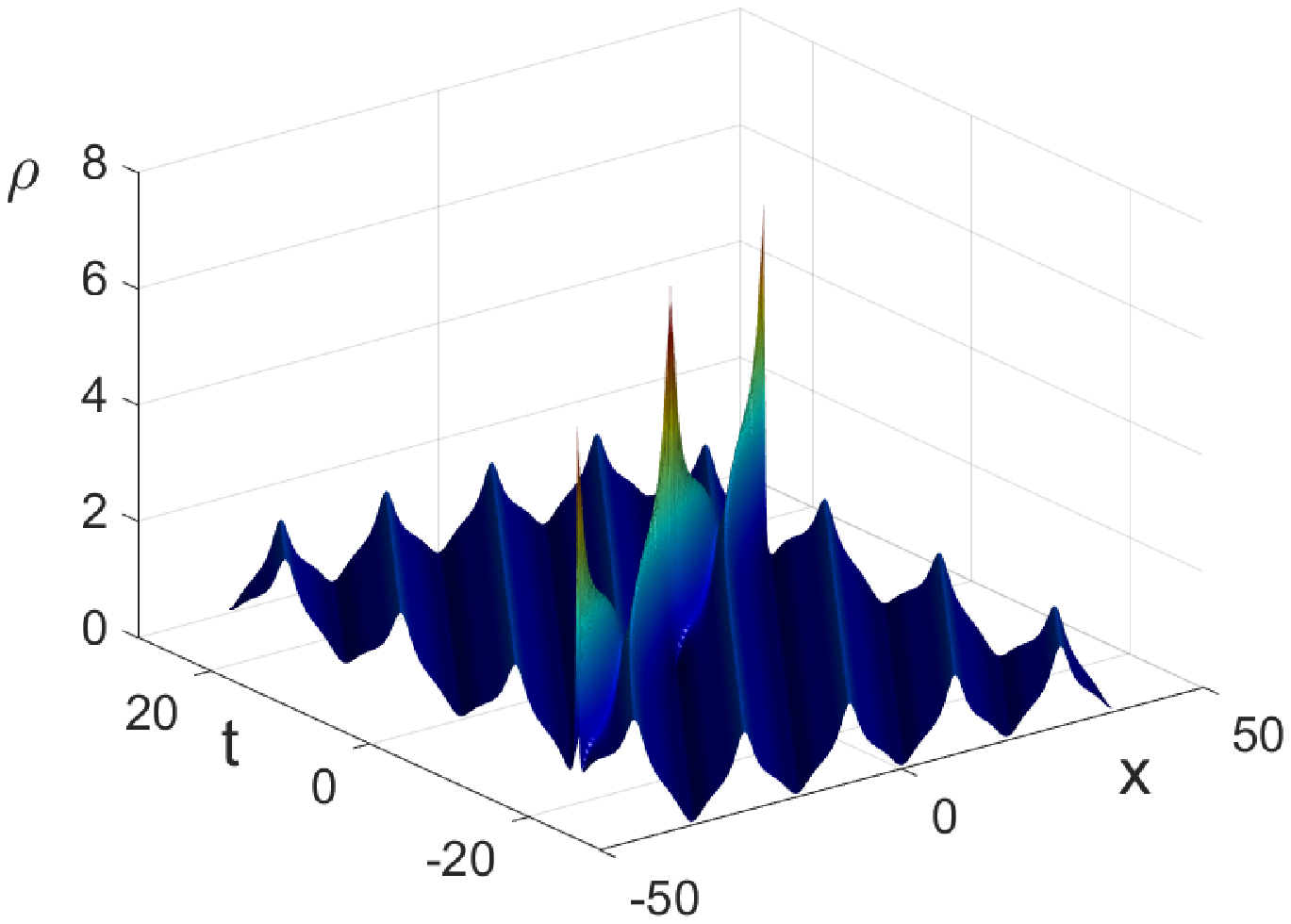}
	\includegraphics[width=0.48\textwidth]{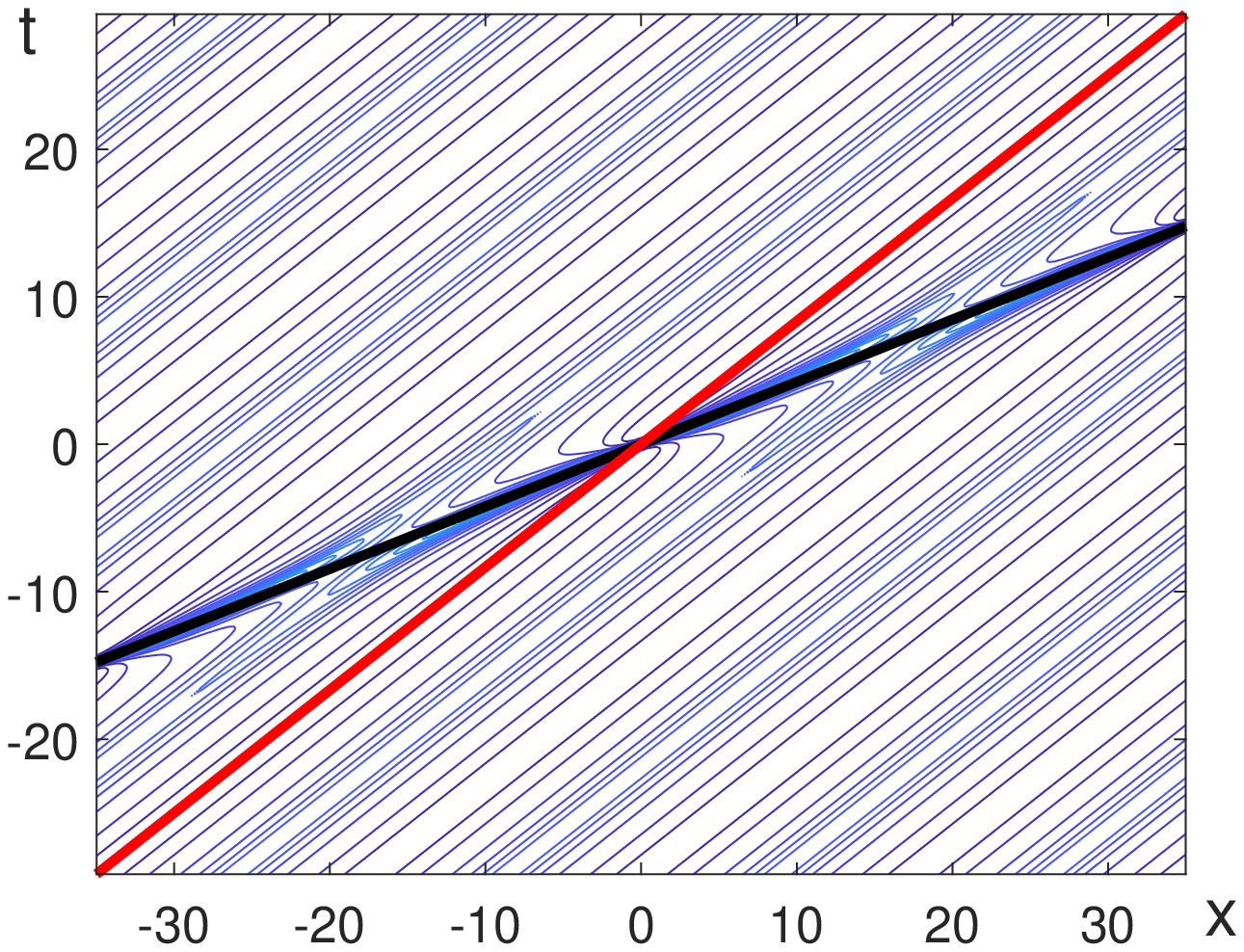} \\
	\includegraphics[width=0.48\textwidth]{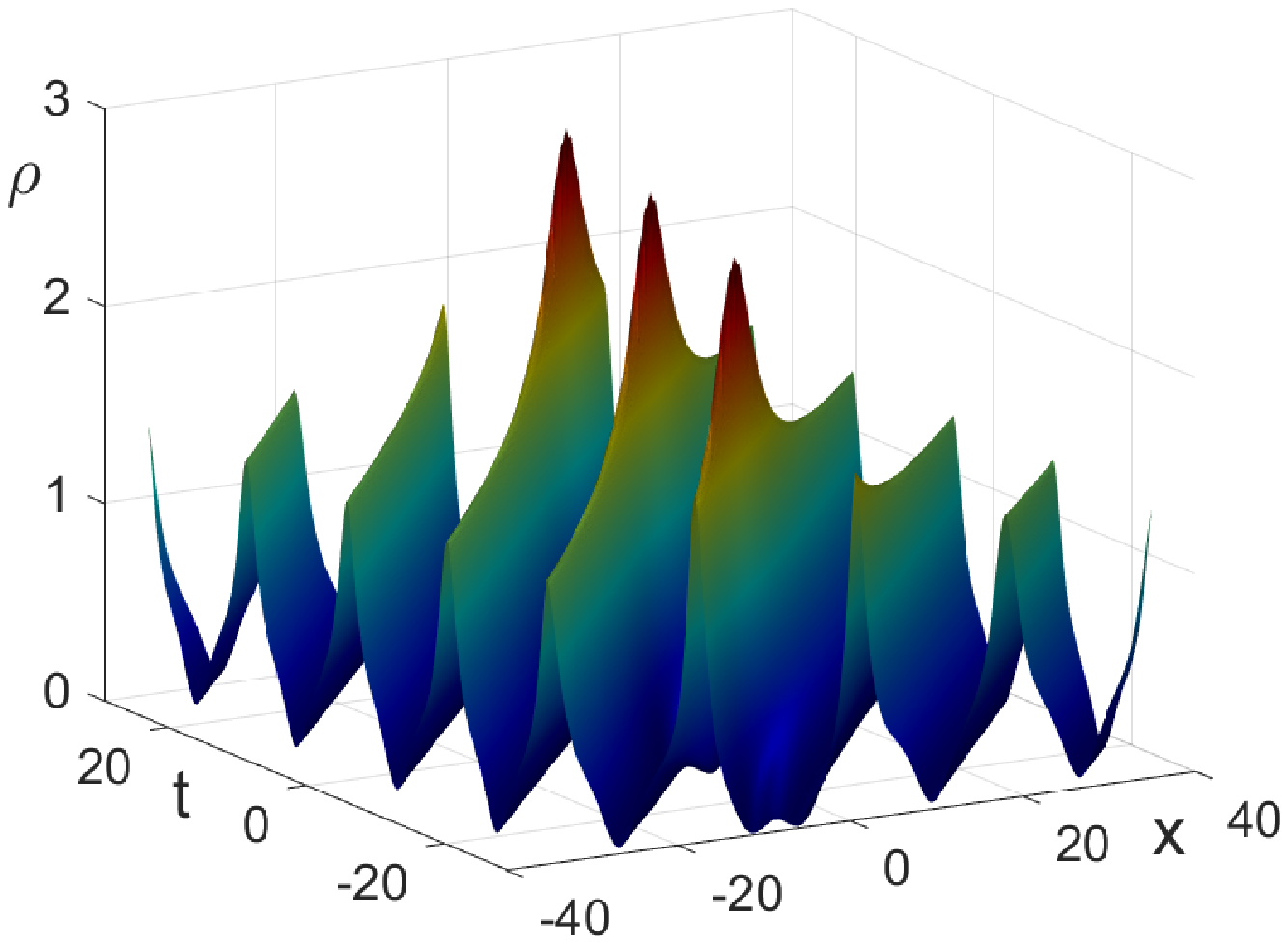}
	\includegraphics[width=0.48\textwidth]{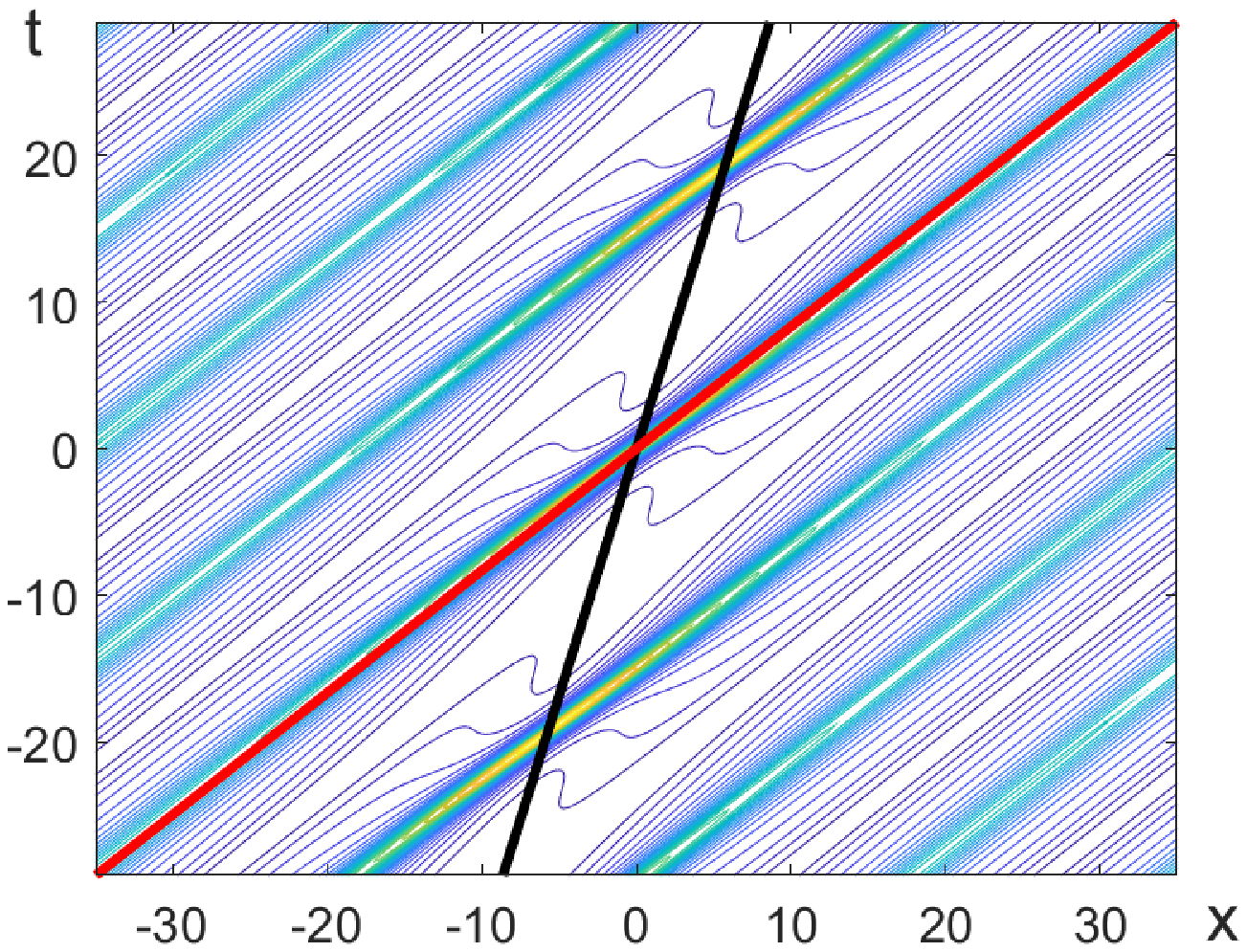} \\
	\caption{New solutions to the DNLS equation 
		which corresponds to the propagation of an algebraic soliton
		on the background of the periodic standing wave (\ref{3.20-simple}): solution surface
		(left) and contour plot (right).
		Top and bottom panels correspond to the solutions obtained by the  transformation (\ref{1foldDT-rogue}) with the eigenvalues
		$\lambda_3 = i \beta_3$ and $\lambda_4 = i \beta_4$
		respectively.} \label{fig-mix2}
\end{figure}

The left panel of Figure \ref{fig-mix1} shows the plot of $\rho$
and $\rho_{\rm tr} := \frac{1}{2} |u_{\rm tr}|^2$
versus $x$ after the transformation (\ref{1foldDT}) for 
the eigenvalue $\lambda_3 = i \beta_3$. 
The transformed wave for the eigenvalue $\lambda_4 = i \beta_4$ 
is a half-period translation of $\rho_{\rm tr}$ for the eigenvalue 
$\lambda_3 = i \beta_3$.
The right panel shows $\rho$ and $\rho_{\rm tr}$ after 
the two-fold transformation (\ref{5.8}) with 
the complex eigenvalue $\lambda_1 = \alpha_1 + i \beta_1$.

Figure \ref{fig-mix2} shows the surface plot of $\hat{\rho} := \frac{1}{2} |\hat{u}|^2$ (left) and the contour plot (right)
after the transformation (\ref{1foldDT-rogue}) associated with
the eigenvalues $\lambda_3 = i \beta_3$ (top) and $\lambda_4 = i \beta_4$ (bottom). The red line on the contour plot shows the line $x + 2ct = 0$
and the black line shows the line (\ref{direction-soliton}).
We can see that the algebraic soliton propagates
along this direction, where the algebraic soliton is modulated
due to interaction with the periodic standing wave.

Similarly to the expressions (\ref{hat-rho-max}) and (\ref{hat-rho-max-2})
for the maxima of algebraic solitons, we obtain 
\begin{equation}
\label{beta-3}
\hat{\rho}_{\rm max} = \frac 14 \left(3 \sqrt{u_1} - \sqrt{u_2} \pm \sqrt{2(\sqrt{\gamma^2 + \eta^2}+\gamma)}\right)^2, 
\end{equation}
where the upper sign is for the eigenvalue
$\lambda_3 = i\beta_3$ and the lower sign is for the eigenvalue $\lambda_4 = i\beta_4$. For the parameters on Fig. \ref{fig-mix2}, we have excellent agreement with $\hat{\rho}_{\rm max} \approx 6.66$ for $\lambda_3 = i \beta_3$ and $\hat{\rho}_{\rm max} \approx 2.76$ for $\lambda_4 = i \beta_4$.

\begin{figure}[htb!]
	\includegraphics[width=0.48\textwidth]{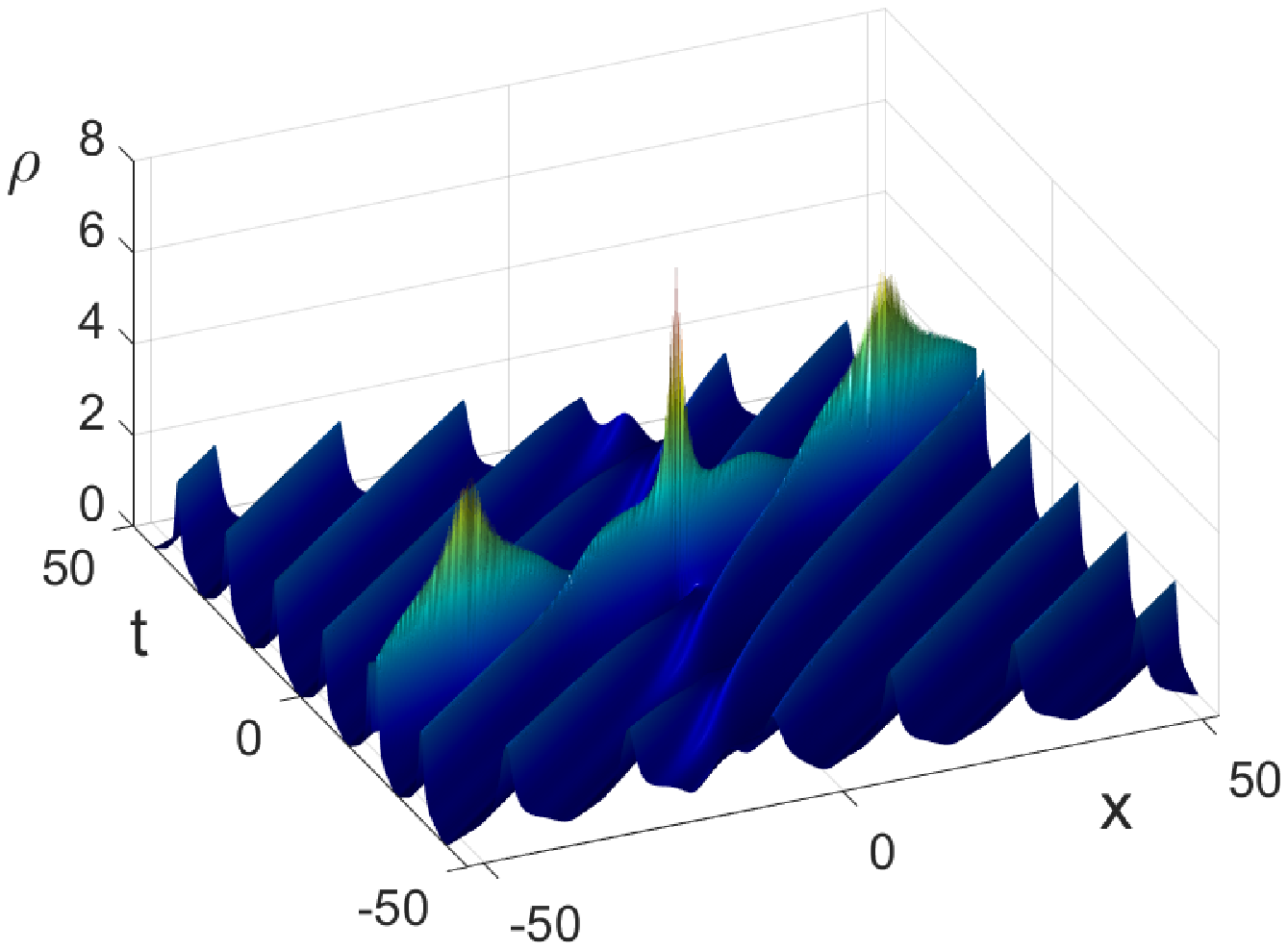}
	\includegraphics[width=0.48\textwidth]{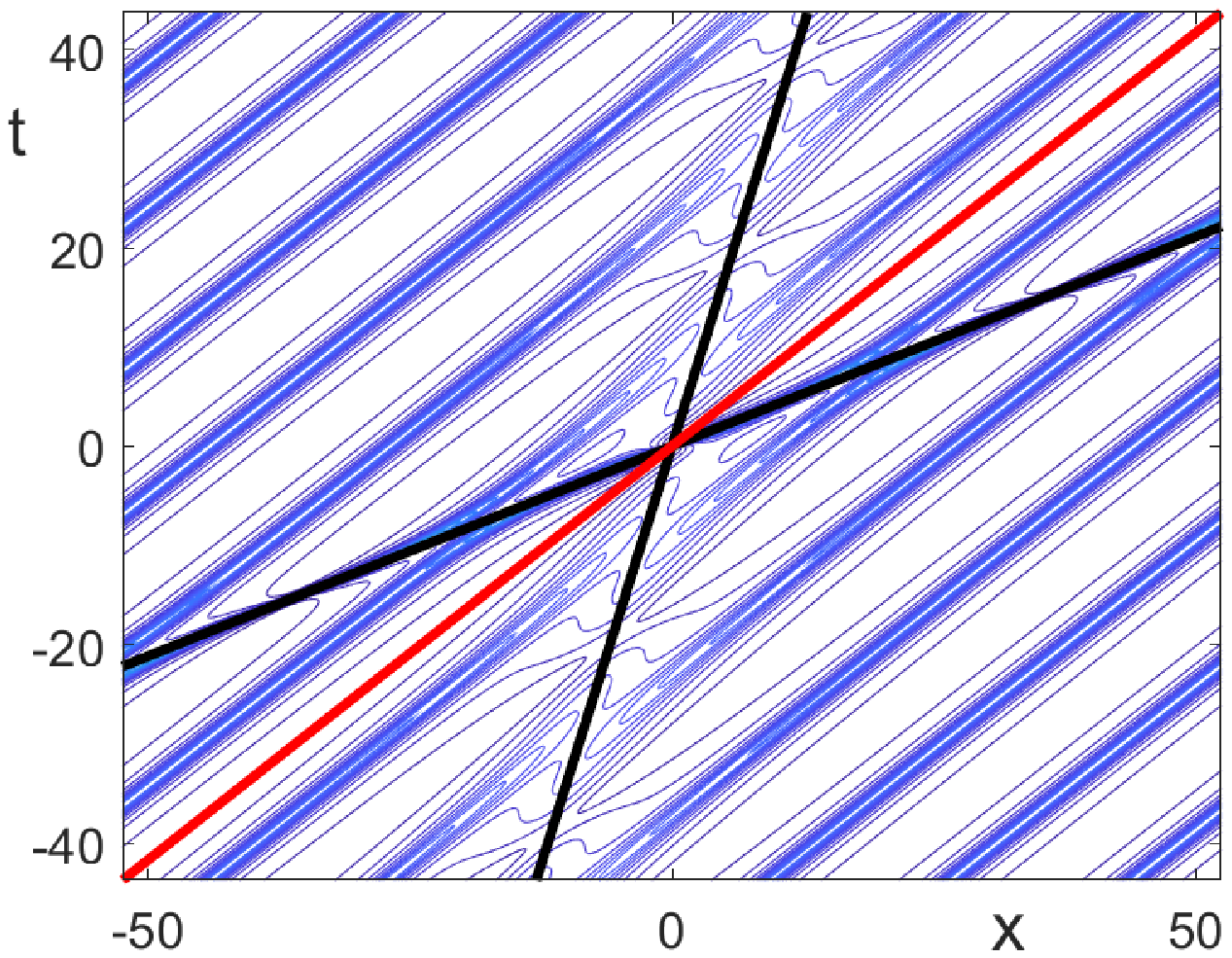}
	\caption{New solutions to the DNLS equation 
		obtained
		by the two-fold transformation with the two eigenvalues
	$\lambda_3 = i \beta_3$ and $\lambda_4 = i \beta_4$
 (left) and contour plot (right).
Two algebraic solitons propagate along the lines shown by
black curves on the right.} \label{fig-mix3}
\end{figure}

Figure \ref{fig-mix3} shows the result of the two-fold transformation
(\ref{5.8a}) associated with the two eigenvalues
$\lambda_3 = i \beta_3$ and $\lambda_4 = i \beta_4$. The background
wave is the same as on the right panel of Fig. \ref{fig-mix1}, that is,
it is a half-period translation of the periodic standing wave (\ref{3.20}).
Two algebraic solitons propagate along the lines (\ref{direction-soliton})
shown by black curves on the contour plots (right) together
with the line $x + 2ct = 0$ shown by the red curve.

\begin{figure}[htb!]
	\begin{center}
	\includegraphics[width=0.6\textwidth]{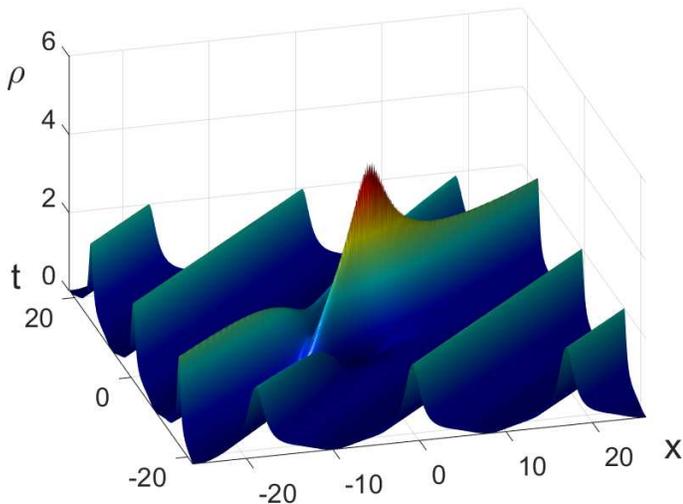}
	\end{center}
	\caption{New solutions to the DNLS equation in variable $\hat{\rho}$
		obtained
		by the two-fold transformation with the quadruplet
		$\lambda_1 = \alpha_1 + i \beta_1$ (right).
	A rogue wave is formed on the background of the
periodic standing wave (\ref{3.20-simple}).} \label{fig-mix4}
\end{figure}

Finally, Figure \ref{fig-mix4} shows the result of the two-fold
transformation (\ref{5.8}) associated with the quadruplet
of complex eigenvalue $\lambda_1 = \alpha_1 + i \beta_1$.
The surface plot of $\hat{\rho}$ indicates that the rogue waves
is fully localized on the background of the periodic standing waves.
This is explained again by the fact that the real and imaginary parts
of the complex-valued equation (\ref{direction-soliton-2})
give two lines intersecting at the only point $(0,0)$.
The maximal amplitude is given by the formula (\ref{max-rogue1})
with $\hat{\rho}_{\rm max} = 8$.

\section{Conclusion}
\label{sec-5}

We have studied the rogue waves and algebraic solitons arising
on the background of periodic standing waves in the derivative NLS
equation. By using comprehensive analysis and numerical visualizations
for selected parameter values, we showed that the modulationally
stable periodic standing waves support propagation of algebraic solitons
along the background, whereas the modulationally unstable periodic
standing waves are associated with the rogue waves localized in space and time.
Although only very few numerical experiments were displayed,
our analysis ensures that the conclusion extends to all
periodic standing waves of the DNLS equation (both sign-definite and sign-indefinite). By using computations reported in the supplementary material, we have derived
exact expressions for the maximal amplitudes of the rogue waves and the
associated algebraic solitons.

This work opens further ways to understanding the rogue wave phenomenon
in the DNLS equation, which is one of the canonical models for dynamics of waves in plasma physics and optics. The recent results complement
the study of rogue waves on the constant-amplitude wave background \cite{YangYang}, maximal amplitudes of hyperelliptic solutions \cite{WrightDNLS},
and modulational instability of periodic standing waves \cite{CPU}.
The next tasks would be to set up physical experiments and to
confirm the modulational instability of the periodic standing waves
and the maximal amplitudes of rogue waves in the way
it was done for in \cite{XuKibler} in hydrodynamical and optical experiments. It is also interesting to understand rogue wave phenomena in a more general setting
of quasi-periodic solutions of the DNLS equation studied in \cite{Chen}.

\vspace{0.25cm}

{\bf Acknowledgements.} This work was supported by the National Natural Science Foundation of China (No. 11971103).

\newpage

\appendix
\section{Proof of one-fold Darboux transformation}
\label{appendix-a}

For convenience, let us denote the solution of the DNLS equation (\ref{dnls}) by $\psi = u$ and the solution of the Lax equations (\ref{lax-equations}) for the eigenvalue $\lambda = \lambda_1$
by $\phi = (p_1,q_1)^T$. Let $T(u,\lambda)$ be the gauge transformation in the form:
\begin{equation}\label{A.1}
T(u,\lambda) = \lambda \left(\begin{array}{cc}
A_1 & 0\\
0 & D_1 \\
\end{array}\right) + \left(\begin{array}{cc}
0 & \lambda_1\\
\lambda_1 & 0 \\
\end{array}\right),
\end{equation}
where
\begin{equation}\label{A.2}
A_1=-\frac{q_1}{p_1}, \qquad D_1 = -\frac{p_1}{q_1},
\end{equation}
so that $A_1 D_1 =1$ and $\det T(u,\lambda) = \lambda^2 - \lambda_1^2$.
We impose the constraints on $T(u,\lambda)$ from the condition
that if $\phi$ is a solution of the Lax equations (\ref{lax-equations}) 
with $u$, 
then $\hat{\phi} = T(u,\lambda) \phi$ is also a solution of the Lax equations
(\ref{lax-equations}) with a new solution 
of the DNLS equation (\ref{dnls}), which we denote by $\hat{u}$. This condition yields
the Darboux equations:
\begin{equation}\label{A.6}
U(\hat{u},\lambda) = (T_x + T U)T^{-1}, \qquad
V(\hat{u},\lambda) = (T_t+TV)T^{-1},
\end{equation}
where $U = U(u,\lambda)$ and $T = T(u,\lambda)$. By using  (\ref{A.1}) and (\ref{A.6}), we obtain the following system in different powers of $\lambda$:
\begin{eqnarray}
\label{A.3-1}
\left\{ \begin{array}{l}
A_{1,x} - \lambda_1 \bar{u} = \lambda_1 \hat{u}, \\	
u A_1 + 2 i \lambda_1 = \hat{u}D_1, \\
\bar{u}D_1 + 2 i \lambda_1 = \bar{\hat{u}} A_1, \\
D_{1,x} + \lambda_1 u = -\lambda_1 \bar{\hat{u}}
\end{array} \right.
\end{eqnarray}
and
\begin{eqnarray}
\label{A.4-1}
\left\{ \begin{array}{l}
i A_1 |u|^2 - 2 \lambda_1 \bar{u} = i A_1 |\hat{u}|^2 + 2 \lambda_1 \hat{u},\\	
A_{1,t} + \lambda_1 (i \bar{u}_x + |u|^2 \bar{u}) = \lambda_1 (i \hat{u}_x - |\hat{u}|^2 \hat{u}),\\
A_1 (i u_x - |u|^2 u) - i \lambda_1 |u|^2 =
D_1 (i \hat{u}_x - |\hat{u}|^2 \hat{u}) + i \lambda_1 |\hat{u}|^2, \\
D_1 (i \bar{u}_x + |u|^2 \bar{u}) + i \lambda_1 |u|^2 =
A_1 (i \bar{\hat{u}}_x + |\hat{u}|^2 \bar{\hat{u}}) - i \lambda_1 |\hat{u}|^2, \\
-i D_1 |u|^2 + 2 \lambda_1 u = -i D_1 |\hat{u}|^2 - 2 \lambda_1 \bar{\hat{u}},\\	
D_{1,t} + \lambda_1 (i u_x - |u|^2 u) = \lambda_1 (i \bar{\hat{u}}_x + |\hat{u}|^2 \bar{\hat{u}}).
\end{array} \right.
\end{eqnarray}

It follows from (\ref{A.2}) and the Lax equations (\ref{lax-equations}) that
\begin{eqnarray}
\label{A.3}
\left\{ \begin{array}{l}
A_{1,x} = \lambda_1 \bar{u} +\lambda_1 u A_1^2 + 2 i \lambda_1^2 A_1,\\
D_{1,x} = -\lambda_1 u - \lambda_1 \bar{u} D_1^2 - 2 i \lambda_1^2 D_1,
\end{array} \right.
\end{eqnarray}
and
\begin{eqnarray}
\label{A.4}
\left\{ \begin{array}{l}
A_{1,t} = 2\lambda_1^3 \bar{u} -\lambda_1 (i \bar{u}_x +|u|^2 \bar{u}) + 4 i \lambda_1^4 A_1 -2i \lambda_1^2 A_1 |u|^2
+2 \lambda_1^3A_1^2 u + \lambda_1 A_1^2(i u_x - |u|^2 u), \\
D_{1,t} = -2\lambda_1^3u -\lambda_1(iu_x-|u|^2u) - 4i \lambda_1^4 D_1 + 2i\lambda_1^2 D_1 |u|^2  -2\lambda_1^3D_1^2\bar{u} +\lambda_1 D_1^2 (i\bar{u}_x+|u|^2\bar{u}).
\end{array} \right.
\end{eqnarray}

Substituting (\ref{A.3}) into (\ref{A.3-1}) shows that the first and fourth equations of system (\ref{A.3-1}) are redundant, whereas the other two equations produce the following transformation formulas:
\begin{eqnarray}
\label{A.8}
\hat{u} = u \frac{q_1^2}{p_1^2} - 2 i \lambda_1 \frac{q_1}{p_1}, \qquad
\bar{\hat{u}} = \bar{u} \frac{p_1^2}{q_1^2} - 2 i \lambda_1 \frac{p_1}{q_1}.
\end{eqnarray}
The complex-conjugate reduction in (\ref{A.8}) is satisfied if $|p_1| = |q_1|$ and $\lambda_1 + \bar{\lambda}_1 = 0$. Hence, $\lambda_1 = i \beta_1 \in i \mathbb{R}$. Selecting $q_1 = -i\bar{p}_1$ as in the reduction (\ref{constraint-red-1}) gives the one-fold transformation formula
(\ref{5.2}) from system (\ref{A.8}).

In view of relations (\ref{A.8}), the first and fifth equations in system (\ref{A.4-1}) are redundant. Substituting (\ref{A.4}) into (\ref{A.4-1}) shows that the second and sixth equations of system (\ref{A.4-1}) are redundant. Finally, the third and fourth equations of system (\ref{A.4-1})
are satisfied when we substitute derivatives of (\ref{A.8}) in $x$ and use
relations (\ref{A.3}).

\section{Proof of two-fold Darboux transformation}
\label{appendix-b}

Let $\phi =(p_1,q_1)^T$ and $\phi =(p_2,q_2)^T$ be two solutions of the Lax equations (\ref{lax-equations}) for eigenvalues $\lambda = \lambda_1$ and $\lambda = \lambda_2$ satisfying $\lambda_1^2\neq \bar{\lambda}_2^2$. Let $T(u,\lambda)$ be the gauge transformation in the form:
\begin{equation}\label{B.1}
T(u,\lambda) =
\left(\begin{array}{cc}
\mathcal{T}_{11}(\lambda) & \mathcal{T}_{12}(\lambda)\\
\mathcal{T}_{21}(\lambda) & \mathcal{T}_{22}(\lambda)\\
\end{array}\right) = \lambda^2 \mathcal{T}_2 + \lambda \mathcal{T}_1+ \mathcal{T}_0,
\end{equation}
where
\begin{equation}
\mathcal{T}_2 = \left(\begin{array}{cc}
A_2 & 0\\
0 & D_2 \\
\end{array}\right),\quad \mathcal{T}_1 = \left(\begin{array}{cc}
0 & B_1\\
C_1 & 0 \\
\end{array}\right),\quad \mathcal{T}_0 = \left(\begin{array}{cc}
A_0 & 0\\
0 & A_0 \\
\end{array}\right), 
\label{B.2}
\end{equation}
with 
$$
A_2 = \frac{\lambda_1\alpha_1 - \lambda_2\alpha_2}{\lambda_1\alpha_2 - \lambda_2\alpha_1},\quad
B_1= \frac{\lambda_2^2-\lambda_1^2}{\lambda_1\alpha_2 - \lambda_2\alpha_1},\quad
C_1 = \frac{\alpha_1\alpha_2(\lambda_2^2-\lambda_1^2)}{\lambda_1\alpha_1 - \lambda_2\alpha_2}
$$
and
$$
A_0 =\lambda_1\lambda_2,\quad A_2D_2=1,\quad \alpha_1= \frac{q_1}{p_1}, \quad 
\alpha_2 = \frac{q_2}{p_2}.
$$
It follows from (\ref{lax-equations}) that $\alpha_k$ for $k = 1,2$ satisfy the following Riccati equations:
\begin{eqnarray}
\alpha_{kx}&=& -\lambda_k \bar{u} +2i\lambda_k^2\alpha_k - \lambda_k u \alpha_k^2, \label{B.3}\\
\alpha_{kt}&=& -2\lambda_k^3\bar{u} +\lambda_k(i\bar{u}_x+|u|^2\bar{u})+2i\lambda_k^2\alpha_k (2\lambda_k^2-|u|^2) \nonumber\\&&-
\alpha_k^2[2\lambda_k^3 u +\lambda_k(iu_x - |u|^2u)].\label{B.4}
\end{eqnarray}
In addition, we check that
\begin{equation}\label{B.5}
\det T(u,\lambda)=(\lambda^2 - \lambda_1^2) (\lambda^2 - \lambda_2^2).
\end{equation}
By using the Darboux equations (\ref{A.6}), we show that the new solution $\hat{u}$ is expressed by
\begin{equation}\label{B.8}
	\hat{u}=A_2^2 u + 2i A_2 B_1,\qquad \bar{\hat{u}} = D_2^2\bar{u} + 2i C_1 D_2.
\end{equation}
Let $T^* := T^{-1} \det T$ be the adjugate matrix of $T(u,\lambda)$ and
\begin{equation}
(T_x+T U)T^*=\left(\begin{array}{cc} \mathcal{F}_{11}(\lambda,u)& \mathcal{F}_{12}(\lambda,u)\\
\mathcal{F}_{21}(\lambda,u)&\mathcal{F}_{22}(\lambda,u)\\
\end{array}\right).\label{B.9}
\end{equation}
It follows from \eqref{matrix-U} and \eqref{B.1} that $\mathcal{F}_{11}(\lambda,u)$ and $\mathcal{F}_{22}(\lambda,u)$ are the sixth-order polynomials in $\lambda$, and $\mathcal{F}_{12}(\lambda,u)$ and $\mathcal{F}_{21}(\lambda,u)$ are the fifth-order polynomials in $\lambda$. It follows from \eqref{B.1} and \eqref{B.2} that for any $k = 1,2$, 
\begin{equation}\label{B.10}
\mathcal{T}_{11}(\lambda_k) = - \mathcal{T}_{12}(\lambda_k) \alpha_k, \quad \mathcal{T}_{11,x}(\lambda_k)= - \mathcal{T}_{12,x}(\lambda_k) \alpha_k - \mathcal{T}_{12}(\lambda_k) \alpha_{kx},
\end{equation}
\begin{equation}\label{B.11}
\mathcal{T}_{21}(\lambda_k) = - \mathcal{T}_{22}(\lambda_k) \alpha_k, \quad \mathcal{T}_{21,x}(\lambda_k)= - \mathcal{T}_{22,x}(\lambda_k) \alpha_k - \mathcal{T}_{22}(\lambda_k) \alpha_{kx}, 
\end{equation}
and
\begin{align}
\mathcal{T}_{11}(-\lambda_k)= \mathcal{T}_{11}(\lambda_k),\quad \mathcal{T}_{12}(-\lambda_k)=- \mathcal{T}_{12}(\lambda_k), \nonumber \\
\mathcal{T}_{21}(-\lambda_k)= -\mathcal{T}_{21}(\lambda_k),\quad  \mathcal{T}_{22}(-\lambda_k) = \mathcal{T}_{22}(\lambda_k).
\label{B.12}
\end{align}
It follows from \eqref{B.3} that $\pm \lambda_1$ and $\pm \lambda_2$ are the roots of $\mathcal{F}_{ij}(\lambda,u)$ for every $i,j = 1,2$. For instance, since 
$$
\mathcal{F}_{11}(\lambda,u)= \mathcal{T}_{22}(-i\lambda^2 \mathcal{T}_{11} - \lambda \bar{u} \mathcal{T}_{12}+\mathcal{T}_{11,x})
- \mathcal{T}_{21} (\lambda u \mathcal{T}_{11} + i \lambda^2 \mathcal{T}_{12} + \mathcal{T}_{12,x}),
$$
we confirm
$$
\mathcal{F}_{11}(\pm \lambda_k,u) = \mathcal{T}_{12}(\lambda_k) \mathcal{T}_{22}(\lambda_k)
(-\alpha_{kx} -\lambda_k \bar{u} +2i\lambda_k^2\alpha_k - \lambda_k u \alpha_k^2)=0.
$$

Dividing both sides of (\ref{B.9}) by $\det{T}$ and using \eqref{B.5} yields
$$
U(\hat{u},\lambda) := \lambda^2 \mathcal{U}_2 + \lambda \mathcal{U}_1,
$$
where the matrixes $\mathcal{U}_1$ and $\mathcal{U}_2$ are independent of $\lambda$. Substituting these expressions into the first Darboux equation (\ref{A.6}) and using $U(u,\lambda) = \lambda^2 U_2 + \lambda U_1$, we obtain
\begin{equation}
\label{B.13} \begin{array} {l}
\lambda^4 \mathcal{T}_2 U_2 + \lambda^3 (\mathcal{T}_1 U_2 + \mathcal{T}_2 U_1) + \lambda^2 (\mathcal{T}_1 U_1 + \mathcal{T}_0 U_2 + \mathcal{T}_{2x}) + \lambda(\mathcal{T}_0 U_1 + \mathcal{T}_{1x})\\
=\lambda^4 \mathcal{U}_2 \mathcal{T}_2 +\lambda^3 (\mathcal{U}_2 \mathcal{T}_1 + \mathcal{U}_1 \mathcal{T}_2) +
\lambda^2(\mathcal{U}_2 \mathcal{T}_0 + \mathcal{U}_1 \mathcal{T}_1) + \lambda \mathcal{U}_1 \mathcal{T}_0.
\end{array}
\end{equation}
The comparison of $\lambda^4$ and $\lambda^3$ on both sides of \eqref{B.13} delivers
$$
\mathcal{U}_2 = \left( \begin{array}{cc}
-i & 0\\
0 & i\\
\end{array} \right), \qquad 
\mathcal{U}_1 = \left( \begin{array}{cc}
0 & A_2^2 u + 2 i A_2 B_1\\
- D_2^2 \bar{u} - 2 i D_2 C_1 & 0\\
\end{array} \right)=: \left( \begin{array}{cc}
0 & \hat{u}\\
-\bar{\hat{u}} & 0\\
\end{array} \right),
$$
which confirms the claim in \eqref{B.8}. Also, the coefficients of $\lambda^2$ and $\lambda$ on both sides of \eqref{B.13} yield four identities
\begin{equation} \label{B.14} \begin{array}{lll}
A_{2x} &=& B_1 \bar{u} + C_1 ( A_2^2 u + 2 i A_2 B_1), \quad D_{2x}= - C_1 u - B_1 (D_2^2 \bar{u} + 2 i D_2 C_1),\\
B_{1x} &=& A_0 (A_2^2 u + 2 i A_2 B_1 - u), \quad C_{1x} = A_0 (\bar{u} - D_2^2 \bar{u} - 2 i D_2 C_1),
\end{array} \end{equation}
which can be verified by using the formulas \eqref{B.2} and \eqref{B.3}.

For the time-evolution equation, let us denote
\begin{equation}
(T_t+T V)T^*=\left(\begin{array}{cc} \mathcal{G}_{11}(\lambda,u) & \mathcal{G}_{12}(\lambda,u)\\
\mathcal{G}_{21}(\lambda,u) &\mathcal{G}_{22}(\lambda,u)\\
\end{array}\right).\label{B.15}
\end{equation}
It is seen from \eqref{matrix-V} and \eqref{B.1} that $\mathcal{G}_{11}(\lambda,u)$ and $\mathcal{G}_{22}(\lambda,u)$ are the eighth-order polynomials in $\lambda$, and $\mathcal{G}_{12}(\lambda,u)$ and $\mathcal{G}_{21}(\lambda,u)$ are the seventh-order polynomials in $\lambda$. It follows from \eqref{B.10} and \eqref{B.11} that
$$
\mathcal{T}_{11,t}(\lambda_k)= - \mathcal{T}_{12,t}(\lambda_k) \alpha_k - \mathcal{T}_{12}(\lambda_k) \alpha_{kt}, \qquad
\mathcal{T}_{21,t}(\lambda_k)= - \mathcal{T}_{22,t}(\lambda_k) \alpha_k - \mathcal{T}_{22}(\lambda_k) \alpha_{kt},
$$
which together with \eqref{B.4} and \eqref{B.12} indicates that $\pm \lambda_1$ and $\pm \lambda_2$ are the roots of the polynomials $\mathcal{G}_{ij}(\lambda,u)$ for all $i,j = 1,2$. For instance, using 
\begin{eqnarray*}
	\mathcal{G}_{11}(\lambda,u)&=& \mathcal{T}_{11}\mathcal{T}_{22}(-2i \lambda^4+ i\lambda^2|u|^2) + \mathcal{T}_{12}\mathcal{T}_{22}[-2\lambda^3 \bar{u} + \lambda (i \bar{u}_x +|u|^2 \bar{u})] + \mathcal{T}_{11,t}\mathcal{T}_{22}\\&&
	-\mathcal{T}_{11}\mathcal{T}_{21} [2\lambda^3 u +\lambda (i u_x -|u|^2 u)]- \mathcal{T}_{12}\mathcal{T}_{21}(2i \lambda^4 - i \lambda^2 |u|^2)- \mathcal{T}_{21}\mathcal{T}_{12,t},
\end{eqnarray*}
we confirm
\begin{eqnarray*}
	\mathcal{G}_{11}(\pm \lambda_k,u) &=& \mathcal{T}_{12}(\lambda_k) \mathcal{T}_{22}(\lambda_k) \left[ 2\alpha_k (2 i \lambda_k^4 - i \lambda_k^2 |u|^2) - \alpha_k^2[2 \lambda_k^3 u + \lambda_k (i u_x - |u|^2u)] \right. \\&&
	\left. - 2\lambda_k^3 \bar{u} + \lambda_k (i \bar{u}_x + |u|^2 \bar{u}) - \alpha_{kt} \right] = 0.
\end{eqnarray*}

Dividing both sides of (\ref{B.15}) by $\det T$ and using \eqref{B.5} yields
$$
V(\hat{u},\lambda) := \lambda^4 \mathcal{V}_4 + \lambda^3 \mathcal{V}_3 + \lambda^2 \mathcal{V}_2 + \lambda \mathcal{V}_1,
$$
where the matrices $\mathcal{V}_{1,2,3,4}$ are independent of $\lambda$. Substituting these expressions into the second Darboux equation (\ref{A.6}) and using $V(u,\lambda) = \lambda^4 V_4 + \lambda^3 V_3 + \lambda^2 V_2 + \lambda V_1$, we obtain
\begin{equation}
\label{B.16} \begin{array} {l}
\lambda^6 \mathcal{T}_2 V_4 +\lambda^5 (\mathcal{T}_2 V_3 + \mathcal{T}_1 V_4) + \lambda^4 (\mathcal{T}_2 V_2 +
\mathcal{T}_1 V_3 + \mathcal{T}_0 V_4) + \lambda^3 (\mathcal{T}_2 V_1 + \mathcal{T}_1 V_2 + \mathcal{T}_0 V_3)\\
+ \lambda^2 (\mathcal{T}_1 V_1 + \mathcal{T}_0 V_2 + \mathcal{T}_{2t}) + \lambda (\mathcal{T}_0 V_1 + \mathcal{T}_{1t})
= \lambda^6 \mathcal{V}_4 \mathcal{T}_2 + \lambda^5 (\mathcal{V}_3 \mathcal{T}_2 + \mathcal{V}_4 \mathcal{T}_1) \\
+ \lambda^4 (\mathcal{V}_2 \mathcal{T}_2 + \mathcal{V}_3 \mathcal{T}_1 + \mathcal{V}_4 \mathcal{T}_0) +
\lambda^3 (\mathcal{V}_1 \mathcal{T}_2 + \mathcal{V}_2 \mathcal{T}_1 + \mathcal{V}_3 \mathcal{T}_0) +
\lambda^2 (\mathcal{V}_1 \mathcal{T}_1 + \mathcal{V}_2 \mathcal{T}_0) + \lambda \mathcal{V}_1 \mathcal{T}_0.
\end{array}\end{equation}
The comparison of $\lambda^6$, $\lambda^5$, $\lambda^4$, and $\lambda^3$ on both sides of (\ref{B.16}) yields
$$
\mathcal{V}_4 = \left( \begin{array}{cc}
-2i & 0\\
0 & 2i\\
\end{array} \right),
$$
$$
\mathcal{V}_3 = \left( \begin{array}{cc}
0 & 2 A_2^2 u + 4 i A_2 B_1\\
-2 D_2^2 \bar{u} - 4 i D_2 C_1 & 0\\
\end{array} \right) = \left( \begin{array}{cc}
0 & 2\hat{u}\\
-2\bar{\hat{u}} & 0\\
\end{array} \right),
$$
$$
\mathcal{V}_2 = \left( \begin{array}{cc}
\mathcal{V}_2^{(11)} & 0\\
0 & \mathcal{V}_2^{(22)}\\
\end{array} \right) = \left( \begin{array}{cc}
i|\hat{u}|^2 & 0\\
0 & -i |\hat{u}|^2\\
\end{array} \right), 
$$
and
$$
\mathcal{V}_1 = \left( \begin{array}{cc}
0 & \mathcal{V}_1^{(12)}\\
\mathcal{V}_1^{(21)} & 0\\
\end{array} \right) = \left( \begin{array}{cc}
0 & i \hat{u}_x - |\hat{u}|^2 \hat{u} \\
i \bar{\hat{u}}_x + |\hat{u}|^2 \bar{\hat{u}} & 0\\
\end{array} \right),
$$
where
\begin{eqnarray*}
\mathcal{V}_2^{(11)} &=& i |u|^2 - 2 D_2 B_1 \bar{u} - C_1(2 u A_2 + 4 i B_1),\\ \mathcal{V}_2^{(22)} &=& - i |u|^2 + 2 u C_1 A_2 + B_1 (2 D_2 \bar{u} + 4i C_1),\\
\mathcal{V}_1^{(12)} &=& A_2^2 (i u_x - |u|^2 u) - 2A_2 (i B_1 |u|^2 - A_0 u) + 2 B_1^2 \bar{u} + 2 (B_1 C_1 -A_0A_2) (u A_2^2 + 2 i A_2 B_1),  \\
\mathcal{V}_1^{(21)} &=& D_2^2 (i \bar{u}_x + |u|^2 \bar{u}) + 2 D_2(i C_1|u|^2 - A_0 \bar{u}) - 2 C_1^2 u
	- 2 (B_1 C_1 -A_0 D_2) (D_2^2 \bar{u} + 2 i C_1 D_2).
\end{eqnarray*}
Note that the equality for $\mathcal{V}_1$ can be confirmed 
in view of \eqref{B.8} and \eqref{B.14}. Furthermore, the comparison of $\lambda^2$ and $\lambda$ on both sides of \eqref{B.16} gives rise to four identities
$$
i B_1 \bar{u}_x + B_1 |u|^2 \bar{u} + i A_0 |u|^2 + A_{2t} = i A_0|u|^2 - 2 D_2 B_1 A_0 \bar{u} - A_0 C_1
(2 A_2 u + 4 i B_1) +C_1 \mathcal{V}_1^{(12)},
$$
$$
i C_1 u_x - C_1 |u|^2 u - i A_0 |u|^2 + D_{2t} = - i A_0 |u|^2 + 2 A_0 A_2 C_1 u + A_0 B_1 (2 D_2 \bar{u} + 4 i C_1)
+ B_1 \mathcal{V}_1^{(21)},
$$
and
$$
i A_0 u_x - A_0|u|^2 u + B_{1t} = A_0 \mathcal{V}_1^{(12)}, \qquad
i A_0 \bar{u}_x + A_0 |u|^2 \bar{u} + C_{1t} = A_0 \mathcal{V}_1^{(21)},
$$
which could be directly verified by using \eqref{B.2} and \eqref{B.4}. Once again, the time-dependent problem confirms validity of the Darboux transformation formula \eqref{B.8}.

If we restrict $\lambda_{1,2}=i\beta_{1,2}$ and $q_{1,2}= -i\bar{p}_{1,2}$ 
in (\ref{B.8}), the we obtain the two-fold Darboux transformation \eqref{5.8a}.
If we restrict $\lambda_2 = \bar{\lambda}_1$, $q_2 = -\bar{p}_1$, and $p_2 = \bar{q}_1$ in (\ref{B.8}), then we obtain the two-fold Darboux transformation \eqref{5.8}.

\section{Proof of (\ref{hat-rho})}
\label{app-C}

For $\lambda_1 = i \beta_1$ with $\beta_1 \in \mathbb{R}$, we substitute \eqref{4.2}, \eqref{4.4}, and \eqref{4.4a}
with $Q_1 = -i\bar{P}_1$ into \eqref{5.2} and obtain
\begin{eqnarray}
\hat{u} = - \frac {\bar{P}_1^2}{P_1^2} \frac {b - \beta_1^4 + \frac{1}{2} \beta_1^2 \rho + \frac{\beta_1^2}{2 \rho} (i \rho' + a)}{b + c \beta_1^2 + \beta_1^4 + \beta_1^2 \rho} R e^{-i\theta}  e^{-8ibt}. \label{a7.6}
\end{eqnarray}
Expressing $\rho'$ from (\ref{3.8}) and (\ref{quartic-Q}) 
for $\hat{\rho} = \frac{1}{2} |\hat{u}|^2$, we reduce (\ref{a7.6}) 
to (\ref{b7.6}), which we can write as $\mathcal{N}_1/\mathcal{D}_1$.
Substituting parameters $(a,b,c,d)$ from (\ref{3.25}), 
the periodic wave $\rho$ from (\ref{3.14}), and the eigenvalue $\beta_1$ from (\ref{3.27a}) into (\ref{b7.6}) yields
\begin{eqnarray*}
	\mathcal{D}_1 &=& \left[(u_2 - u_4) (b + c \beta_1^2 + \beta_1^4 + u_1 \beta_1^2) + (u_1 - u_2) (b + c \beta_1^2 + \beta_1^4 + u_4 \beta_1^2) {\rm sn}^2(\nu x;k) \right]^2\\
	&=& \frac18 \beta_1^2 (\sqrt{u_1} + \sqrt{u_2})^2 (\sqrt{u_1} + \sqrt{u_4})^2 (\sqrt{u_2} + \sqrt{u_4})^2 \times \\ &&
	\left[(\sqrt{u_1} + \sqrt{u_3}) (\sqrt{u_2} - \sqrt{u_4}) + (\sqrt{u_1} - \sqrt{u_2}) (\sqrt{u_3} + \sqrt{u_4})
	{\rm sn}^2(\nu x;k)\right]^2,
\end{eqnarray*}
and 
\begin{eqnarray*}
	\mathcal{N}_1 &=& {\rm sn}^4(\nu x;k) (u_1 - u_2)^2 [\beta_1^2 (a + u_4^2) (b - c \beta_1^2 - \beta_1^4) + 2 d \beta_1^4 + u_4[b^2 + \beta_1^4(2b +a - c^2) + \beta_1^8]]\\&&
	+ {\rm sn}^2(\nu x;k) (u_1 - u_2) (u_2 - u_4)[2 \beta_1^2 (a + u_1 u_4) (b - c \beta_1^2 - \beta_1^4) + 4d \beta^4_1 \\&& + (u_1 + u_4) [b^2 + \beta_1^4 (2b +a - c^2) +\beta_1^8]] + (u_2 - u_4)^2 [
	\beta_1^2 (a+ u_1^2) (b - c \beta_1^2 - \beta_1^4) \\&& + 2d\beta_1^4 + u_1 [b^2 + \beta_1^4 (2b +a - c^2) +\beta_1^8]]\\
	&=& \frac{\beta_1^2}{32} (\sqrt{u_1} + \sqrt{u_2})^2 (\sqrt{u_1} + \sqrt{u_4})^2 (\sqrt{u_2} + \sqrt{u_4})^2 \times \\&& [{\rm sn}^2(\nu x;k) (\sqrt{u_1} - \sqrt{u_2}) (\sqrt{u_3} + \sqrt{u_4}) +
	(\sqrt{u_2} - \sqrt{u_4}) (\sqrt{u_1} + \sqrt{u_3})] \times \\ &&
	[{\rm sn}^2(\nu x;k) (\sqrt{u_1} - \sqrt{u_2}) (\sqrt{u_3} + \sqrt{u_4}) (\sqrt{u_1} + \sqrt{u_2} + \sqrt{u_3}- \sqrt{u_4})^2 \\&& + (\sqrt{u_2} - \sqrt{u_4}) (\sqrt{u_1} + \sqrt{u_3})
	(\sqrt{u_1} - \sqrt{u_2} - \sqrt{u_3}- \sqrt{u_4})^2].
\end{eqnarray*}
Canceling the common factors of $\mathcal{D}_1$ and $\mathcal{N}_1$ yields 
the quotient:
\begin{eqnarray*}
	\hat{\rho} &=& 
	\frac{1}{4} (\sqrt{u_1} + \sqrt{u_2} + \sqrt{u_3}- \sqrt{u_4})^2 \\
	&& 
	- \frac{(\sqrt{u_1} + \sqrt{u_3}) (\sqrt{u_2} - \sqrt{u_4}) (\sqrt{u_1}-\sqrt{u_4}) (\sqrt{u_2} + \sqrt{u_3})}{(\sqrt{u_1} + \sqrt{u_3}) (\sqrt{u_2} - \sqrt{u_4}) + (\sqrt{u_1} - \sqrt{u_2}) (\sqrt{u_3} + \sqrt{u_4})
	{\rm sn}^2(\nu x;k)},
\end{eqnarray*}
which is equivalent to \eqref{hat-rho} with notations in \eqref{roots-v1-v4}.

\section{Proof of (\ref{5.10})}
\label{app-E}

Substituting \eqref{4.2} into \eqref{5.8} yields
\begin{eqnarray}
\hat{u} =  \left(\frac{\bar{\lambda}_1|P_1|^2 + \lambda_1|Q_1|^2}{\lambda_1 |P_1|^2 + \bar{\lambda}_1 |Q_1|^2}\right)^2 \left[ R -
\frac{2i (\lambda_1^2 - \bar{\lambda}_1^2) P_1\bar{Q}_1}
{\bar{\lambda}_1 |P_1|^2 + \lambda_1 |Q_1|^2} \right] e^{i \theta}.
\label{5.9}
\end{eqnarray}
so that 
\begin{eqnarray}
\label{expr-app-E}
	\hat{\rho} = \frac{1}{2} \left| R - \frac{2i (\lambda_1^2 - \bar{\lambda}_1^2) P_1\bar{Q}_1}
	{\bar{\lambda}_1 |P_1|^2 + \lambda_1 |Q_1|^2} \right|^2,
\end{eqnarray}
where $\hat{\rho} := \frac{1}{2} |\hat{u}|^2$. In what follows, we obtain 
explicit expressions for $|P_1|^2$, $|Q_1|^2$, and $P_1 \bar{Q}_1$.

By substituting (\ref{3.6}) into (\ref{4.4}) and (\ref{4.4a}), we obtain
\begin{eqnarray*}
	&&	2 |\lambda_1|^2 |\lambda_1^2-\bar{\lambda}_1^2|^2 (|P_1|^4 + |Q_1|^4) \\
	&&  = 4 d - 4 a c + 2 a (\lambda_1^2 + \bar{\lambda}_1^2) + (4b + 4 |\lambda_1|^4 - 2 c (\lambda_1^2 + \bar{\lambda}_1^2)) R^2 - \frac{1}{2} (\lambda_1^2 + \bar{\lambda}_1^2) R^4
\end{eqnarray*}
and
\begin{eqnarray*}
	|\lambda_1|^4 |\lambda_1^2-\bar{\lambda}_1^2|^2 |P_1|^2 |Q_1|^2 & = &
	b^2 - cb (\lambda_1^2 + \bar{\lambda}_1^2) + b (\lambda_1^4 + \bar{\lambda}_1^4) + c^2 |\lambda_1|^4 - c |\lambda_1|^4 (\lambda_1^2 + \bar{\lambda}_1^2) + |\lambda_1|^8 \\ &&
	- \frac{1}{2} b (\lambda_1^2 + \bar{\lambda}_1^2) R^2
	+ c |\lambda_1|^4 R^2 - \frac{1}{2} |\lambda_1|^4 (\lambda_1^2 + \bar{\lambda}_1^2) R^2 + \frac{1}{4} |\lambda_1|^4 R^4.
\end{eqnarray*}
Combining these two expressions together yields
\begin{eqnarray*}
	&& 4 |\lambda_1|^2 |\lambda_1^2-\bar{\lambda}_1^2|^2 (|P_1|^2 + |Q_1|^2)^2 \\
	&& =
	8 d - 8 a c + 4 a (\lambda_1^2 + \bar{\lambda}_1^2)
	+ 8 |\lambda_1|^{-2} b [b - c (\lambda_1^2 + \bar{\lambda}_1^2) + (\lambda_1^4 + \bar{\lambda}_1^4)] \\
	&& + 8 c^2 |\lambda_1|^2 - 8 c |\lambda_1|^2 (\lambda_1^2 + \bar{\lambda}_1^2) + 8 |\lambda_1|^6 + (8b + 8 |\lambda_1|^4 - 4 c (\lambda_1^2 + \bar{\lambda}_1^2)) R^2 \\ 
	&& - 4 b |\lambda_1|^{-2} (\lambda_1^2 + \bar{\lambda}_1^2) R^2
	+ 8 c |\lambda_1|^2 R^2 - 4 |\lambda_1|^2 (\lambda_1^2 + \bar{\lambda}_1^2) R^2 \\
	&& -(\lambda_1^2 + \bar{\lambda}_1^2) R^4 + 2 |\lambda_1|^2 R^4.
\end{eqnarray*}
When $a$, $b$, $c$, and $d$ are expressed from (\ref{3.25}) and (\ref{configuration-1}), it shows that each term in the right-hand side has a common factor $(\lambda_1 - \bar{\lambda}_1)^2$. When the common factor is canceled, we obtain the following compact expression:
\begin{eqnarray*}
	4 |\lambda_1|^2 (\lambda_1 + \bar{\lambda}_1)^2 (|P_1|^2 + |Q_1|^2)^2 & = &
	(\lambda_1^2 + \bar{\lambda}_1^2 - \lambda_2^2 - \bar{\lambda}_2^2 + 2 |\lambda_1|^2 - 2 |\lambda_2|^2)^2 \\ &&
	+ 2 (\lambda_1^2 + \bar{\lambda}_1^2 + \lambda_2^2 + \bar{\lambda}_2^2 + 2 |\lambda_1|^2 + 2 |\lambda_2|^2) R^2 + R^4.
\end{eqnarray*}
Expressing $\lambda_1$ and $\lambda_2$ by (\ref{configuration-1}) and (\ref{3.27}) and using $\rho:= \frac{1}{2} R^2$,
we rewrite this expression in the following form:
\begin{equation}
\label{4.4-1}	
|\lambda_1|^2 (\lambda_1 + \bar{\lambda}_1)^2 (|P_1|^2 + |Q_1|^2)^2
= (\rho - u_3) (\rho - u_4).
\end{equation}
Similarly, we obtain
\begin{eqnarray*}
	&& 4 |\lambda_1|^2 |\lambda_1^2-\bar{\lambda}_1^2|^2 (|P_1|^2 - |Q_1|^2)^2 \\
	&& =
	8 d - 8 a c + 4 a (\lambda_1^2 + \bar{\lambda}_1^2)
	- 8 |\lambda_1|^{-2} b [b - c (\lambda_1^2 + \bar{\lambda}_1^2) + (\lambda_1^4 + \bar{\lambda}_1^4)] \\
	&&  - 8 c^2 |\lambda_1|^2 + 8 c |\lambda_1|^2 (\lambda_1^2 + \bar{\lambda}_1^2) - 8 |\lambda_1|^6 + (8b + 8 |\lambda_1|^4 - 4 c (\lambda_1^2 + \bar{\lambda}_1^2)) R^2 \\ &&
	+ 4 b |\lambda_1|^{-2} (\lambda_1^2 + \bar{\lambda}_1^2) R^2
	- 8 c |\lambda_1|^2 R^2 + 4 |\lambda_1|^2 (\lambda_1^2 + \bar{\lambda}_1^2) R^2 \\
	&&  -(\lambda_1^2 + \bar{\lambda}_1^2) R^4 - 2 |\lambda_1|^2 R^4,
\end{eqnarray*}
where each term in the right-hand side has now a common factor $(\lambda_1 + \bar{\lambda}_1)^2$.  When the common factor is canceled, we obtain the following compact expression:
\begin{eqnarray*}
	4 |\lambda_1|^2 (\lambda_1 - \bar{\lambda}_1)^2 (|P_1|^2 -|Q_1|^2)^2 & = &
	(\lambda_1^2 + \bar{\lambda}_1^2 - \lambda_2^2 - \bar{\lambda}_2^2 - 2 |\lambda_1|^2 + 2 |\lambda_2|^2)^2 \\ &&
	+ 2 (\lambda_1^2 + \bar{\lambda}_1^2 + \lambda_2^2 + \bar{\lambda}_2^2 - 2 |\lambda_1|^2 - 2 |\lambda_2|^2) R^2 + R^4,
\end{eqnarray*}
from which we obtain with the help of (\ref{configuration-1}) and (\ref{3.27}):
\begin{equation}
\label{4.4-2}	
|\lambda_1|^2 (\lambda_1 - \bar{\lambda}_1)^2 (|P_1|^2 - |Q_1|^2)^2
= (\rho - u_1) (\rho - u_2).
\end{equation}
By substituting (\ref{3.14})  and (\ref{3.13}) into (\ref{4.4-1}) and (\ref{4.4-2}) and extracting the square root, we derive the following expressions:
\begin{equation}\label{4.4-3-1}
|\lambda_1| (\lambda_1 + \bar{\lambda}_1) (|P_1|^2 + |Q_1|^2)
= \frac{(u_2-u_4)\sqrt{(u_1-u_3)(u_1-u_4)}{\rm dn}(\nu x;k)}
{(u_2-u_4)+(u_1-u_2){\rm sn}^2(\nu x;k)},
\end{equation}
and
\begin{equation}\label{4.4-3-2}
i |\lambda_1| (\lambda_1 - \bar{\lambda}_1) (|P_1|^2 - |Q_1|^2)
= \frac{(u_1-u_2)\sqrt{(u_1-u_4)(u_2-u_4)}{\rm sn}(\nu x;k){\rm cn}(\nu x;k)}
{(u_2-u_4)+(u_1-u_2){\rm sn}^2(\nu x;k)},
\end{equation}
where $\alpha_1$ and $\beta_1$ are assumed to be positive. The particular sign in (\ref{4.4-3-2}) has been chosen due to the following expression obtained from (\ref{4.4}):
	\begin{eqnarray*}
		i |\lambda_1|^2 (\lambda_1^2-\bar{\lambda}_1^2) (|P_1|^4 - |Q_1|^4)
		=	-\frac{d \rho}{d x},
	\end{eqnarray*}
	which implies that the sign of $i (\lambda_1^2-\bar{\lambda}_1^2) (|P_1|^4 - |Q_1|^4)$ is the same as the sign of ${\rm sn} (\nu x;k){\rm cn} (\nu x;k)$.

By adding and subtracting (\ref{4.4-3-1}) and  (\ref{4.4-3-2}), we obtain the desired expressions:
\begin{eqnarray}
\nonumber
&& 2 |\lambda_1| (\lambda_1 |P_1|^2+\bar{\lambda}_1|Q_1|^2) \\
&& = \sqrt{(u_1-u_4)(u_2-u_4)} \frac{2 \nu {\rm dn}(\nu x;k) - i(u_1-u_2)
	{\rm sn}(\nu x;k){\rm cn}(\nu x;k)}
{(u_2-u_4)+(u_1-u_2){\rm sn}^2(\nu x;k)}
\label{expression-1}
\end{eqnarray}
and
\begin{eqnarray}
\nonumber
&& 2 |\lambda_1| (\bar{\lambda}_1 |P_1|^2+\lambda_1|Q_1|^2) \\
&& =
\sqrt{(u_1-u_4)(u_2-u_4)} \frac{2 \nu {\rm dn}(\nu x;k) + i(u_1-u_2)
	{\rm sn}(\nu x;k){\rm cn}(\nu x;k)}
{(u_2-u_4)+(u_1-u_2){\rm sn}^2(\nu x;k)}.
\label{expression-2}
\end{eqnarray}

It follows from (\ref{4.4}) that
\begin{eqnarray*}
	&&	4 |\lambda_1|^2 |\lambda_1^2-\bar{\lambda}_1^2|^2 P_1^2 \bar{Q}_1^2 \\
	&& = -2 \left(\frac{dR}{dx}\right)^2  + 2 i \frac{dR}{dx} \left( (\lambda_1^2 + \bar{\lambda}_1^2 - c) R - \frac{1}{4} R^3 + \frac{a}{R} \right) \\
	&&
	+ 4 d - 4 a c + 2 a (\lambda_1^2 + \bar{\lambda}_1^2) + (4b + 4 |\lambda_1|^4 - 2 c (\lambda_1^2 + \bar{\lambda}_1^2)) R^2 - \frac{1}{2} (\lambda_1^2 + \bar{\lambda}_1^2) R^4,
\end{eqnarray*}
Substituting (\ref{3.25}) and (\ref{3.27})
and expressing $\rho = \frac{1}{2} R^2$ yield
\begin{eqnarray*}
	4 |\lambda_1|^2 |\lambda_1^2-\bar{\lambda}_1^2|^2 P_1^2 \bar{Q}_1^2
	& = & - \frac{1}{\rho} \left(\frac{d \rho}{dx}\right)^2  - i  \frac{1}{\rho}  \frac{d \rho}{dx} (\rho + \sqrt{u_1u_2}) (\rho - \sqrt{u_3u_4}) \\
	&& -(\lambda_1 + \bar{\lambda}_1)^2 (\rho - u_1) (\rho - u_2)
	-(\lambda_1 - \bar{\lambda}_1)^2 (\rho - u_3) (\rho - u_4).
\end{eqnarray*}
In order to see that the right-hand side is a complete square, we use  (\ref{3.27}) and write
\begin{eqnarray*}
	4 |\lambda_1|^2 |\lambda_1^2-\bar{\lambda}_1^2|^2 P_1^2 \bar{Q}_1^2
	& = & \frac{(u_1 - u_4) (u_2 - u_4) Z}{[u_1(u_2-u_4) + u_4 (u_1-u_2) {\rm sn}^2(\nu x;k)][(u_2-u_4) + (u_1-u_2) {\rm sn}^2(\nu x;k)]^3},
\end{eqnarray*}
where
\begin{eqnarray*}
	Z &=& -(u_1-u_3) (u_1-u_4) (u_1-u_2)^2 (u_2-u_4)^2  {\rm dn}^2(\nu x;k)
	{\rm sn}^2(\nu x;k) {\rm cn}^2(\nu x;k) \\
	&&
	+ i \sqrt{(u_1-u_3) (u_2-u_4)} (u_1-u_2)  {\rm dn}(\nu x;k)
	{\rm sn}(\nu x;k) {\rm cn}(\nu x;k) \\
	&& \times
	[ (u_1 + \sqrt{u_1u_2})(u_2-u_4) + (u_4+\sqrt{u_1u_2})(u_1-u_2)  {\rm sn}^2(\nu x;k)] \\
	&& \times [(u_1 - \sqrt{u_3u_4}) (u_2-u_4) + (u_4 - \sqrt{u_3u_4}) (u_1-u_2)  {\rm sn}^2(\nu x;k)] \\
	&& + \frac{1}{2} (\sqrt{-u_4} + \sqrt{-u_3})^2 (u_1-u_2)^2
	{\rm sn}^2(\nu x;k) {\rm cn}^2(\nu x;k) \\
	&& \times  [u_1(u_2-u_4) + u_4 (u_1-u_2) {\rm sn}^2(\nu x;k)][(u_2-u_4) + (u_1-u_2) {\rm sn}^2(\nu x;k)] \\
	&& + \frac{1}{2} (\sqrt{u_1} + \sqrt{u_2})^2 (u_1-u_3)(u_2-u_4) {\rm dn}^2(\nu x;k) \\
	&& \times [u_1(u_2-u_4) + u_4 (u_1-u_2) {\rm sn}^2(\nu x;k)][(u_2-u_4) + (u_1-u_2) {\rm sn}^2(\nu x;k)].
\end{eqnarray*}
Long but straightforward computations yield
\begin{eqnarray*}
	Z &=& \frac{1}{2}  (\sqrt{u_1} + \sqrt{u_2})^2 \left\{
	2\nu {\rm dn}(\nu x;k) [\sqrt{u_1} (u_2-u_4) {\rm cn}^2(\nu x;k) +
	\sqrt{u_2} (u_1-u_4) {\rm sn}^2(\nu x;k)] \right.\\
	&&  + i (\sqrt{u_1}-\sqrt{u_2}) {\rm sn}(\nu x;k){\rm cn}(\nu x;k)\times \\ && \left.
	[(u_1-\sqrt{u_3u_4}) (u_2-u_4) + (u_4-\sqrt{u_3u_4}) (u_1-u_2) {\rm sn}^2(\nu x;k)] \right\}^2.
\end{eqnarray*}
Extracting the negative square root yields the final expression: 
\begin{eqnarray}
2 |\lambda_1| (\lambda_1 + \bar{\lambda}_1) P_1\bar{Q}_1 = - \sqrt{(u_1 - u_4) (u_2 - u_4)}  \frac{\mathcal{N}_2}{\mathcal{D}_2},
\label{4.4-7}
\end{eqnarray}
where
\begin{eqnarray*}
	\mathcal{N}_2 & := & 2 \nu
{\rm dn}(\nu x;k) [\sqrt{u_1} (u_2-u_4) {\rm cn}^2(\nu x;k) +
\sqrt{u_2} (u_1-u_4) {\rm sn}^2(\nu x;k)] \\
&&  + i (\sqrt{u_1}-\sqrt{u_2}) {\rm sn}(\nu x;k){\rm cn}(\nu x;k)\\
&&
\times 
[(u_1-\sqrt{u_3u_4}) (u_2-u_4) + (u_4-\sqrt{u_3u_4}) (u_1-u_2) {\rm sn}^2(\nu x;k)]
\end{eqnarray*}
and
\begin{eqnarray*}
	\mathcal{D}_2 := \sqrt{[u_1(u_2-u_4) + u_4 (u_1-u_2) {\rm sn}^2(\nu x;k)][(u_2-u_4) + (u_1-u_2) {\rm sn}^2(\nu x;k)]^3}.
	\end{eqnarray*}
We chose the negative root by the continuity argument from the degenerate case $u_1 = u_2$ ($k=0$), for which the periodic wave (\ref{3.14}) becomes the constant-amplitude solution $\rho(x) = u_1$. Indeed, it follows from (\ref{4.4-3-1}) and  (\ref{4.4-3-2}) that if $u_1 = u_2$ ($k=0$), then
	\begin{equation*}
	2|\lambda_1| (\lambda_1 + \bar{\lambda}_1) |P_1|^2 =	2|\lambda_1| (\lambda_1 + \bar{\lambda}_1) |Q_1|^2 = \sqrt{(u_1 - u_3)(u_1 - u_4)} \label{4.4a-4}
	\end{equation*}
	which coincides with the expression
	\begin{equation*}
	2|\lambda_1| (\lambda_1 + \bar{\lambda}_1) P_1 \bar{Q}_1 = -\sqrt{(u_1 - u_3)(u_1 - u_4)}.
	\label{4.3a-4}
	\end{equation*}
	obtained from (\ref{4.4-7}) with the negative sign.
	Note that $P_1 = -Q_1$ is $x$-independent in the degenerate case.

Combining together (\ref{expression-2}) and (\ref{4.4-7}) yields
\begin{eqnarray*}
	&&	\rho -
	\frac{i (\lambda_1^2 - \bar{\lambda}_1^2) R P_1\bar{Q}_1}
	{\bar{\lambda}_1 |P_1|^2 + \lambda_1 |Q_1|^2} \\
	&& =
	\frac{-\sqrt{u_1 u_2} \sqrt{(u_1-u_3) (u_2-u_4)} {\rm dn} (\nu x;k) + i \sqrt{u_3 u_4} (u_1 - u_2) {\rm sn} (\nu x;k) {\rm cn} (\nu x;k)} {\sqrt{(u_1-u_3) (u_2-u_4)} {\rm dn} (\nu x;k) + i (u_1 - u_2) {\rm sn} (\nu x;k) {\rm cn} (\nu x;k)}.
\end{eqnarray*}
Substituting into (\ref{expr-app-E}) yields 
\begin{eqnarray}
\hat{\rho}(x) &=& \frac{u_1 u_2 (u_1-u_3) (u_2-u_4) {\rm dn}^2 (\nu x;k) + u_3 u_4 (u_1 - u_2)^2 {\rm sn}^2 (\nu x;k) {\rm cn}^2 (\nu x;k)} {\rho \left[ (u_1-u_3) (u_2-u_4) {\rm dn}^2 (\nu x;k) + (u_1 - u_2)^2 {\rm sn}^2 (\nu x;k) {\rm cn}^2 (\nu x;k) \right]} \nonumber \\
&=& \frac{u_2 (u_1-u_3) - u_3 (u_1 - u_2) {\rm sn}^2 (\nu x;k)}{ (u_1-u_3) - (u_1 - u_2) {\rm sn}^2 (\nu x;k)},
\end{eqnarray}
which is nothing but (\ref{5.10}).

\section{Proof of (\ref{3.3.7})}
\label{app-CC}

Substituting parameters ($a,b,c,d$) from (\ref{3.25}), 
the periodic wave $\rho$ from (\ref{3.20}), and the eigenvalue 
$\beta_3$ from (\ref{a3.16})
into (\ref{b7.6}) yields
\begin{equation}\label{3.4.2}
\hat{\rho} = \frac{\mathcal{N}_3}{\mathcal{D}_3},
\end{equation}
where
\begin{eqnarray}
\mathcal{D}_3 &=& \left[ (b+ c \beta_3^2 + \beta_3^4)[1+ \delta + (\delta -1){\rm cn} (\mu x;k)] \right. \nonumber \\
&& \left. 
+ \beta_3^2[u_1\delta + u_2 +(u_1\delta - u_2){\rm cn} (\mu x;k)]\right]^2 \label{3.4.3}
\end{eqnarray}
and
\begin{eqnarray}
\mathcal{N}_3 &=& [\beta_3^2 a (b - c \beta_3^2 - \beta_3^4) + 2 d \beta_3^4] [1+ \delta + (\delta -1){\rm cn} (\mu x;k)]^2 \nonumber\\&& +
[b^2 +\beta_3^4(2b + a - c^2) + \beta_3^8][1+ \delta + (\delta -1){\rm cn} (\mu x;k)]\nonumber \\&& \times
[u_1\delta + u_2 +(u_1\delta - u_2){\rm cn} (\mu x;k)] \nonumber\\&&
+\beta_3^2 (b - c \beta_3^2 - \beta_3^4)[u_1\delta + u_2 +(u_1\delta - u_2){\rm cn} (\mu x;k)]^2 
\label{3.3.4}
\end{eqnarray}

After canceling the common factors in $\mathcal{D}_3$ and $\mathcal{N}_3$, we arrive at the expression
\begin{eqnarray}
\hat{\rho} = \frac{1}{4} \left(\sqrt{u_1} + \sqrt{u_2} -
	\sqrt{2(\gamma+ \sqrt{\gamma^2+\eta^2})}\right)^2  + \frac{A (1 - {\rm cn} (\mu x;k))}{B (1 - {\rm cn} (\mu x;k)) + C (1 + {\rm cn} (\mu x;k))},
\label{3.3.5}
\end{eqnarray}
where
\begin{eqnarray*}
	A & = & \sqrt{2(\gamma+ \sqrt{\gamma^2+\eta^2})}\left(\sqrt{u_1} + \sqrt{u_2}\right)
	\left(u_2 - \sqrt{2u_2(\gamma+ \sqrt{\gamma^2+\eta^2})} + \sqrt{\gamma^2 + \eta^2}\right), \\
	B & = & u_2 - \sqrt{2u_2(\gamma+ \sqrt{\gamma^2+\eta^2})} + \sqrt{\gamma^2 + \eta^2}, \\
	C & = & \delta\left(u_1 + \sqrt{2u_1(\gamma+ \sqrt{\gamma^2+\eta^2})} + \sqrt{\gamma^2 + \eta^2}\right).
\end{eqnarray*}
This expression yields (\ref{3.3.7}) with (\ref{3.3.6}) and (\ref{3.3.6a}).


\section{Proof of (\ref{a7.5})}
\label{app-F}

The explicit expressions for $\lambda_1 |P_1|^2 + \bar{\lambda}_1 |Q_1|^2$ and $P_1 \bar{Q}_1$ at the periodic wave (\ref{3.20}) are obtained similarly to the derivation explained in the Appendix \ref{app-E}.
By using \eqref{3.20}-\eqref{3.18-1}, \eqref{a3.16}, \eqref{4.4-1}, and \eqref{4.4-2}, we obtain
\begin{eqnarray*}
	|\lambda_1|^2 (\lambda_1 + \bar{\lambda}_1)^2 (|P_1|^2 + |Q_1|^2)^2 =
	\frac {4[(u_2-\gamma)^2+\eta^2]{\rm dn}^2 (\mu x;k)} {[1+\delta +(\delta-1){\rm cn} (\mu x;k)]^2},
\end{eqnarray*}
and
\begin{eqnarray*}
	|\lambda_1|^2 (\lambda_1 - \bar{\lambda}_1)^2 (|P_1|^2 - |Q_1|^2)^2 =
	\frac {-\delta (u_1 - u_2)^2 {\rm sn}^2 (\mu x;k)} {[1+\delta +(\delta-1){\rm cn} (\mu x;k)]^2},
\end{eqnarray*}
which result in
\begin{equation}\label{a6.15}
|\lambda_1| (\lambda_1 + \bar{\lambda}_1) (|P_1|^2 + |Q_1|^2)
=\frac {2 \sqrt {(u_2-\gamma)^2+\eta^2} {\rm dn} (\mu x;k)} {1+\delta +(\delta-1){\rm cn} (\mu x;k)},
\end{equation}
and
\begin{equation}\label{a6.16}
i |\lambda_1| (\lambda_1 - \bar{\lambda}_1) (|P_1|^2 - |Q_1|^2)
=\frac {\sqrt{\delta} (u_1 - u_2) {\rm sn} (\mu x;k)} {1+\delta +(\delta-1){\rm cn} (\mu x;k)},
\end{equation}
where we have extracted the negative squared root analogous to the equation \eqref{4.4-3-2}.
With the help of \eqref{a6.15} and \eqref{a6.16}, we come to the required expressions for the periodic standing waves \eqref{3.20}
\begin{equation}\label{a6.17}
2 |\lambda_1| (\lambda_1 |P_1|^2+\bar{\lambda}_1|Q_1|^2) =
\frac {2 \sqrt {(u_2-\gamma)^2+\eta^2} {\rm dn} (\mu x;k) - i \sqrt{\delta} (u_1 - u_2) {\rm sn} (\mu x;k)}
{1+\delta +(\delta-1){\rm cn} (\mu x;k)},
\end{equation}
and
\begin{equation}\label{a6.18}
2 |\lambda_1| (\bar{\lambda}_1 |P_1|^2+\lambda_1|Q_1|^2) =
\frac {2 \sqrt {(u_2-\gamma)^2+\eta^2} {\rm dn} (\mu x;k) + i \sqrt{\delta} (u_1 - u_2) {\rm sn} (\mu x;k)}
{1+\delta +(\delta-1){\rm cn} (\mu x;k)}.
\end{equation}

On the other hand, we obtain from \eqref{4.4} after a lengthy calculation:
\begin{eqnarray}
	\sqrt{2} |\lambda_1| |\lambda_1^2-\bar{\lambda}_1^2| P_1 \bar{Q}_1
	= -\frac{\mathcal{N}_4}{\mathcal{D}_4}, 
	\label{a6.19}
\end{eqnarray}
where 
\begin{eqnarray*}
	\mathcal{N}_4 &=& {\rm dn} (\mu x;k)
	(\sqrt{u_1} + \sqrt{u_2}) \sqrt{(u_2 - \gamma)^2 + \eta^2} [(\delta \sqrt{u_1} - \sqrt{u_2}) {\rm cn} (\mu x;k) +  \delta \sqrt{u_1} + \sqrt{u_2}] \\
		&& + \frac {i}{2} (u_1 - u_2) {\rm sn} (\mu x;k) \sqrt{\delta} \left[
	u_2 + u_1 \delta - (1 + \delta) \sqrt{\gamma^2 +\eta^2} \right. \\ 
		&& 
	\left.  + {\rm cn} (\mu x;k) \left[ u_1 \delta - u_2 +(1- \delta) \sqrt{\gamma^2 +\eta^2} \right] \right],\\
	\mathcal{D}_4 &=& [1+\delta +(\delta-1){\rm cn} (\mu x;k)]^{\frac{3}{2}} [u_2 (1 - {\rm cn} (\mu x;k)) + u_1 \delta (1 + {\rm cn} (\mu x;k))]^{\frac{1}{2}},
\end{eqnarray*}
where we have extracted the negative squared root thanks to the same reason as \eqref{4.4-7}.

By applying (\ref{a6.18}) and (\ref{a6.19}), we obtain
\begin{eqnarray*}
	\rho -
	\frac{i (\lambda_1^2 - \bar{\lambda}_1^2) R P_1\bar{Q}_1}
	{\bar{\lambda}_1 |P_1|^2 + \lambda_1 |Q_1|^2} =
	\frac {-2 \sqrt{u_1u_2[(u_2 - \gamma)^2 + \eta^2]} {\rm dn} (\mu x;k) + i (u_1- u_2) \sqrt {\delta(\gamma^2 + \eta^2)} {\rm sn} (\mu x;k)} {2 \sqrt{(u_2 - \gamma)^2 + \eta^2} {\rm dn} (\mu x;k) + i \sqrt{\delta} (u_1- u_2) {\rm sn} (\mu x;k)},
\end{eqnarray*}
which yields
\begin{eqnarray}
\hat{\rho}(x) &=& \frac {4 u_1u_2[(u_2 - \gamma)^2 + \eta^2] {\rm dn}^2 (\mu x;k) +
	\delta (u_1- u_2)^2 (\gamma^2 + \eta^2) {\rm sn}^2 (\mu x;k)} {4 [(u_2 - \gamma)^2 + \eta^2] {\rm dn}^2 (\mu x;k) + \delta (u_1- u_2)^2 {\rm sn}^2 (\mu x;k)} \nonumber \\
&=& \frac {u_1 \delta (1 - {\rm cn} (\mu x;k)) + u_2 (1 + {\rm cn} (\mu x;k))} {1 + {\rm cn} (\mu x;k) +
	\delta (1 - {\rm cn} (\mu x;k))}, \nonumber
\end{eqnarray}
and hence the explicit expression (\ref{a7.5}).

\section{Proof of (\ref{hat-rho-max}) and (\ref{hat-rho-max-2})}
\label{app-G}

We select $c_1 = 0$ so that $\chi_1(0,0) = 0$ in (\ref{chi}). Resorting to \eqref{3.25}, \eqref{3.14}, \eqref{3.27a}, \eqref{4.2}, \eqref{4.4}, and \eqref{4.4a}, we have
\begin{eqnarray}
|\hat{u}(0,0)| &=& \left| u(0,0) - 2i \beta_1 \frac{p_1(0,0)}{\bar{p}_1(0,0)} \right| \nonumber \\
&=& \left| \frac{-i \beta^2_1 \rho'(0) + 3 \beta_1^2 \rho^2(0) + 2 \rho(0) (3 \beta_1^4 + 2 c \beta_1^2 +b) - a \beta_1^2}
{\sqrt{2 \rho(0)}(b + c \beta_1^2 + \beta_1^4 + \beta_1^2 \rho(0))}
\right|, \label{G.1}
\end{eqnarray}
which yields (\ref{hat-rho-max}) since $\rho(0)= u_1$, $\rho'(0) = 0$, and $\hat{\rho} = \frac{1}{2} |\hat{u}|^2$.

Since the expression \eqref{G.1} applies to other purely imaginary eigenvalues, 
we can replace $\rho(0)= u_1$ and $\beta_1$ with $\rho(0)= u_3$ and $\beta_3$ in \eqref{G.1} and obtain \eqref{hat-rho-max-2}.

\section{Proof of (\ref{4.7b}) and (\ref{4.7b-2})}
\label{app-I}

For the derivation of (\ref{4.7b}), it follows from (\ref{3.25}) and (\ref{3.27a}) that 
\begin{equation*}
\beta^2_1(d - a c - a \beta_1^2) = (b + c \beta_1^2 + \beta_1^4)^2.
\end{equation*}
Substituting this relation into (\ref{4.7a}) yields
\begin{eqnarray}
\frac{\partial \chi_1}{\partial t} &=& \frac{2 \beta_1^2 (\beta_1^2-\beta_2^2) (\beta_1^2 d - a \beta_1^2 (c+\beta_1^2) + 2 \beta_1^2 (c \beta_1^2 + \beta_1^4 + b) \rho + \beta_1^4 \rho^2)}
{(b + c \beta_1^2 + \beta_1^4 + \beta_1^2 \rho)^2} \nonumber \\
&=& \frac{2 \beta_1^2 (\beta_1^2-\beta_2^2) \left[(b + c \beta_1^2 + \beta_1^4)^2 + 2 \beta_1^2 (c \beta_1^2 + \beta_1^4 + b) \rho + \beta_1^4 \rho^2\right]}
{(b + c \beta_1^2 + \beta_1^4 + \beta_1^2 \rho)^2} \nonumber \\
&=& 2 \beta_1^2 (\beta_1^2-\beta_2^2), \label{I.2}
\end{eqnarray}
which coincides with \eqref{4.7b}.

For the derivation of (\ref{4.7b-2}), we substitute 
(\ref{3.25}), (\ref{4.4}), and (\ref{4.4a}) into (\ref{z-4.7-1}) and obtain
\begin{eqnarray}
\frac{\partial \chi_1}{\partial t} = \frac{\mathcal{N}_5}{\mathcal{D}_5},
\label{z-4.7-2}
\end{eqnarray}
where
\begin{eqnarray*}
	\mathcal{D}_5 &=& (|p_1|^2+|q_1|^2)^2, \\
	\mathcal{N}_5 &=& \frac{4}{\bar{\lambda}^2_1}(b - c \bar{\lambda}_1^2 + \bar{\lambda}_1^4 - \bar{\lambda}_1^2 \rho)
(\rho + c - \lambda_1^2 - \bar{\lambda}_1^2)  \\ 
&& + \frac{1}{\bar{\lambda}_1(\lambda_1 + \bar{\lambda}_1)}[2(\lambda_1^2 + \bar{\lambda}_1^2 +|\lambda_1|^2)
(\rho^2 + 2 (c - 2 \bar{\lambda}_1^2)\rho -a) + 2 \bar{\lambda}_1^2 \rho^2  \\ 
&& + 4(c \bar{\lambda}_1^2 - 2b)\rho + 4ac - 2 a \bar{\lambda}_1^2 - 4d].
\end{eqnarray*}
It follows from \eqref{4.4-3-1} that
\begin{equation}\label{I.3}
\mathcal{D}_5 =\frac {2 (u_2 - u_4)^2(u_1 - u_4)(u_1 - u_3){\rm dn}^2 (\nu x;k)}{|\lambda_1|^2(\sqrt{-u_3}+\sqrt{-u_4})^2}.
\end{equation}
On the other hand, it follows from \eqref{3.14} and \eqref{3.27} that 
\begin{eqnarray}
\mathcal{N}_5 &=&
\frac{2i (\sqrt{u_1}+\sqrt{u_2})(u_1-u_4)(u_2 - u_4)(\sqrt{u_1}+\sqrt{u_2} - i(\sqrt{-u_3}+\sqrt{-u_4}))}{(\sqrt{-u_3}+\sqrt{-u_4})(\sqrt{u_1}+\sqrt{u_2} +i(\sqrt{-u_3}+\sqrt{-u_4}))}  \nonumber \\ &&  \times \left[ u_2 u_3 {\rm cn}^2 (\nu x;k) + u_4 (u_2 {\rm sn}^2 (\nu x;k) - u_3)
- u_1( u_2 - u_3 {\rm sn}^2 (\nu x;k) - u_4 {\rm cn}^2 (\nu x;k))\right] \nonumber \\ &=&
\frac{2i (\sqrt{u_1}+\sqrt{u_2})(u_1-u_4)(u_2 - u_4)^2(\sqrt{u_1}+\sqrt{u_2} - i(\sqrt{-u_3}+\sqrt{-u_4}))}{(\sqrt{-u_3}+\sqrt{-u_4})(\sqrt{u_1}+\sqrt{u_2} +i(\sqrt{-u_3}+\sqrt{-u_4}))}  \nonumber \\ && 
\times (u_1-u_3)\left[ 1+ \frac{(u_1 - u_2)(-u_3 + u_4)}
{(u_1 - u_3)(u_2 - u_4)} {\rm sn}^2 (\nu x;k)\right] \nonumber \\ &=&
\frac{2i (\sqrt{u_1}+\sqrt{u_2})(u_1 - u_3)(u_1-u_4)(u_2 - u_4)^2}{(\sqrt{-u_3}+\sqrt{-u_4})(\sqrt{u_1}+\sqrt{u_2} +i(\sqrt{-u_3}+\sqrt{-u_4}))} \nonumber \\&&
\times (\sqrt{u_1}+\sqrt{u_2} - i(\sqrt{-u_3}+\sqrt{-u_4})){\rm dn}^2 (\nu x;k). \label{I.4}
\end{eqnarray}
A simple calculation for $\mathcal{N}_5/\mathcal{D}_5$ via \eqref{I.3} and \eqref{I.4} yields \eqref{4.7b-2}.

\section{Proof of (\ref{max-rogue1}) and (\ref{max-rogue2})}
\label{app-J}

We select $c_1 = 0$ so that $\chi_1(0,0) = 0$ in \eqref{chi-2}. Resorting to \eqref{5.8}, \eqref{4.2}, \eqref{z-4.5-1}, we have
\begin{eqnarray}
|\hat{u}(0,0)| = \left| R(0) +
\frac{2i (\lambda_1^2 - \bar{\lambda}_1^2) P_1(0,0) \bar{Q}_1(0,0)}
{\bar{\lambda}_1 |P_1(0,0)|^2 + \lambda_1 |Q_1(0,0)|^2} \right|. \label{J.1}
\end{eqnarray}
It follows from \eqref{expression-1} and \eqref{4.4-7} that
\begin{equation}\label{J.2}
2 |\lambda_1| (\lambda_1 |P_1(0,0)|^2+\bar{\lambda}_1|Q_1(0,0)|^2)= \sqrt{(u_1-u_3)(u_1-u_4)}
\end{equation}
and
\begin{equation}\label{J.2a}
2 |\lambda_1| (\lambda_1 + \bar{\lambda}_1) P_1(0,0) \bar{Q}_1(0,0) = -\sqrt{(u_1-u_3)(u_1-u_4)}.
\end{equation}
Substituting $\lambda_1$ from \eqref{3.27}, $R(0) =\sqrt{2u_1}$, as well as relations \eqref{J.2} and (\ref{J.2a}) into \eqref{J.1} yields \eqref{max-rogue1}. 

Repeating computations for eigenvalue $\lambda_2$ in \eqref{3.27} yields
\begin{equation}\label{J.3}
2 |\lambda_2| (\lambda_2 |P_2(0,0)|^2+\bar{\lambda}_2|Q_2(0,0)|^2)= \sqrt{(u_1-u_3)(u_1-u_4)}
\end{equation}
and
\begin{equation}\label{J.4}
2 |\lambda_2| (\lambda_2 + \bar{\lambda}_2) P_2(0,0)\bar{Q}_2(0,0) = -\sqrt{(u_1-u_3)(u_1-u_4)}
\end{equation}
which yields \eqref{max-rogue2}.
\end{document}